\documentclass[12pt]{article}
\usepackage{amsfonts,amsmath,amssymb,graphicx,color} 
\newcommand{\kar}[2]{
 \begin{minipage}{25.mm}
 \begin{center}
 \includegraphics[scale=0.35]{#1}\\
 #2 \\ \hspace*{.5mm}
 \end{center}
 \end{minipage}
 }

\newcommand{\beq}[1]{
\marginpar{\small\textsf{#1}}
\begin{equation}\label{#1}}
\newcommand{\eeq}{\end{equation}}
\newcommand{\bea}[1]{
\marginpar{\small\textsf{#1}}
\begin{eqnarray}\label{#1}}
\newcommand{\eea}{\end{eqnarray}}
\begin{document}

\begin{titlepage}
\begin{center}
{\Large \bf
Exclusive photoproduction of a heavy vector meson  
in 
QCD
}
\vspace{1 cm}
\par

{\sc
D.~Yu.~Ivanov
}$^{1}$,
{\sc
A.~Sch\"afer
}$^{2}$,
{\sc
L.~Szymanowski
}$^{3}$
and
{\sc
G.~Krasnikov
}$^{2,4}$
\\[0.5cm]
\vspace*{0.1cm} ${}^1$ {\it
Institute of Mathematics, 630090 Novosibirsk, Russia
                       } \\[0.2cm]
\vspace*{0.1cm} ${}^2${\it
Institut f\"ur Theoretische Physik, Universit\"at
Regensburg, \\ D-93040 Regensburg, Germany
                       } \\[0.2cm]
\vspace*{0.1cm} ${}^3$ {\it
Soltan Institute for Nuclear Studies,
Hoza 69,\\ 00-681 Warsaw, Poland
                       } \\[0.2cm]
\vspace*{0.1cm} ${}^4$ {\it
Department of Theoretical Physics, St.Petersburg State University, \\
198904,
St. Petersburg, Russia
                       } \\[1.0cm]
\vspace{0.6cm}
\bigskip
\centerline{\large \em
\today
}
\bigskip
\vskip1.2cm
{\bf
Abstract:
\\[10pt]} \parbox[t]{\textwidth}{
The process of exclusive heavy vector
meson photoproduction, $\gamma p \to V p$, is studied
in the framework of QCD factorization. The mass of the produced
meson, $V=\Upsilon$ or $J/\Psi$, provides a hard scale for the process.
We demonstrate, that in the heavy quark limit and at the
one-loop order in perturbation theory, the amplitude factorizes in a
convolution of a perturbatively calculable hard-scattering amplitude with
the generalized parton densities and the nonrelativistic
QCD  matrix element $\langle O_1\rangle_{V}$. We evaluate the hard
scattering amplitude at one-loop order and compare the data with theoretical
predictions using an available model for generalized parton distributions.
}
\end{center}
\end{titlepage}

\section{Introduction}
\setcounter{equation}{0}

The process of elastic production of heavy quarkonium in 
photon-proton collisions,  
\begin{equation}
\gamma p \to V p  \,\, , \quad \mbox{where} \quad 
V=J/\Psi \,\, \mbox{or} \,\, \Upsilon \, ,
\label{process}
\end{equation}
was studied in fix target \cite{Binkley:1981kv,Denby:1983az} 
and in HERA collider experiments both for the
case of a real photon in the initial state (photoproduction) 
\cite{Aid:1996dn,Breitweg:1997rg,Breitweg:1998ki,Adloff:2000vm,
Chekanov:2002xi}
and for the
case when the meson is produced by a highly virtual photon
(electroproduction) \cite{Aid:1996ee,Breitweg:1998nh}. 
The primary motivation for the strong interest in this process (and in the
similar process of light vector meson electroproduction) is that 
it can potentially serve to constrain the gluon density in a proton.
On the theoretical side, the large mass of the heavy quarks provides a hard
scale for the process which justifies the application of QCD factorization 
methods that allow to separate the contributions to the amplitude 
coming from different scales. 

The first step in this direction was
made by M. Ryskin \cite{Ryskin:1992ui} who expressed the amplitude 
of exclusive heavy meson production in terms of the gluon density 
and, accordingly, 
predicted that the cross section, which is proportional to the square 
of the gluon density, grows fastly with energy. 
Electroproduction of light vector mesons was studied later in
\cite{Brodsky:1994kf}, where it was shown that in this case 
the amplitude factorizes in terms
of a perturbative hard scattering coefficient function 
and nonperturbative quantities:
a meson distribution amplitude and a gluon density in a proton. 
Again,  an increase of the cross section with energy was predicted.  
The data from HERA appear to be in accord with 
these predictions.   

The early approaches to factorization 
in exclusive vector meson 
production \cite{Ryskin:1992ui,Brodsky:1994kf} were based on the use
of leading double $\ln\left(1/x\right)\ln Q^2$ approximation 
and were designed for the
description of the process at high energies 
(in the diffractive or small $x$ kinematics). 
Later on it was understood that in the scaling limit, 
$Q^2\to \infty$ and $x=Q^2/W^2$ fixed,
deeply virtual meson electroproduction \cite{Radyushkin:1996ru} 
and Compton scattering (DVCS) \cite{Ji:1996ek,Ji:1996nm,Radyushkin:1996nd}
processes may be studied within the QCD collinear factorization method. 
The proof of factorization for meson electroproduction was provided
in \cite{Collins:1996fb}.
Due to nonvanishing momentum transfer in the $t-$ channel the amplitude 
of this 
deeply virtual exclusive process factorizes in terms of 
generalized parton distributions (GPDs) rather than the ordinary 
parton densities which enter the QCD description 
of inclusive deep inelastic scattering and the other hard inclusive processes.
GPDs extend the forward parton distributions and the nucleon
electromagnetic formfactors to the nonforward kinematics of 
the electroproduction processes, they encode much richer 
information about the dynamics 
of a nucleon than the conventional parton distributions.
This additional information can be presented e.g. in terms of
spacial distributions of energy, spin ... within a nucleon
\cite{Burk,Diehl,Ji}. 
By now the studies of deeply virtual exclusive processes and GPDs
have developed in a very dynamical field, for recent reviews 
see, \cite{Goeke:2001tz,Diehl:2003ny}.  
  
Another  QCD approach to exclusive meson production at high energies
is related to $k_\perp-$ (or high energy) factorization
\cite{Catani:1990eg,Collins:1991ty}, it is based on the BFKL
method \cite{Kuraev:fs,Balitsky:ic}. 
In this scheme  large logarithms of energy 
$\ln\left(1/x\right)$ are resumed and amplitudes
are given by an overlap integral of the $k_\perp$ dependent
(unintegrated) gluon density and the hard scattering kernel.
High energy factorization can 
be formulated  also in terms of color 
dipoles \cite{Nikolaev:1990ja,Mueller:1993rr}.
Although these approaches to hard diffractive processes 
are very promising, their firm foundation, unfortunately, remains
limited due to the use of the leading $\ln\left(1/x\right)$ approximation.
A few years ago the BFKL formalism was extended to the next-to-leading
order \cite{Fadin:1998py}, 
but the generalization of the $k_\perp-$ factorization
scheme or the dipole approach to this order remains still 
a matter of debate.
An extended overview of different approaches to this problem may be found in 
\cite{Andersson:2002cf}.

Most of the theoretical studies 
\cite{Nemchik:1994fp,Ryskin:1995hz,Frankfurt:1997fj,Martin:1999rn,
Frankfurt:1998yf,Hufner:2000jb,Caldwell:2001ky,Gotsman:2001ic,Ma:2001yf}
of process (\ref{process}) were 
performed in the framework of $k_\perp -$ factorization or dipole 
approaches. Despite of great progress and evident success in the description
of the data the theoretical uncertainties remain poorly understood.
In particular, it is believed that the account of skewedness, i.e. the effect
of different parton momentum fractions, is very important for the
kinematic range available in the experiment. But since this effect is 
beyond leading $\ln\left(1/x\right)$, its model independent 
implementation into $k_\perp -$ factorization scheme or dipole formalism 
remains a challenge for theory.
      
In this paper we study process (\ref{process}) in the heavy quark limit 
in the collinear factorization approach. The physics behind 
collinear factorization is the separation of scales. 
The mass of the heavy quark, $m$, provides a hard scale. A photon
fluctuates into the heavy quark pair at small transverse distances $\sim 1/m$,
which are much smaller than the ones $\sim 1/\Lambda$ related to 
any nonperturbative hadronic scale $\Lambda$.
We will show by explicit calculation that to leading 
power in $1/m$ counting and  one--loop order in perturbation 
theory the amplitude is given by the convolution of the perturbatively            
calculable hard scattering amplitude and nonperturbative quantities.
The latter are
gluon and quark GPDs and the nonrelativistic QCD (NRQCD)
\cite{Bodwin:1994jh} matrix element $\langle O_1\rangle_V$ which 
parametrizes in our case an essential nonrelativistic dynamics of a heavy meson 
system. 
This means that two firmly 
founded QCD approaches, namely collinear factorization 
and NRQCD, can be combined to construct a model free 
description of heavy meson photoproduction which is free of any 
high energy approximation and may be used also in the kinematic domain 
where the energy of the photon nucleon collision, $W$, is of order of the 
meson mass, $M$. We evaluate the hard scattering amplitude at 
next-to-leading order. This allows to reduce the scale 
dependence, which is especially 
important at high energies, since in this case (i.e. in the  
small $x$ region) the 
dependence of the gluon distribution on the scale is very strong.

The factorization theorem \cite{Collins:1996fb} for meson electroproduction
expresses the amplitude in a form containing a meson light-cone distribution
amplitude. Its application to the production of a heavy meson is restricted to
the region of very large virtualities, $Q^2\gg m^2$, where the mass of the  
heavy quark may be completely neglected. In contrast, 
in photoproduction 
or electroproduction at moderate virtualities the heavy quark 
mass provides a hard scale
and the nonrelativistic nature of heavy meson is important. 
In this case, according to NRQCD which 
provides a systematic nonrelativistic expansion, a factorization formalism 
must be constructed in terms of matrix elements of 
NRQCD operators. They are characterized by their different scaling 
behavior with respect to $v$, the typical 
velocity of the heavy quark. In the leading approximation only the  
matrix element $\langle O_1
\rangle_V$ contributes, which describes in NRQCD the leptonic meson decay
rate \cite{Bodwin:1994jh}
\begin{equation}
\Gamma[V\to l^+l^-]=\frac{2e_q^2\pi\alpha^2}{3}
\frac{\langle O_1\rangle_V }{m^2} 
\left( 1-\frac{8\alpha_S}{3\pi}\right)^2 \, .
\label{decay}
\end{equation}
Here $\alpha$ is the fine-structure constant and $m$ and $e$ are
the pole mass and the electric charge of the heavy quark ($e_c=2/3$,
$e_b=-1/3$). Equation (\ref{decay}) includes the one-loop
QCD correction \cite{Barbieri:1975ki} and $\alpha_S$ is the strong coupling
constant.

The leading relativistic correction to the meson decay rate and to
the photoproduction process (\ref{process}) scales 
$\sim \langle  v^2 \rangle$, see \cite{Bodwin:1994jh}.
It is expressed through the matrix element of an additional NRQCD operator.
Since for a nonrelativistic Coulomb system $v\sim \alpha_S$, the
relativistic effect is less important than the one-loop perturbative 
correction.
The relativistic correction ($\sim \langle v^2 \rangle $) to the result 
\cite{Ryskin:1992ui}
for heavy meson production was studied in \cite{Hoodbhoy:1996zg}, 
see also \cite{Vanttinen:1998en}.
Despite the fact
that $\langle v^2 \rangle_{J/\Psi}\sim 0.2\div 0.25$, the
relativistic effect was found to be rather small. On the cross
section level it amounts to 
$7\%$ for $J/\Psi$ and it should be even smaller for $\Upsilon$
production. 

We will neglect relativistic corrections and consider the process 
(\ref{process}) in  leading order of the relativistic expansion. 
In this case 
all essential information about the 
quarkonium structure is encoded
in one NRQCD matrix element. 
In potential models it can be related to the value of the radial wave
function at the origin,
\begin{equation}
\langle O_1 \rangle_V=\frac{N_c}{2\pi}|R_S(0)|^2+{\cal O}(v^2) \ ,
\label{poten}
\end{equation}
here $N_c=3$ for QCD. Due to the relation to potential
models this scheme of calculation is often called in the literature the
static or non-relativistic approximation. However, 
one should notice that 
using NRQCD it can be improved in a systematic and rigorous way
calculating relativistic and perturbative corrections. In this paper we will
concentrate on the one-loop perturbative correction. 
Our main result is that
for this process the collinear factorization method is compatible at
one-loop level with the relativistic expansion. This allows us to obtain
unambiguous predictions. We found that QCD corrections are large.
They change
not only the overall normalization but may affect, also, the predictions for the
dependence of the cross section on energy.

Our presentation is organized as follows, In Section 2 we introduce the
notations, discuss the factorization procedure and give the predictions for 
the amplitude in  leading order (LO). Section 3 is devoted to the detailed
derivation of the hard-scattering amplitude at  next-to-leading order 
(NLO). Our method is similar to the one 
we used recently \cite{ISz} 
for the calculation of light vector meson
electroproduction in NLO. It is based on the use of dispersion relations and 
the low energy theorem for the radiation of a soft gluon, the non-abelian
generalization of the theorem known in QED \cite{Low:1958sn}. 
In Section 4 we present a  numerical analysis.
In the concluding section we summarize and  discuss 
open questions. 

\section{Factorization and the amplitude at LO}

The kinematics of heavy vector meson photoproduction is shown in Fig. 1.
The momenta of the incoming photon, incoming nucleon, outgoing nucleon
and the produced meson 
are $q$, $p$, $p^\prime$ and $K$, respectively. 
In the leading order of the 
relativistic expansion  the meson mass can be taken
as  twice the
heavy quark pole mass, $K^2=M^2$ and $M=2m$.
The photon and nucleon are on the
mass shell, 
$q^2=0$,
$p^2=p^{\prime 2}=m_N^2$, where $m_N$ is the proton mass.
The photon
polarization is described by the vector
$e_\gamma$, $(e_\gamma q)=0$. 
The invariant c.m. energy is $s_{\gamma p}=(q+p)^2=W^2$. We define
\begin{eqnarray}
&&
\Delta=p^\prime -p \, , \ \ P=\frac{p+p^\prime}{2} \, , \ \ t=\Delta^2 \, ,
\nonumber \\
&&
(q-\Delta )^2=K^2=M^2 \, , \ \ \zeta =\frac{M^2}{W^2} \, .
\label{not1}
\end{eqnarray}
In our case the 
variable $\zeta$ has a similar meaning as $x_{Bj}$ in
the electroproduction process.
 
We introduce two light-cone vectors
\begin{equation}
n_{+}^2=n_{-}^2=0 \, , \ \ n_+ n_- = 1 \, .
\label{not2}
\end{equation}
Any vector $a$ is decomposed as 
\begin{equation}
a^\mu=a^+n_+^\mu+a^-n_-^\mu+a_\perp \, , \ \ a^2=2a^+a^- - \vec a^2 \, .
\label{not3}
\end{equation}
We choose the light cone vectors in a similar way as in Ji's notation,
namely
\begin{eqnarray}
&&
q=\frac{(W^2-m_N^2)}{2(1+\xi)W}\, n_- \, ,
\nonumber \\
&&
p=(1+\xi)W\, n_+ + \frac{m_N^2}{2(1+\xi)W}\, n_- \, ,
\nonumber \\
&&
p^\prime=(1-\xi)W\, n_+ 
+\frac{(m_N^2+\vec \Delta^2)}{2(1-\xi)W}\, n_- +\Delta_\perp \, ,
\nonumber \\
&&
\Delta=-2\, \xi \, W\, n_+ +\left(\frac{\xi\, m_N^2}{(1-\xi^2)W}+\frac{\vec
\Delta^2}{2\,(1-\xi)W}
\right)n_- +\Delta_\perp \, .
\label{not33}
\end{eqnarray}
We are interested in the kinematic region where
the invariant transfered momentum, 
\begin{equation}
t=\Delta^2=-\left(\frac{4\, \xi^2}{1-\xi^2}m_N^2+\frac{1+\xi 
}{1-\xi}\vec \Delta^2\right) \, ,
\label{not4}
\end{equation} 
is small, much smaller than $m$. 
In the scaling limit the 
variable $\xi$ which parametrizes the 
plus component of the momentum transfer equals $\xi=\zeta/(2-\zeta )$.   

\begin{figure}
\begin{center}
\scalebox{0.8}{
\input{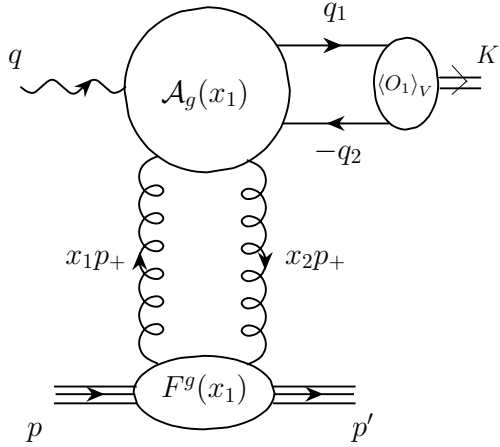}}
\end{center}
\caption[]{\small
Kinematics of heavy vector meson photoproduction.
}
\label{fig:1}
\end{figure}

The amplitude of quarkonium bound state production 
can be derived from the matrix element which describes the production of 
the on-shell heavy quark pair, $q_1^2=q_2^2=m^2$,
$q_1+q_2=K$, with a small relative momentum. 
The explicit equations providing the projection 
onto quarkonium states with 
different quantum numbers may be found in \cite{Braaten:2002fi}.
For the $S-$wave, spin-triplet case, which we are interested in,  
the procedure corresponds to neglecting the relative momentum of the pair,
$q_1=q_2=K/2$, and the replacement of the quark spinors by
\begin{equation}
v_i(q_2)\, \bar u_j(q_1)\to \frac{\delta_{ij}}
{4 N_c}\left(\frac{\langle O_1\rangle_V}{m}\right)^{1/2}
\!\not\! e_{V}^*\left(\not\! K + M \right) \, .
\label{spinors}
\end{equation} 
Here the indices $i,j$ parametrize the color state of the pair,
and the  vector 
$e_V$ describes the polarization  of the produced vector meson,
$(e_V e_V^*)=-1$ and $(K e_V)=0$.

Collinear factorization states that to leading twist
accuracy, i.e. neglecting the contributions which are suppressed by powers of
$1/m$, the amplitude can be calculated in the form suggested by Fig.~1:
\begin{equation}
{\cal M}=\left(\frac{\langle O_1\rangle_V}{m}\right)^{1/2}\!
\sum_{p=g,q,\bar q}\, \int\limits^1_0 dx_1 \, A^p_H(x_1,\mu_F^2)\, 
{\cal F}^p_\zeta (x_1,t,\mu_F^2) \ .
\label{factor}
\end{equation}
Here $ {\cal
F}^p_\zeta (x_1,\mu_F^2)$ is the gluon or quark GPD in Radyushkin's notation
\cite{Radyushkin:1996ru}; 
$x_1$ and $x_2=x_1-\zeta$ are the plus momentum fractions of the
emitted and the absorbed partons, respectively. 
$A^p_H(x_1,\mu_F^2)$ is the
hard-scattering amplitude and $\mu_F$ is the (collinear) factorization
scale. By definition, GPDs only involve small transverse momenta,
$k_\perp < \mu_F$, and the hard-scattering amplitude is calculated
neglecting the parton transverse momenta. Since quarkonium consists of heavy
quarks, it can by produced in LO only
by gluon exchange. The Feynman diagrams which describe the LO gluon 
hard-scattering amplitude are shown in Fig.~2.  
The contribution of the light quark exchange to quarkonium photoproduction
starts in collinear factorization at NLO, it is shown in Fig.~3. 
Since in this paper we consider the leading 
helicity non-flip amplitude, in eq. (\ref{factor})
the hard-scattering amplitudes $A^p_H(x_1,\mu_F^2)$ do not depend on $t$.
The account of this dependence would lead to the power suppressed, 
$\sim t/m$, contribution.

\begin{figure}
\begin{center}
\begin{tabular}{cc}
\includegraphics[scale=0.5]{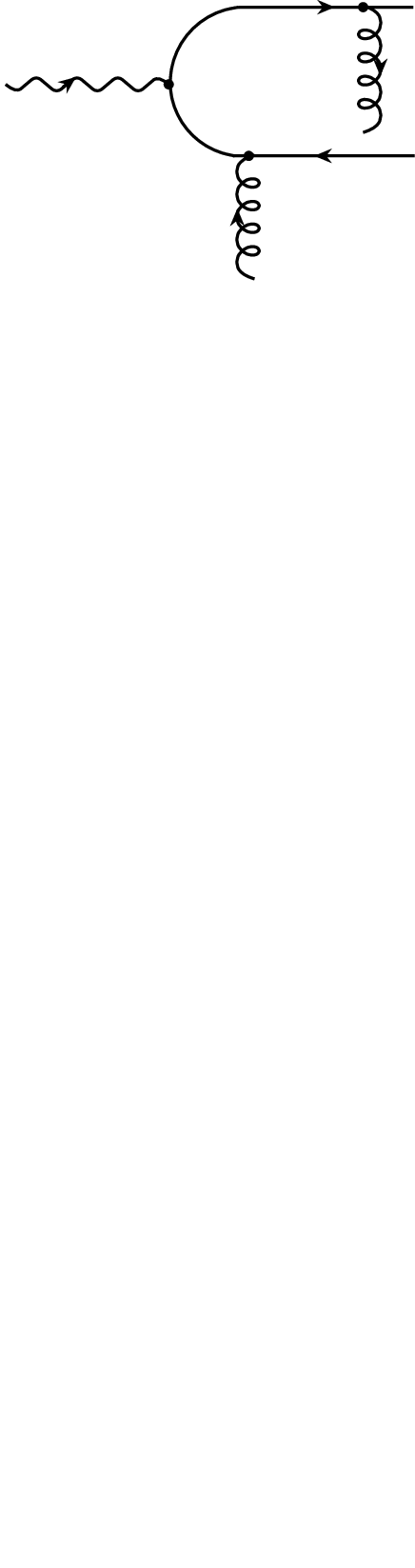}&
\includegraphics[scale=0.5]{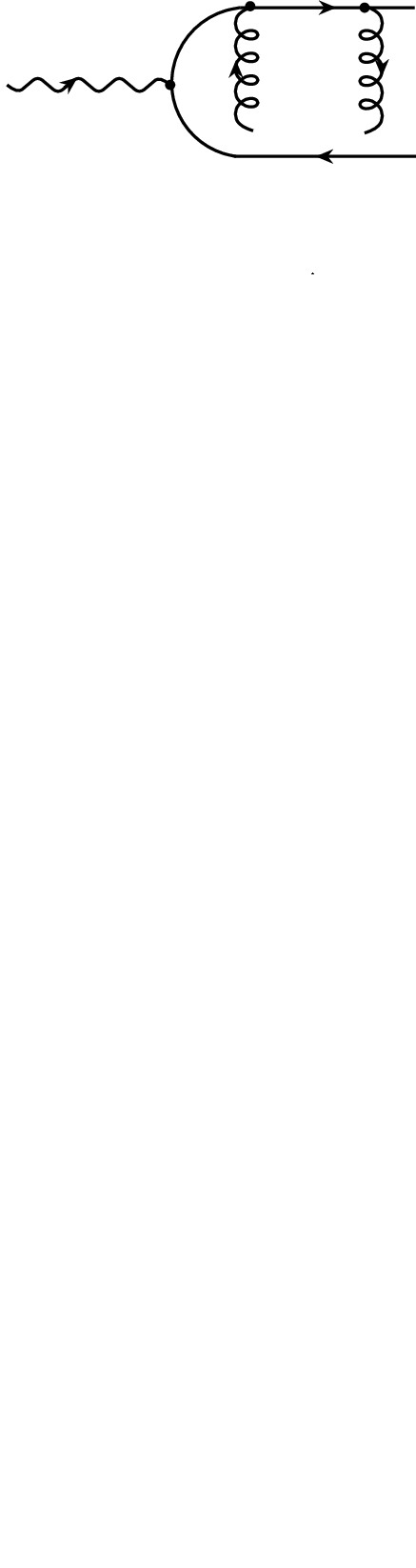}\\ (a) & (b)\\[1mm]
\includegraphics[scale=0.5]{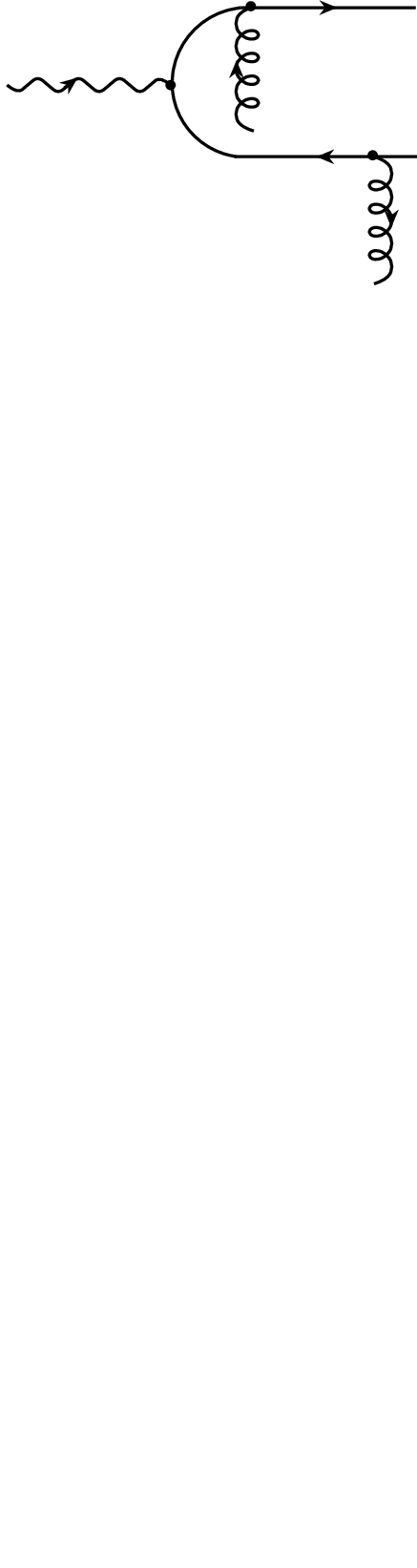}&
\includegraphics[scale=0.5]{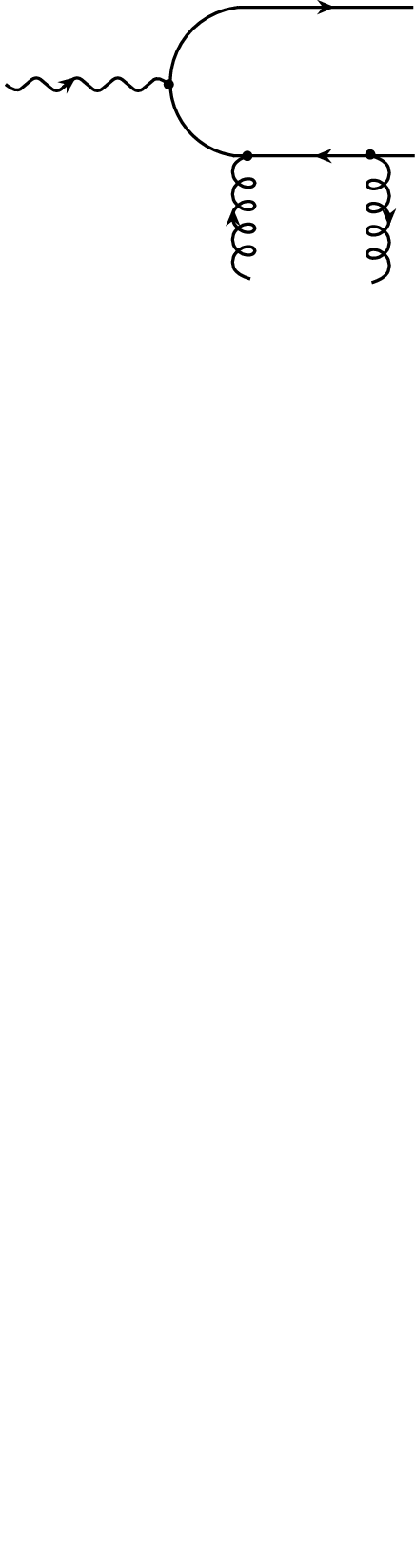}\\ (c) & (d)\\[1mm]
\includegraphics[scale=0.5]{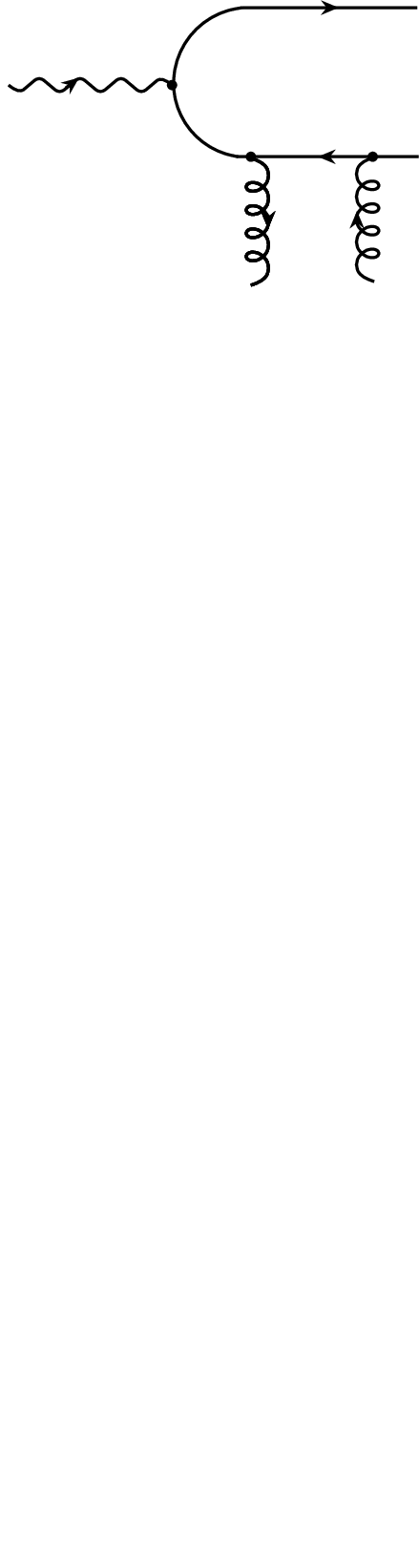}&
\includegraphics[scale=0.5]{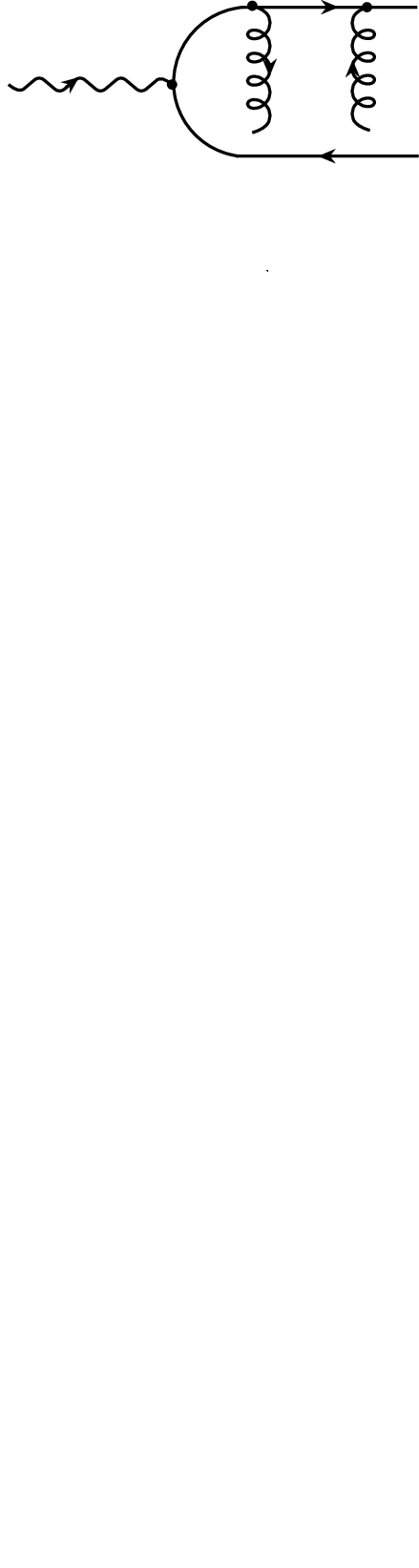}\\ (e) & (f)
\end{tabular}
\end{center}
\caption{\small The hard-scattering amplitude at LO.}
\label{fig:2}
\end{figure}

The momentum fraction $x_1$, $0\leq x_1\leq 1$, 
is defined with respect
to the momentum of the incoming proton. It is convenient to introduce  
the variable $x$, $-1\leq x\leq 1$, which parametrizes parton momenta
with respect to the symmetric momentum $P=(p+p^\prime)/2$. The relation 
between the different variables is 
\begin{equation}
x_1=\frac{x+\xi}{1+\xi} \, , \ \ x_2=\frac{x-\xi}{1+\xi} \, .
\label{not5}    
\end{equation}
In terms of symmetric variable $x$ the factorization formula
reads
\begin{eqnarray}
&&
 {\cal M}
=\frac{4\pi \sqrt{4\pi\alpha}\, e_q (e^*_V e_\gamma )}
{N_c \, \xi}\left(\frac{\langle O_1 \rangle_V}{m^3}\right)^{1/2}
 \int\limits^1_{-1} dx
\left[\, T_g( x,\xi)\, F^g(x,\xi,t)+
T_q (x,\xi) F^{q,S} (x,\xi,t) \, 
\right] \, ,
\nonumber \\
&&
F^{q,S} (x,\xi,t)=\sum_{q=u,d,s}  F^q (x,\xi,t) \, .
\label{fact1}
\end{eqnarray}
Here the dependence of the GPDs and the hard-scattering amplitudes 
on $\mu_F$ is suppressed for shortness.
In the quark contribution the sum runs over all light flavors, see Fig.~3.

\begin{figure}
\begin{center}
\scalebox{0.8}{
\input{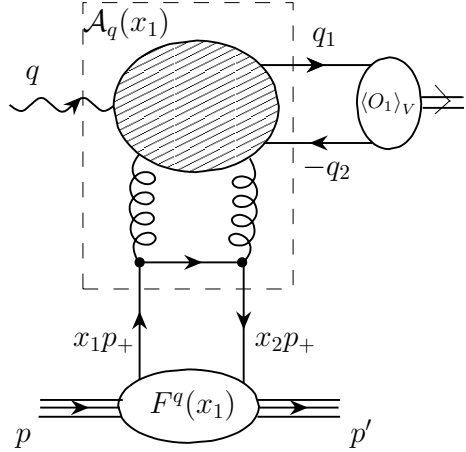}}
\end{center}
\caption[]{\small
The light quark contribution to  heavy meson photoproduction. 
}
\label{fig:3}
\end{figure}

GPDs are defined as the matrix element of the renormalized 
light-cone quark 
and gluon operators:
\begin{eqnarray}
&&
F^q (x,\xi,t)=\frac{1}{2}\int\frac{d\lambda}{2\pi}
e^{i x (P z)}\langle
p^\prime |\bar q \left(-\frac{z}{2}\right)\not \! n_-
q \left(\frac{z}{2}\right)
|p\rangle|_{z=\lambda n_-}
\nonumber \\
&&
=\frac{1}{2(Pn_- )}\left[
{\cal H}^q (x,\xi,t)\, \bar u(p^\prime)\not \! n_- u(p)+
{\cal E}^q (x,\xi,t)\, \bar
u(p^\prime)\frac{i\sigma^{\alpha\beta}n_{-\alpha}\Delta_\beta}{2\,m_N}  u(p)
\right] \, ,
\label{qGPD}
\end{eqnarray}
\begin{eqnarray}
&&
F^g (x,\xi,t)=\frac{1}{(Pn_-)}\int\frac{d\lambda}{2\pi}
e^{i x (P z)}\,
n_{-\alpha}n_{-\beta}\,\langle
p^\prime |G^{\alpha\mu} \left(-\frac{z}{2}\right)
G^{\beta}_\mu \left(\frac{z}{2}\right)
|p\rangle|_{z=\lambda n_-}
\nonumber \\
&&
=\frac{1}{2(Pn_- )}\left[
{\cal H}^g (x,\xi,t)\, \bar u(p^\prime)\not \! n_- u(p)+
{\cal E}^g (x,\xi,t)\, \bar
u(p^\prime)\frac{i\sigma^{\alpha\beta}n_{-\alpha}\Delta_\beta}{2\,m_N} 
u(p)
\right] \, .
\label{gGPD}
\end{eqnarray}
In both cases the insertion of the path-ordered gauge factor between the
field operators is implied. In the l.h.s. 
of eqs.~(\ref{qGPD}),
(\ref{gGPD}) the dependence of GPDs on the normalization point $\mu_F$
is suppressed for shortness.  
In the forward limit, $p^\prime=p$, the contributions proportional to  
the functions ${\cal E}^q (x,\xi,t)$ and ${\cal E}^g (x,\xi,t)$ vanish,
and
the distributions ${\cal H}^q (x,\xi,t)$ and ${\cal H}^g (x,\xi,t)$ reduce
to the ordinary quark and gluon densities: 
\begin{eqnarray}
&&
{\cal H}^q (x,0,0)=q(x) \ \ \mbox{for} \ \ x>0 \, ,
\nonumber \\
&&
{\cal H}^q (x,0,0)=-\bar q(-x) \ \ \mbox{for} \ \ x<0 \, ;
\nonumber \\
&&
{\cal H}^g (x,0,0)=\, x \, g(x) \ \ \mbox{for} \ \ x>0 \, .
\label{reduct}
\end{eqnarray}
Note that the gluon GPD is an even function of $x$, 
${\cal H}^g (x,\xi,t)={\cal H}^g(-x,\xi,t)$.

The definition of the gluon distribution (\ref{gGPD}) 
involves a field strength
tensor and, therefore, 
is valid in any gauge. But to evaluate the gluon hard-scattering
amplitude, it is convenient to consider the light-cone gauge  
$n_- A = 0$. In this gauge the parton picture which is behind
the collinear factorization formalism appears at the level of 
the individual
diagram. One can calculate the contributions of each gluon diagram
separately by  
considering photon scattering of on-shell gluons with zero
transverse momentum and the physical, transverse, polarizations.
These gluonic amplitudes have to be multiplied by the light-cone 
matrix element of two gauge field operators, which has the form
\cite{Radyushkin:1996ru}
\begin{multline}\label{g_d}
\int\frac{d\lambda (Pn_-)}{2\pi}\, e^{i x(Pz)}
\langle p\prime | A^a_{\mu}\left( -\frac{z}{2}\right) 
A^b_{\nu}\left(\frac{z}{2}\right)| p\rangle |_{z=\lambda n_-} = \\
\frac{\delta^{ab}}{N_c^2-1}  \left( \frac{-g^{\perp}_{\mu \nu}}{2(1+\epsilon)} \right)
\frac{F^g(x,\xi,t)}{(x-\xi+i\varepsilon)(x+\xi-i\varepsilon)}\, .
\end{multline}
Here $a,b$ are the gluon color indices,
$g^{\perp}_{\mu \nu} = g_{\mu \nu} 
- n_{+ \mu}n_{- \nu} - n_{- \mu} n_{+ \nu}$. 
The factor $2(1+\epsilon)$ counts a number of transverse dimensions within the regularisation method with the dimension
$D=2+2(1+\epsilon)$. It can be understood as making an average over the number of transverse polarisation states available to the gluons in $D$-dimensions, see also eq. \,(10) in \cite{MPSVW98}.  This prescription is in accordance with conventional definition of the evolution kernels needed for the subtraction of collinear divergences\footnote{We are grateful to 
Kornelija Passek-Kumericki and Dieter M\"uller for the discussion of these issues.}.
The
$i \varepsilon$ prescription for the poles in the r.h.s. of eq.~(\ref{g_d})
is important since 
corresponding singularities lie within the integration
domain and contribute to the imaginary part of the amplitude.  
In simple terms the sign of $i\varepsilon$   
can be understood in this case as 
due to the substitution $s\to s+i\varepsilon$, 
or $\xi\to\xi-i\varepsilon$. But one should notice that 
such an argumentation may not work
for more complicated processes which have in their physical regions the 
absorptive parts in variables other than the energy. 
For an 
example and an extended discussion of this issue see \cite{Braun:2002wu}.  
In the case of meson photo- and 
electroproduction the correct sign of $i\varepsilon$
is given by eq.~(\ref{g_d}).

The gluon  and the quark  hard-scattering amplitudes
$T_g(x,\xi)$ and  $T_q(x,\xi)$
describe the partonic subprocesses
\begin{equation}
{\cal A}_g={\cal A}_{\gamma G\to (\bar Q Q) G}
\label{gA}
\end{equation}
and
\begin{equation}
{\cal A}_q={\cal A}_{\gamma q\to (\bar Q Q) q} \, ,
\label{qA}
\end{equation}
respectively. Here $Q$ and $q$ denote the heavy and light quark.
\begin{eqnarray}
&&
 T_g(x,\xi)=\frac{\xi}{(x-\xi+i\varepsilon)(x+\xi-i\varepsilon)(1+\epsilon)}
{\cal A}_g\left(\frac{x-\xi+i\varepsilon}{2\xi}\right) \, ,
\nonumber \\
&&
T_q( x,\xi)={\cal A}_q\left(\frac{x-\xi+i\varepsilon}{2\xi}\right) \, .
\label{gAT}
\end{eqnarray}
In the first relation the
factor $\xi/((x-\xi+i\varepsilon)(x+\xi-i\varepsilon)(1+\epsilon))$ 
in front of the gluon amplitude comes from the 
parametrization of the gluon matrix element in the light-cone gauge
eq.~(\ref{g_d}).

Partonic amplitudes depend on two independent dimensionful variables, 
the partonic subenergy $\tilde s=x_1s$ and the meson mass $M^2=\zeta s$.
Being dimentionless quantities the partonic amplitudes
 can be expressed as a function of the 
ratio
\begin{equation}
y=\frac{\tilde s-M^2}{M^2}=\frac{x_2}{\zeta}=\frac{x-\xi}{2\xi} \, .
\label{ro}
\end{equation} 
This convention is adopted in eq.~(\ref{gAT}).

Another Mandelstam variable for partonic subprocess is $\tilde u=M^2-\tilde
s=-x_1s$. The exchange between the 
two channels, $\tilde s\leftrightarrow \tilde
u$, corresponds to the replacements $x_1\leftrightarrow -x_2$,
or $y\leftrightarrow -(1+y)$, or $x \leftrightarrow -x$. Hard scattering
amplitudes and GPDs possess definite symmetry properties which are closely 
related to charge conjugation invariance. A photon and a vector
meson have the same $C-$ parities, which selects $C-$even exchange 
in the $t-$channel. For gluons only a $C-$even GPD exists at leading twist, 
which is an even function of $x$, as thus also 
the gluon hard-scattering amplitude is even in $x$,
$T_g(x,\xi)=T_g(-x,\xi)$. For the quark there exist both 
$C-$even and $C-$odd GPDs, and $F^q$ has no definite
symmetry under the exchange $x \leftrightarrow -x$. But since the 
quantum
numbers of the photon and vector meson select 
the $C-$even exchange in the $t-$channel,
the quark
hard-scattering amplitude obeys $T_q(x,\xi)=-T_q(-x,\xi)$.
Therefore only the
$C-$even (singlet) component of the quark GPD,
$F^{q(+)}=F^q(x,\xi,t)-F^q(-x,\xi,t)$, contributes to 
(\ref{fact1}). 

Next, we have  to evaluate the partonic amplitudes ${\cal A}_g$ and ${\cal
A}_q$. We will use the dimensional regularization method, with
$D=4+2\epsilon$ dimensions, in order to regularize the ultraviolet (UV)
and infrared (IR) singularities which appear at the intermediate steps
of the calculation. 

At lowest order there exists only the gluon contribution. 
${\cal A}_g$ is given
by 6 tree diagrams shown in Fig.~2. A simple calculation gives the result 
\begin{eqnarray}
&& 
{\cal A}_g^{(0)}(y)=\alpha_S \, , 
\label{LOg} \\
&&
{\cal A}_q^{(0)}(y)=0 \, .
\label{LOq}
\end{eqnarray}

\section{The hard-scattering amplitudes at NLO}

At LO the gluonic amplitude is a constant, it is a tree amplitude which 
has no singularities. At NLO the one-loop gluon and quark 
partonic amplitudes develop
a branch cut singularities
along the lines $[0,+\infty )$ and $(-\infty,-1]$
in the complex plane of variable $y$, see Fig.~4. 
We will use a method based on the dispersion representation 
in order to simplify the calculation of
these one-loop amplitudes. 
Deforming the integration contour as shown
in Fig.~4 one arrives at a representation of the amplitude which allows to
reconstruct it as a function of the 
variable $y$ 
from its discontinuities along the cuts $[0,+\infty )$ and $(-\infty ,-1]$.
Thanks to the symmetry properties of 
the partonic amplitudes discussed above 
the contribution of the brunch cut $(-\infty,-1]$
to the dispersion integral may be expressed in terms of the 
discontinuity at $[0,+\infty )$. 

We will start with the quark contribution, then 
we present the 
more complicated calculation of the gluonic amplitude. 
After that we
discuss the renormalization and the subtraction of the collinear
singularities which lead, finally, to  the finite results for the 
hard-scattering amplitudes at NLO.      

\begin{figure}
\begin{center}
\begin{tabular}{cc}
\includegraphics[scale=0.6]{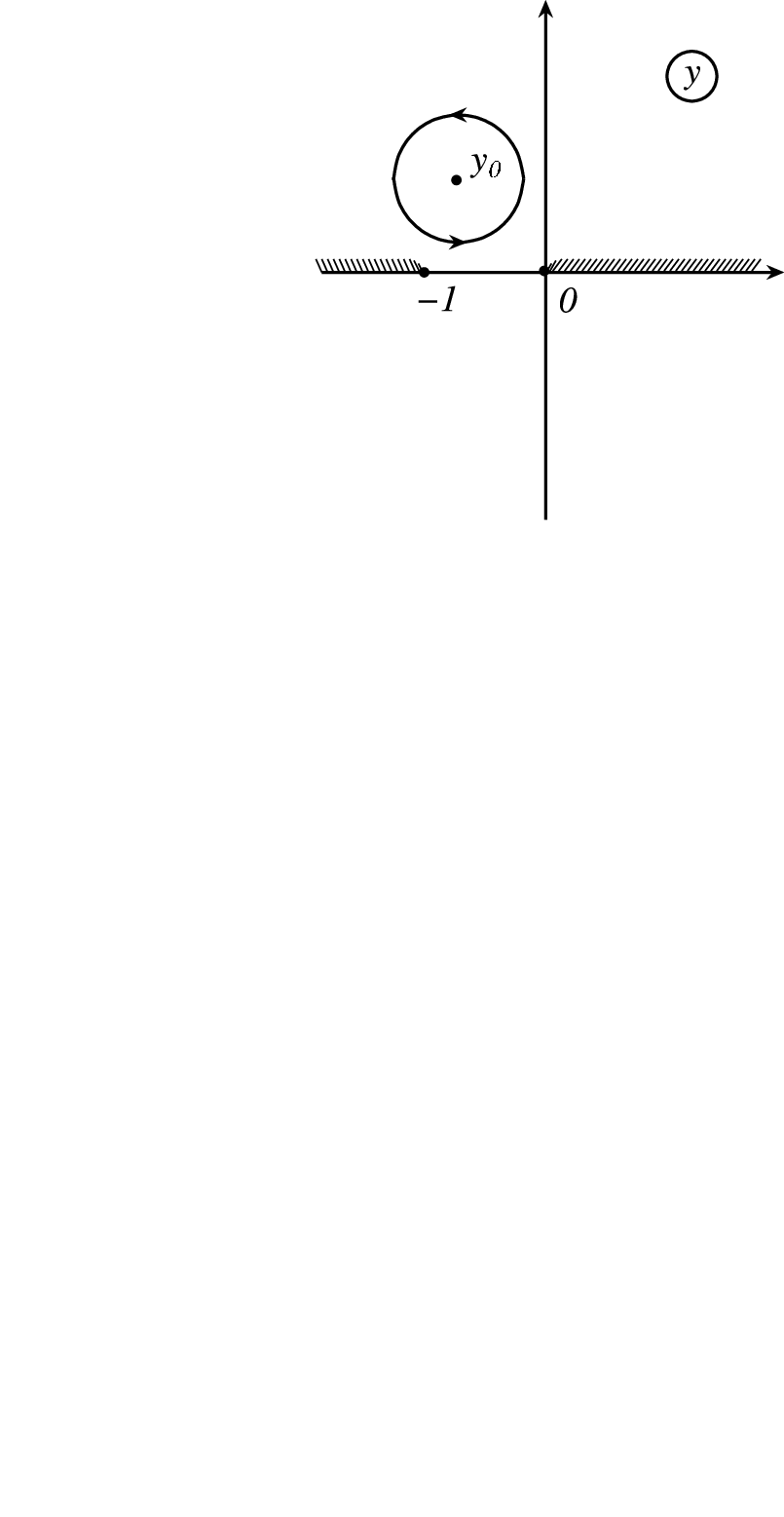}&
\includegraphics[scale=0.6]{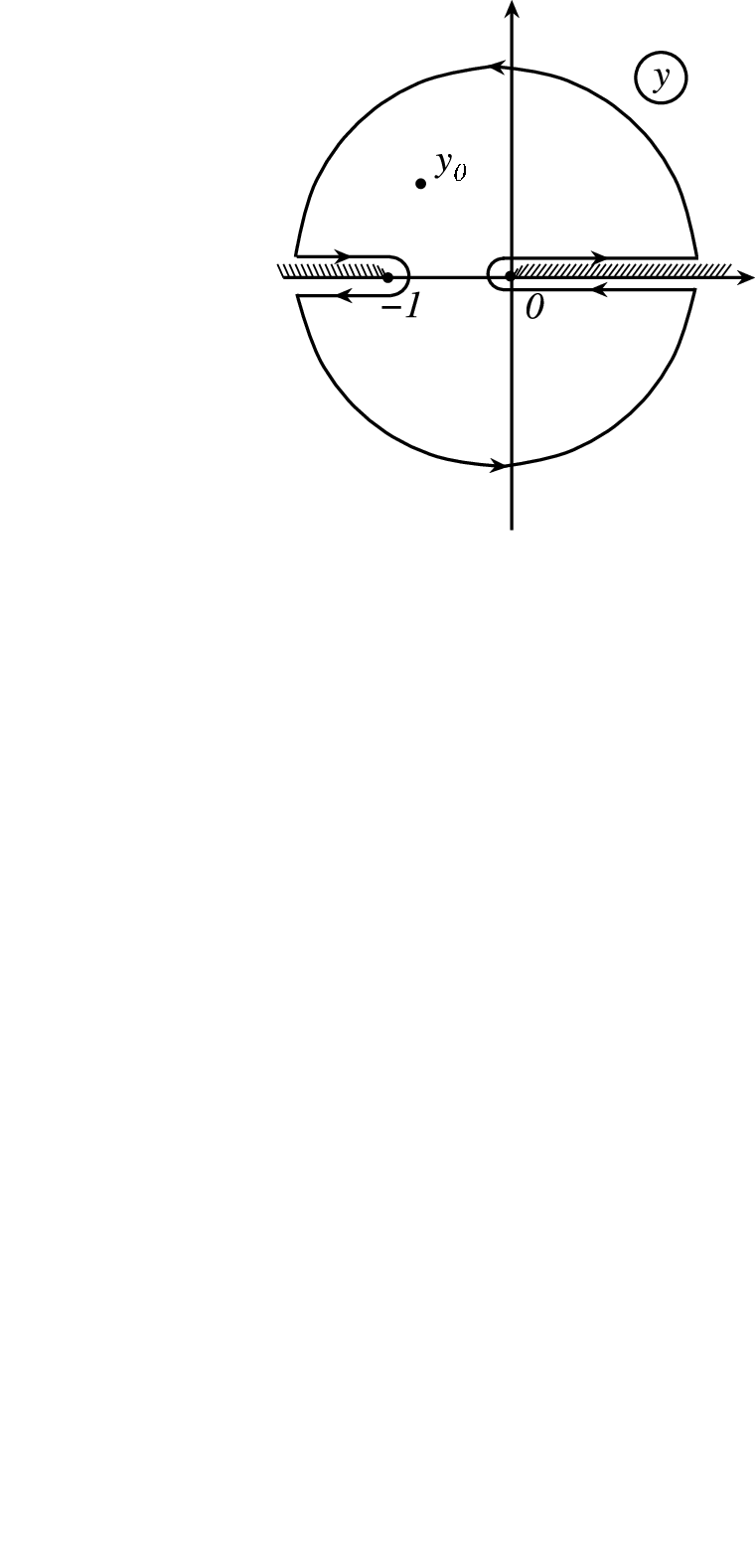}
\end{tabular}
\end{center}
\caption[]{\small
The analytical properties of the partonic  amplitudes at NLO
in the complex plane of $y=x_2/\zeta$.
 }
\label{fig:4}
\end{figure}

\subsection{The quark contribution}

The dispersion representation for the quark NLO amplitude 
${\cal A}_q^{(1)}(y)$ reads
\begin{equation}
{\cal A}_q^{(1)}(y)=
\frac{1}{\pi}\int\limits_0^\infty dz\, 
Im \, {\cal A}_q^{(1)}(z)\left(\frac{1}{z-y}-\frac{1}{z+y+1} \right) \, .
\label{disQ}
\end{equation}
Here $Im \, {\cal A}_q^{(1)}(z)$ stands for the imaginary part of the quark
amplitude in the $\tilde s-$ channel of the quark subprocess. Using the
crossing
symmetry property, ${\cal A}_q(y)=-{\cal A}_q(-1-y)$, 
the contribution of the $\tilde u-$ channel discontinuity was expressed in
terms of $Im \, {\cal A}_q^{(1)}(z)$, it is given by the second term on the 
r.h.s. of eq.~(\ref{disQ}). For the quark amplitude one can use
the unsubtracted dispersion relation, eq.~(\ref{disQ}). 
${\cal A}_q^{(1)}(z)\sim const$ at large $z$, 
but due to cancellation between $\tilde s-$ and $\tilde
u-$ channel contributions the sum of two terms 
in the brackets vanishes at large $z$ as  
$\sim 1/z^2$ while each individual term vanishes as $1/z$. Thus 
  the dispersion integral 
is convergent at the upper limit. In other words a subtraction constant is
not compatible with the symmetry properties of the quark amplitude.

\begin{figure}
\begin{center}
\includegraphics[scale=0.45]{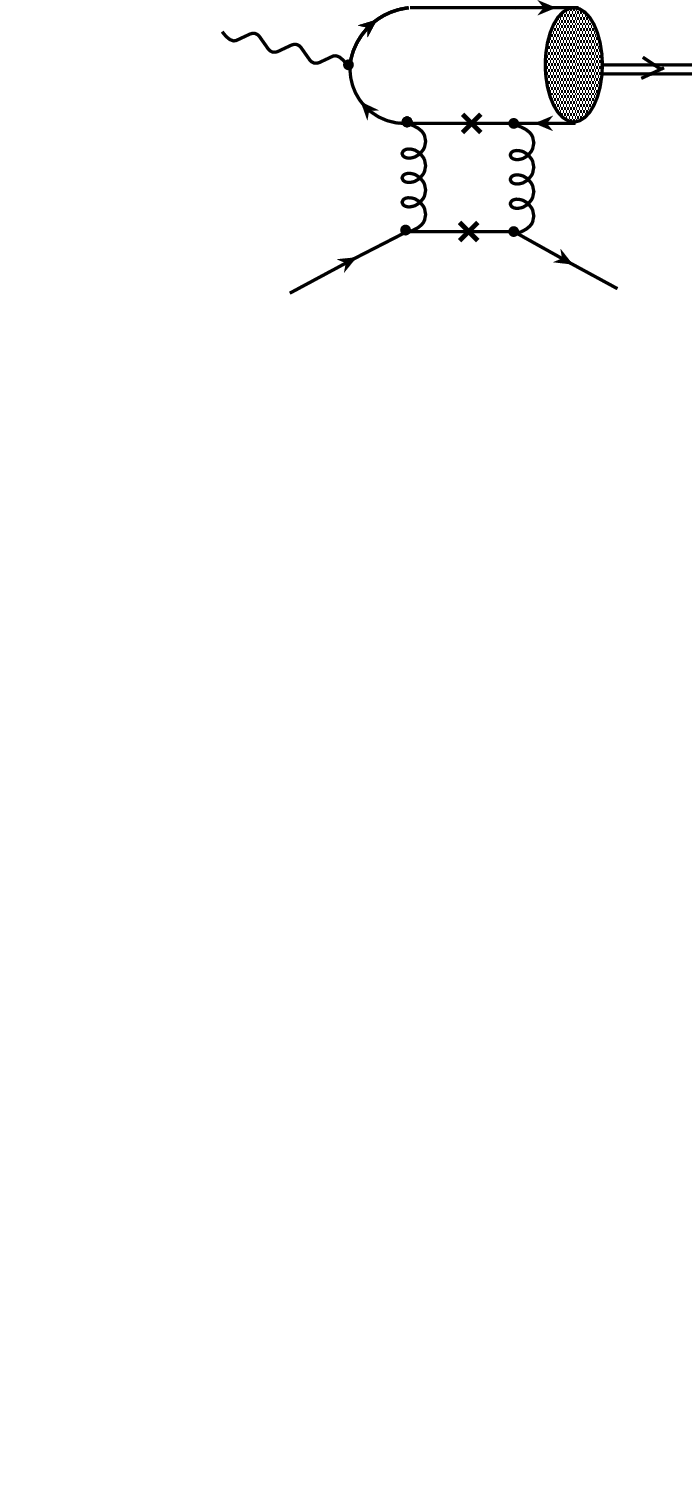}
\includegraphics[scale=0.45]{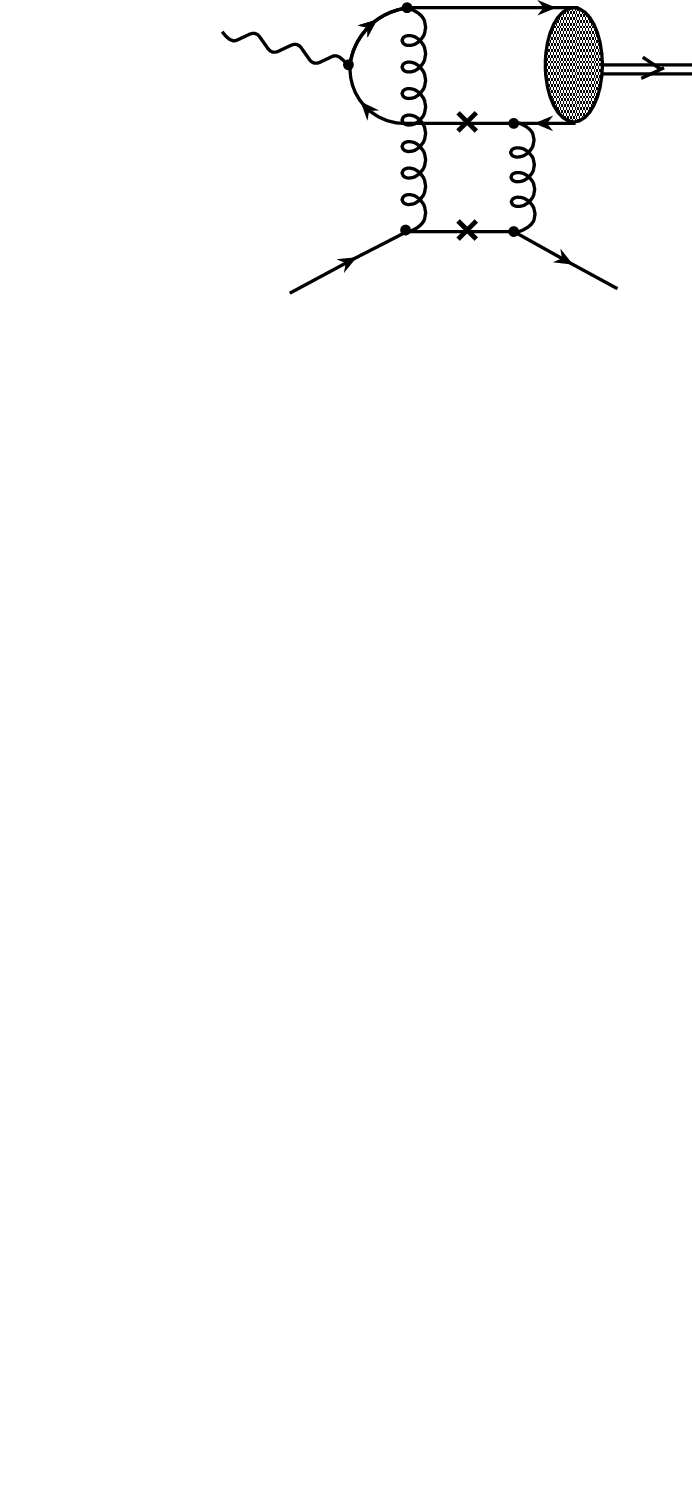}
\includegraphics[scale=0.45]{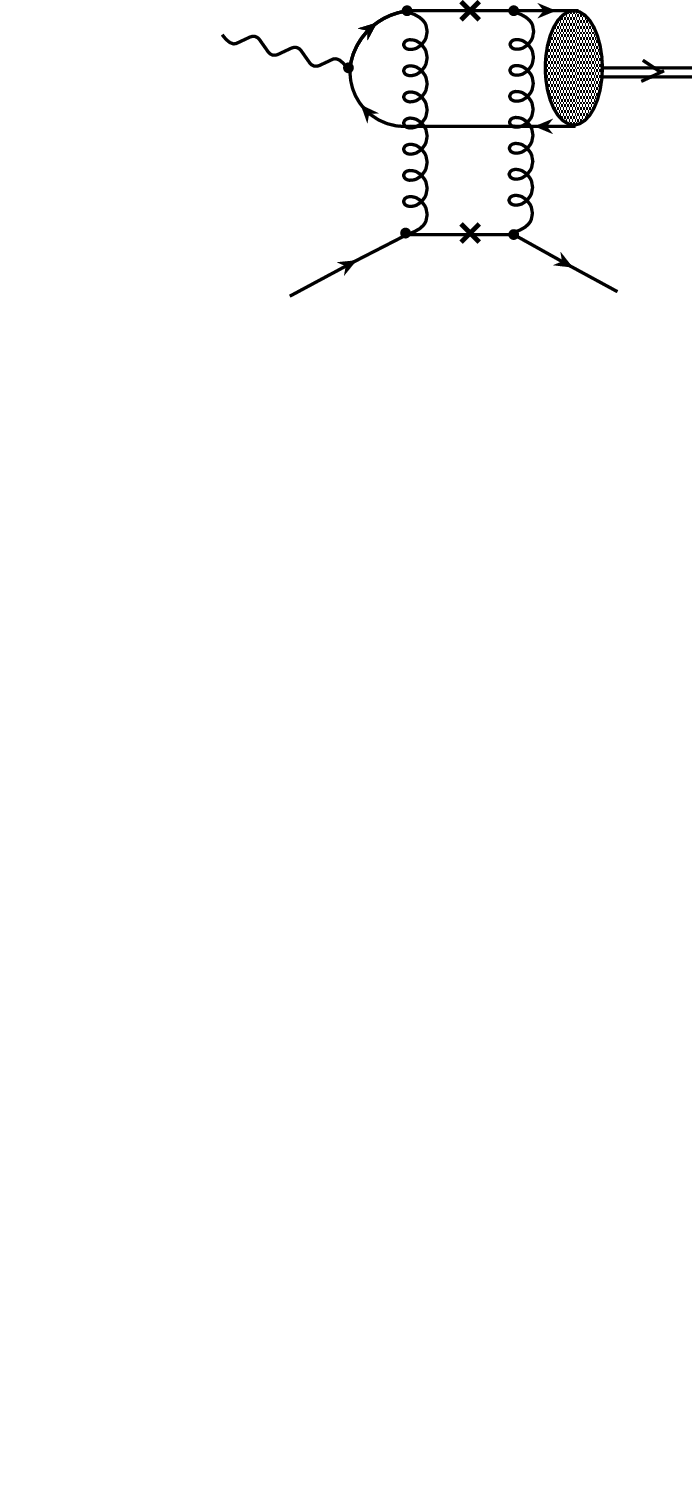}
\includegraphics[scale=0.45]{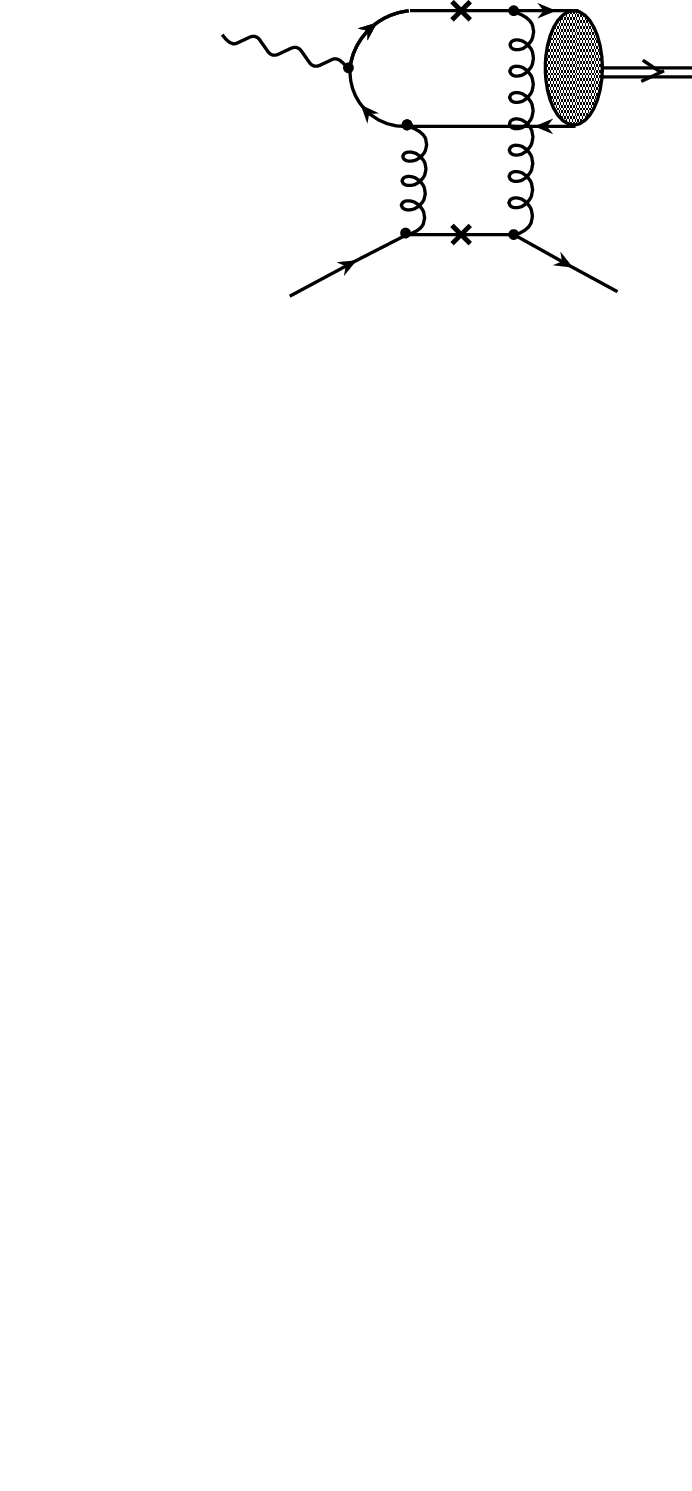}
\end{center}
\caption[]{\small The $\tilde s$- channel cut diagrams for the quark
amplitude.}
\label{fig:5}
\end{figure}

Among the 6 diagrams which contribute to the NLO quark amplitude
only 4 diagrams have a discontinuity in the $\tilde s-$ channel.
They are shown in Fig.~5. It is sufficient to calculate the first  
two diagrams which contain a cut of the light quark and the heavy
antiquark lines. The line of the heavy quark in these diagrams
is not cut since 
it enters directly into the meson vertex and, therefore, is effectively
on the mass shell. The other two diagrams in Fig.~5 describe 
the heavy quark cut, their
contribution is identical to that one of the first two diagrams.

We present the quark amplitude in the form
\begin{equation}
{\cal A}_q^{(1)}(y)=\frac{\alpha_S^2
\, C_F}{(4\pi)^{1+\epsilon}
\Gamma(1+\epsilon)}\left(\frac{4m^2}{\mu^2}\right)^\epsilon{\cal I}_q(y) \, , 
\label{qAA}
\end{equation}
here $\Gamma(\dots)$ is the Euler gamma function and
$C_F=(N_c^2-1)/(2N_c)=4/3$ is the color factor, $\mu$ is a scale introduced
by dimensional regularization.   
Calculating the imaginary part we find
\begin{equation}
\frac{1}{\pi} \,  Im\, {\cal I}_q (y) =
  2 \left(
  \frac{y^2}{1+2y}
  \right)^{\epsilon}
  \left(
  -\frac{1+2y}{1+y}\frac{1}{\epsilon}-
  \frac{1}{1+2y} +
  \frac{3+8y(1+y)}{4y(1+y)}\ln(1+2y)
  \right) \, .
\end{equation}
Then, inserting this equation into the dispersion integral (\ref{disQ}) 
we obtain the following expression for the quark amplitude. 
\begin{multline}
 {\cal I}_q(y) =
  \frac{2}{\epsilon}(1+2y)\left(
   \frac{\ln(-y)}{1+y}-\frac{\ln(1+y)}{y}
  \right)-
  \pi^2\frac{13\,(1+2y)}{24\,y\,(1+y)}+\frac{4\ln2}{1+2y}+\\
  2\frac{\ln(-y)+\ln(1+y)}{1+2y}+
  2(1+2y)\left(
   \frac{\ln^2(-y)}{1+y}-\frac{\ln^2(1+y)}{y}
  \right)+\\
  \frac{3-4y+16y(1+y)}{2y(1+y)}Li_2(1+2y)-
  \frac{7+4y+16y(1+y)}{2y(1+y)}Li_2(-1-2y)\, ,
\label{qBB}
\end{multline}
where 
\begin{equation}
Li_2(z)=-\int\limits^z_0\frac{dt}{t}\ln (1-t)\, .
\label{Li2}
\end{equation}

\subsection{The gluon contribution}

The analysis of the gluon contribution follows the same lines as
for the quark case. However, one has to take into account
that the 
gluonic amplitude is symmetric under crossing, 
${\cal A}_g(y)={\cal A}_g(-1-y)$, and
that the asymptotics of  ${\cal A}_g^{(1)}(y)$ at large $y$ is ${\cal
A}_g^{(1)}(y)\sim y$. Therefore we need a dispersion representation of 
${\cal A}_g^{(1)}(y)$ with one subtraction. It is convenient to perform this 
subtraction at $y=0$, the point where the second gluon carries zero energy,
since the calculation of the amplitude in this point may be considerably
simplified making use of a low energy theorem for the radiation of a soft
gluon. The dispersion representation for the gluonic amplitude reads
\begin{equation}
{\cal A}_g^{(1)}(y)-{\cal A}_g^{(1)}(0)=\frac{1}{\pi}
\int\limits^\infty_0 dz \, Im \, {\cal A}_g^{(1)}(z)
\left( \frac{y}{z(z-y)}-\frac{y}{(z+y)(z+y+1)}\right) \, .
\label{disG}
\end{equation}
The second term in the brackets represents the contribution of the $\tilde
u-$ channel cut. Due to cancelation between the $\tilde
s-$ and the $\tilde u-$ channel contributions the term in the brackets
vanishes as $\sim 1/z^3$ rather than as $\sim 1/z^2$ which makes the
dispersion
integral convergent. Therefore 
one needs only one subtraction, not two. 
This can also be expressed in the following manner:
the term 
linear in $y$ of the subtraction polynomial is absent, because 
it is not compatible with the symmetry
property of the gluonic amplitude.

It is convenient to introduce the auxiliary quantity ${\cal I}_g(y)$ defined
by
\begin{equation}
{\cal A}_g^{(1)}(y)=\frac
 {\alpha_S^2 }
 {(4\pi)^{1 + \epsilon} \Gamma(1+\epsilon)}
\left(\frac{4 m^2}{\mu^2}\right)^{\epsilon}{\cal I}_g(y) \, .
\label{conv}
\end{equation}
The imaginary part of the gluonic amplitude may be represented as sum of
three different contributions
\begin{equation}
Im\, {\cal I}_g(z)=Im\, {\cal I}_g^{(\bar Q Q)}(z)
+Im\, {\cal I}_g^{(Q g)}(z)+Im\, {\cal I}_g^{(\bar Q g)}(z)\, .
\label{three}
\end{equation}
Here $Im\, {\cal I}_g^{(\bar Q Q)}(z)$ represents the sum of 10 
diagrams having a $\bar Q Q$ cut in the intermediate state, see  
Figs.~6 and 7. $Im\, {\cal
I}_g^{(Q g)}(z)$ gives the contribution of the 24 heavy quark gluon cut
diagrams and $Im\, {\cal I}_g^{(\bar Q g)}(z)$ is the contribution to the
imaginary part coming from the 24 cut diagrams with the heavy antiquark and
the gluon in the intermediate state shown in Figs.~8 and 9. 
The latter two contributions are equal,
\begin{equation}
Im\, {\cal I}_g^{(Q g)}(z)=Im\, {\cal I}_g^{(\bar Q g)}(z)\, ,
\label{equal}
\end{equation}
therefore it is enough to calculate only one of them, 
say, $Im {\cal I}_g^{(\bar Q g)}(z)$. We define two contributions 
\begin{equation}
{\cal I}_g(y)-{\cal I}_g(0)={\cal I}_g^{(\bar Q Q)}(y)
+ 2 \, {\cal I}_g^{(\bar Q g)}(y)\, ,
\label{three1}
\end{equation}
in accordance with eq.~(\ref{disG}),
the decomposition
of the imaginary part eq.~(\ref{three}), and
eq.~(\ref{equal}).

\begin{figure}
\begin{center}
\scalebox{0.7}{
\input{g_qbarq.pstex_t}}
\end{center}
\caption[]{\small
The contribution of the $\bar Q Q$ intermediate state to the gluonic amplitude.}
\label{fig:6}
\end{figure}
\subsubsection{$\bar Q Q$- and $\bar Q g$-cut contributions}
The calculation of the  $\bar Q Q$-cut diagrams shown in Figs.~6 and 7
gives
\begin{equation}
  \frac{1}{\pi}Im\, {\cal I}_g^{\bar Q Q}(y) =
    ( y )^\epsilon
   \Theta_g^{\bar Q Q}(y) \, ,
\end{equation}
where
\begin{eqnarray}
&&
\Theta_g^{\bar Q Q}(y) =
  -\frac{\sqrt{y(1+y)}}{y (1 + y)} \left(
     c_1 \frac{7}{2} + c_2 \big( \frac{3}{y} + 1 \big)
     \right)
\nonumber \\
&&
+
  \frac{{\rm arctanh}\sqrt{\frac{y}{1+y}}}{y(1+y)} \left(
     c_1\big(-\frac{3}{2} +2 y\big) +
     c_2 \big( \frac{3}{y} + 6 + 2y \big)
     \right) \, .
\label{iqq}
\end{eqnarray}
Here for shortness
we denote two independent color structures by
\begin{equation}
c_1=C_F \ , \ \ c_2=C_F-\frac{C_A}{2}=-\frac{1}{2N_c} \, . 
\label{colorS}
\end{equation}

Inserting this result to dispersion integral (\ref{disG}) we obtain
\begin{multline}
{\cal I}_g^{\bar Q Q}(y) =
-5 c_1-\frac{3+2y(1+y)}{y(1+y)}c_2+
 \pi\frac{\sqrt{-y(1+y)}}{y (1 + y)}\left(
   \frac{7}{2} c_1-3 c_2
 \right)+\\
 \pi^2\left(
 \frac{3-4y(1+y)}{8y(1+y)}c_1-\frac{3+y(1+y)(9-y(1+y))}{4y^2(1+y)^2}c_2
 \right)+\\
 2 c_2 \frac{\sqrt{-y(1+y)}}{y (1 + y)}\left(
  \frac{1+4y}{1+y}\arctan \sqrt{\frac{-y}{1+y}}+
  \frac{3+4y}{y}\arctan \sqrt{\frac{1+y}{-y}}
  \right)\\ 
-\frac{\arctan^2 \sqrt{\frac{-y}{1+y}}}{2y(1+y)} \left(
 (7+4y)c_1 - 2\frac{1+2y-2y^2}{1+y}c_2\right)\\
-\frac{\arctan^2 \sqrt{\frac{1+y}{-y}}}{2y(1+y)} \left(
 (3-4y)c_1 - 2\frac{3+6y+2y^2}{y}c_2\right) \, .
\label{IQQ}
\end{multline}

Some words about the calculation of integral (\ref{disG}) for 
${\cal I}_g^{\bar Q Q}$ are in order. Since 
$Im\, {\cal I}_g^{\bar Q Q}(z)\sim
z^{\epsilon-1/2}$ at small $z$, 
the contribution of the region $z\leq \delta$ (where
$\delta \ll 1$) to dispersion integral is of the order
\begin{equation}
\sim \int\limits^\delta_0 dz \,
z^{\epsilon-\frac{3}{2}}=\frac{\delta^{\epsilon-\frac{1}{2}}}
{\epsilon-\frac{1}{2}}
|_{\epsilon\to 0} 
\to 
-\frac{2}{\sqrt{\delta}} \, .
\label{ddd}
\end{equation}
However, this contribution to ${\cal I}_g^{\bar Q Q}$, which is singular for
  $\delta\rightarrow 0$ cancels
 with the one coming from the region $z\geq \delta$ and we arrive at
the finite result given by eq.~(\ref{IQQ}). 

The appearance of  integrals like (\ref{ddd}) is related to a   
phenomenon well known in quarkonium physics. The gluon exchange
between the nonrelativistic quark pair contains the Coulomb like
instantaneous contribution. 
In the NRQCD formalism its
contribution has to be subtracted from the hard part of the amplitude.
Let us discuss the corresponding counterterm.

In a frame where the $Q \bar Q$ system is at rest the momenta of the heavy 
quarks are: 
\begin{equation}
q_1=(m+\varepsilon,\vec p)\, , \ \ \ 
q_2=(m+E-\varepsilon , - \vec p) \, , 
\label{qq}
\end{equation}
where $E$ denotes the nonrelativistic energy of the pair. The LO amplitude 
has the form
\begin{equation}
{\cal M}^{LO}=C\int d\vec p \, \Psi (\vec p) \ .
\label{LOa}
\end{equation}
Here C is some factor and $\Psi (\vec p)$ is the nonrelativistic wave
function of the $Q\bar Q$ system in momentum representation. 
The integral (\ref{LOa}) is proportional
to the value of the wave function at the origin
\begin{equation}
\int d\vec p \, \Psi (\vec p)\sim R_S(0) \, .
\label{origin}
\end{equation}

Now consider the $\alpha_S$ correction. 
The momenta of the quarks after the gluon exchange are   
\begin{equation}
q_1^\prime =(m+\varepsilon^\prime,\vec p^{\; \prime })\, , \ \ \ 
q_2^\prime =(m+E-\varepsilon^\prime , - \vec p^{\; \prime} ) \, .
\label{qqprime}
\end{equation}
For the nonrelativistic system the energy and the momentum variables scale
as: $E,\varepsilon,\varepsilon^\prime \sim mv^2$; $|\vec p|,|\vec
p^{\; \prime}|\sim m v$. With  NLO accuracy the amplitude can therefore be written 
as
follows
\begin{equation}
{\cal M}^{NLO}=C\int d\vec p \, \Psi (\vec p)
\left(
1-\frac{\alpha_S C_F}{2\pi^2(2\pi)^{2\epsilon}}
\int d \vec p^{\; \prime} 
\frac{1}{(\vec p-\vec p^{\; \prime})^2[E-\frac{\vec
p^{\; \prime 2}}{m}+i0]}+ {\cal O}(\alpha_S v^0) 
\right)\, .
\label{CoulombC}
\end{equation}
The first term on the r.h.s. of (\ref{CoulombC}) is the LO contribution,
the second and the third terms represent the NLO correction. The later
is  
finite at $v\to 0$.
The second term of (\ref{CoulombC}) scales $\sim \alpha_S C_F/v$, 
it comes from the instantaneous
Coulomb exchange. Eq.~(\ref{CoulombC})  can be easily 
derived considering the integral over the
loop momentum,
$d^{4+2\epsilon}q_1^\prime=d \varepsilon^\prime d\vec p^{\; \prime}$, and 
using the nonrelativistic limit for the
quark propagators. After 
integration over the loop energy $\varepsilon^\prime$ we arrive at the 
expression given above for the 
Coulomb contribution. It can be recognized as the  
exchange potential responsible for the formation of a nonrelativistic 
meson bound state. Indeed, the 
Schr\"odinger equation in momentum representation reads
\begin{equation}
(E-\frac{\vec
p^{\; \prime 2}}{m})\Psi (\vec p^{\; \prime})=-\frac{\alpha_S
C_F}{2\pi^2(2\pi)^{2\epsilon}}
\int d \vec p \, \frac{\Psi (\vec p)}{(\vec p-\vec p^{\; \prime})^2} \, .
\label{Schrodinger}
\end{equation} 
The second term in Eq.~(\ref{CoulombC}), the Coulomb counterterm, 
integrated over $d \vec p$ produces the LO contribution, $C \int d \vec
p^{\; \prime} \, \Psi (\vec p^{\; \prime})$, 
which is already taken into account in the
first term. Therefore the Coulomb counterterm has to be subtracted from
Eq.~(\ref{CoulombC}). After that one can put  the quark pair on the mass
shell; $v,E\to 0$, and $\vec p\to 0$. 

The advantage of using dimensional regularization is that  
the quark pair may be put on the mass shell even before the subtraction of
the Coulomb counterterm. Since at $E= 0$, and $\vec p= 0$ the Coulomb counterterm
 becomes the 
scaleless integral, $\sim \int d\vec p^{\; \prime} /\vec p^{\; \prime
4}\rightarrow 0$, it has to be put equal 
to zero according to the rules of the dimensional regularization method. 
That means that the Coulomb counterterm is zero in this 
scheme.\footnote{We are grateful to 
Maxim Kotsky for the discussion of this issue.}
The price to be paid for the simplification
is the appearance of
integrals like (\ref{ddd}). They have to be treated as described above.   
We encountered integrals of this kind 
also in the calculation of ${\cal I}_g (0)$.  

\begin{figure}
\begin{center}
\begin{tabular}{cccc}
\includegraphics[scale=0.45]{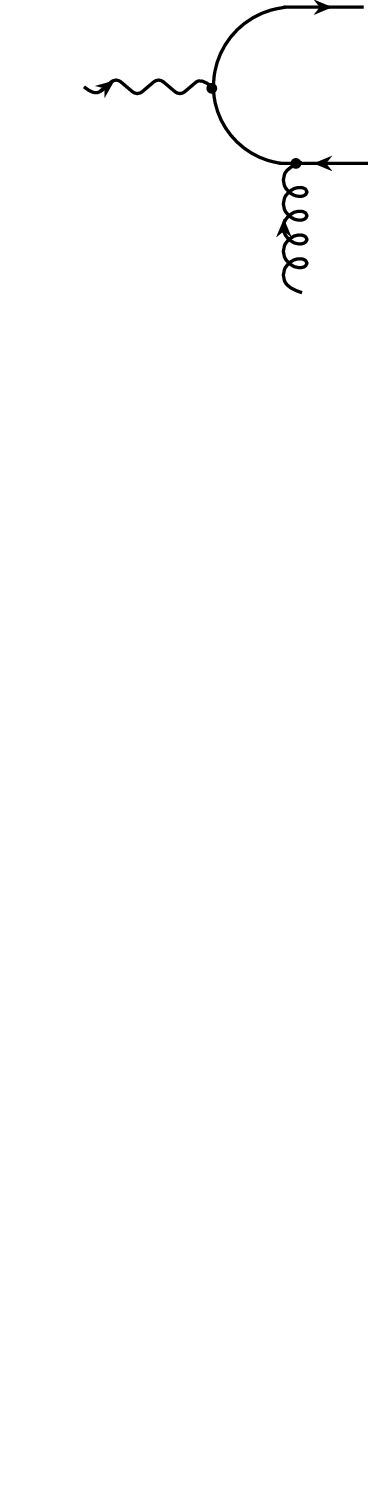}~~~~~~&
\includegraphics[scale=0.45]{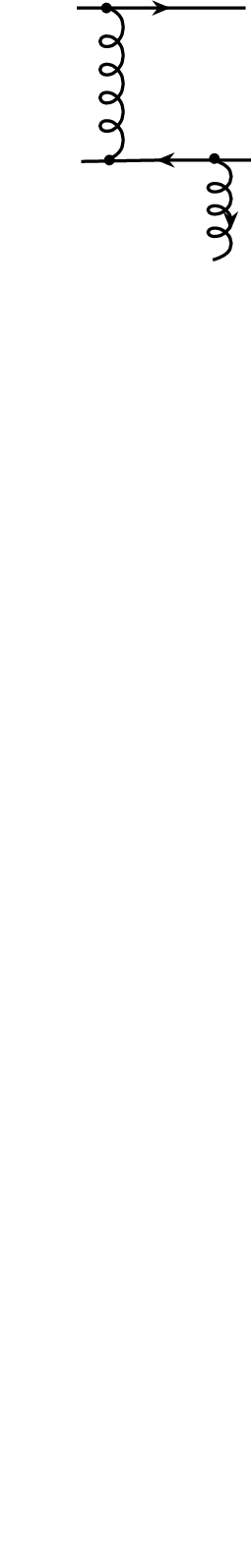}&
\includegraphics[scale=0.45]{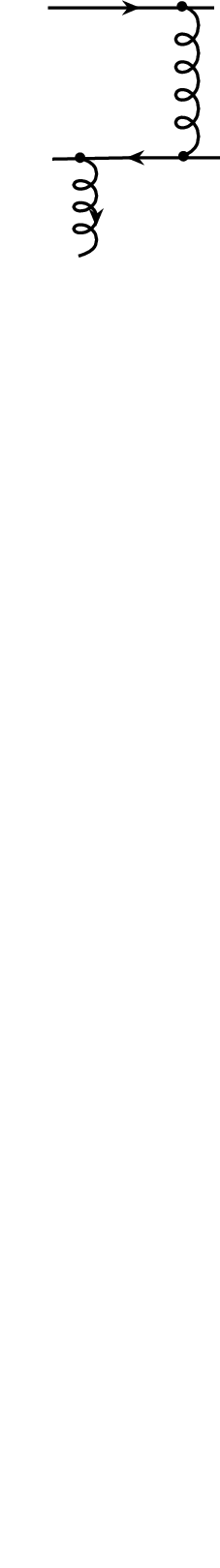}\\ $L_1$ & $R_1$ & $R_2$ \\[2mm]
\includegraphics[scale=0.45]{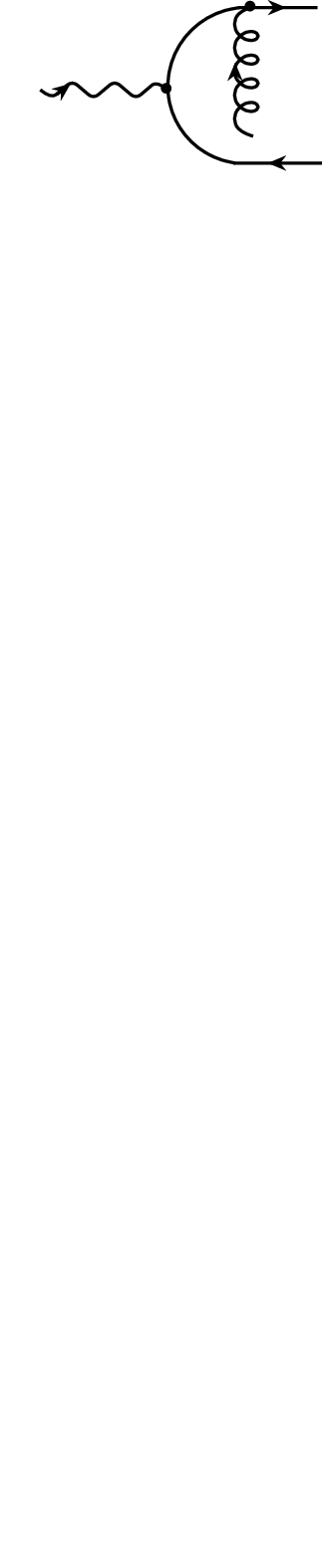}~~~~~~&
\includegraphics[scale=0.45]{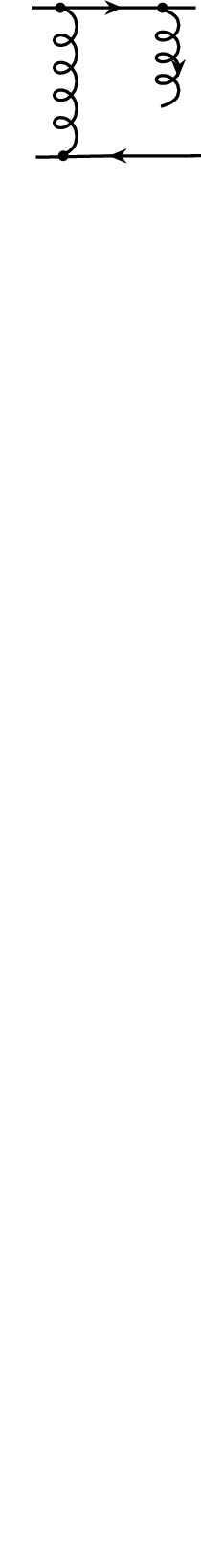}&
\includegraphics[scale=0.45]{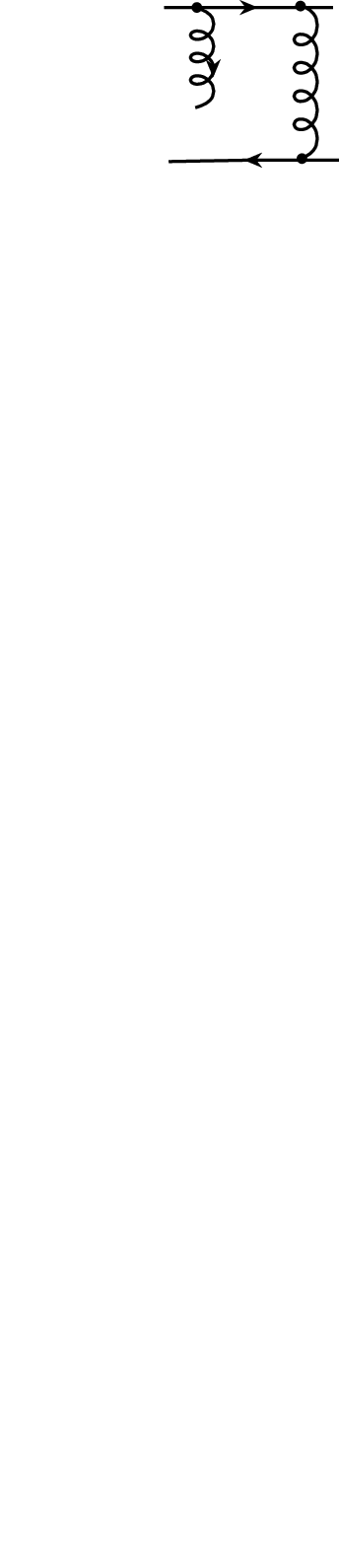}&
\includegraphics[scale=0.45]{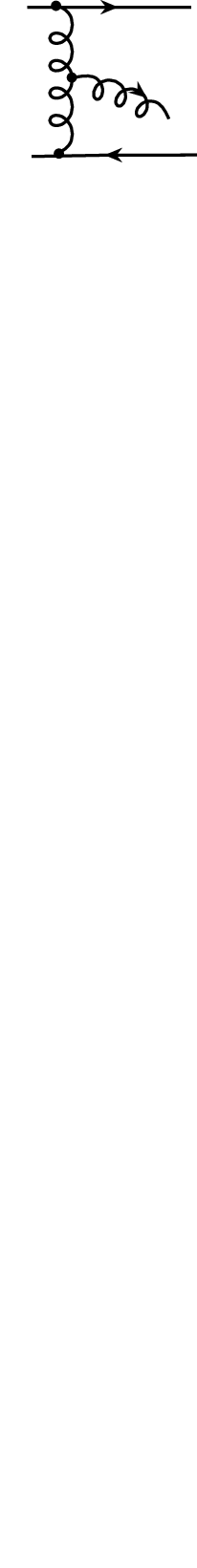}\\ $L_2$ & $R_3$ & $R_4$ & $R_5$
\end{tabular}
\end{center}
\caption[]{\small
The left and the right effective vertices for the $\bar Q Q$-cut.
 }
\label{fig:7}
\end{figure}

Now we proceed to the calculation of ${\cal I}_g^{\bar Q g}$.
The imaginary part related with the $\bar Q g$-cut presented in Figs.~8 and 9
reads
\begin{equation}
 \frac{1}{\pi}  Im\, {\cal I}_g^{\bar Q g}(y) =
    \left(
   \frac{y^2}{1+2y}
            \right)^\epsilon  
  \Theta_g^{\bar Q g}(y) \, ,
\end{equation}
where
\begin{multline} \Theta_g^{\bar Q g}(y) =
  - 2 \frac{1+2y(1+y)}{1+y}\left(\frac{c_1-c_2}{\epsilon}\right)
       -\frac{5c_1-4c_2}{4} - 4(c_1-c_2)y+ \frac{3c_2}{y}
   +\frac{3c_1-c_2}{2(1+y)}
      \\
   +
  \frac{5c_1}{4(1+2y)} - \frac{c_1}{4(1+2y)^2} +
   \Big(
   \frac{3c_1-4c_2}{2} + 4(c_1-c_2)y+
   \frac{ 9c_1-22c_2}{8y} +\frac{5c_1-2c_2}{8(1+y)}
  \\
    -\frac{c_1}{2(1+2y)}
   -\frac{3c_2}{4 y^2}
   -\frac{c_1-2c_2}{4(1+y)^2}
  \Big)
   \ln(1+2y) -\frac{1}{6}(c_1-c_2)(45-2\pi^2) \epsilon \, . 
\end{multline}
Expanding $\Theta_g^{\bar Q g}(y)$ in $\epsilon$
one needs to keep, in the limit of small $y$, the terms which are 
up to linear in $\epsilon$, since in the 
dispersion integral (\ref{disG})
they produce the contribution  $\sim \epsilon^0$.    
Calculating the dispersion integral with $Im\, 
{\cal I}_g^{(\bar Q g)}$ we obtain
\begin{multline}
 {\cal I}_g^{\bar Q g}(y)=
 \frac{c_1 - c_2}{\epsilon^{2}}
+ \frac{c_1-c_2}{4 \epsilon}
  \Big\{
   1 + 8(1 + 2y(1 + y))(
    \frac{\ln(-y)}{1 + y} - \frac{\ln(1 + y)}{y}
    )
  \Big\} \\
-\frac{c_1}{4}+ c_2 \frac{3 + 7 y (1 + y)}{2 y (1 + y)} \\
-\pi^2\left[c_1
    \frac{2 + y (1 + y) (43 + 100 y (1 + y))}{96 y^2(1 + y)^2}
      -c_2
    \frac{8 + y (1 + y) (47 + 61 y (1 + y))}{48 y^2(1 + y)^2}\right] \\
-\left[
  c_1 \frac{1 + 2 y (1 + y) (5 + 14 y (1 + y))}{2 y (1 + y)(1 + 2 y)^2}+
  c_2\frac{1 + 2 y (1 + y)}{2 y (1 + y)}
  \right]\ln(2) \\
+ 2(c_1-c_2)\big(1+2y(1+y)\big)
    \left(
     \frac{\ln^2(-y)}{1+y}-\frac{\ln^2(1+y)}{y}
    \right) \\
+ a_1(y)\ln(-y) + a_1(-1-y)\ln(1+y)  \\
+ a_2(y)Li_2(1+2y) + a_2(-1-y)Li_2(-1-2y) \, ,
\label{IgQq}
\end{multline}
where the functions $a_1$ and $a_2$ are given by the following expressions:
\begin{multline}
a_1(y) =
    \frac{c_1}{4}
     \left(
      5 + 16 y - \frac{6}{1 + y} + \frac{1}{( 1 + 2 y)^2} - \frac{5}{1 + 2 y}
     \right)
      \\ - \frac{c_2}{2}
     \left(
      2 + \frac{3}{y} + 8 y - \frac{1}{1 + y}
     \right) \, ,
\label{a1y}
\end{multline}
\begin{multline}
a_2(y) =
    \frac{c_1}{8}
     \left(
      12 + \frac{9}{y} + 64\, y - \frac{2}{( 1 + y )^2} 
       + \frac{21}{1 + y} - \frac{4}{1 + 2 y}
     \right)
    \\ -\frac{c_2}{4}
     \left(
      8 + \frac{3}{y^2} + \frac{11}{y} + 32\,
           y - \frac{2}{( 1 + y)^2} + \frac{9}{1 + y}
     \right) \, .
\label{a2y}
\end{multline}

\begin{figure}
\begin{center}
\scalebox{0.7}{
\input{g_gbq.pstex_t}}
\end{center}
\caption[]{\small
The contribution of the $\bar Q g$ intermediate state to the gluonic
amplitude.
 }
\label{fig:8}
\end{figure}

\begin{figure}
\begin{center}
\begin{tabular}{cccc}
\includegraphics[scale=0.35]{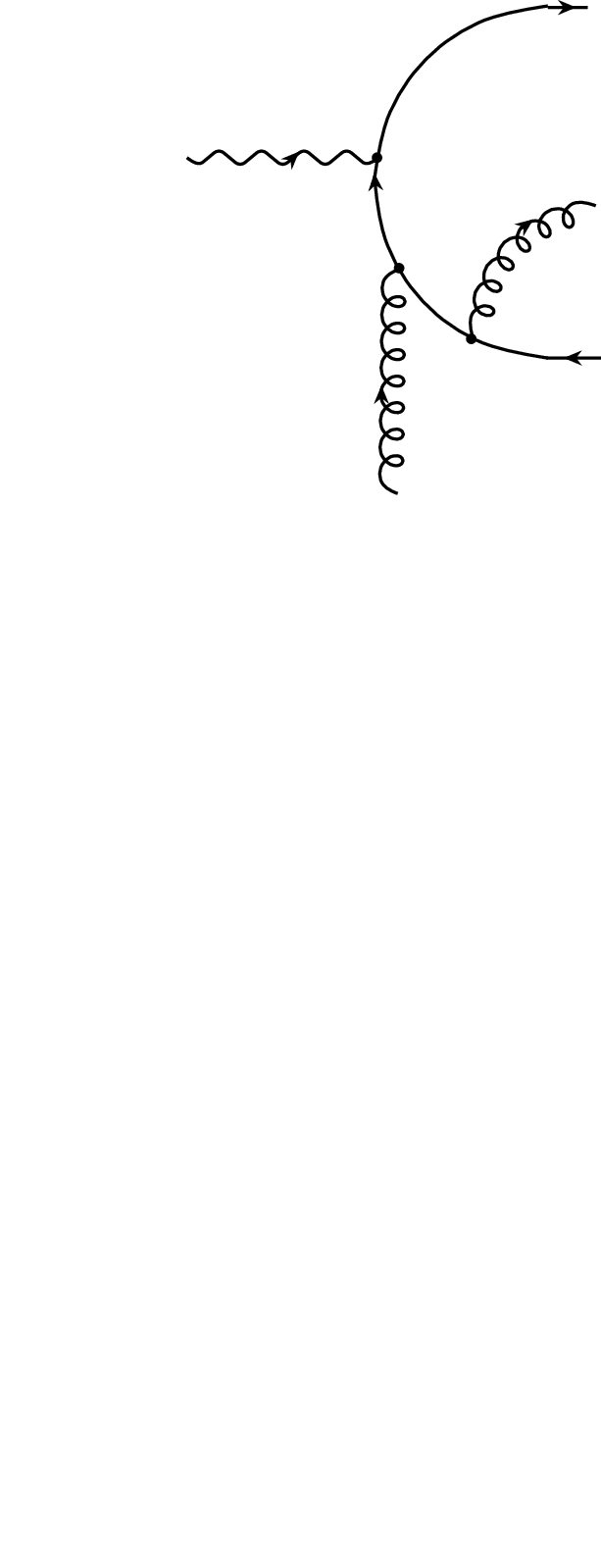}&
\includegraphics[scale=0.35]{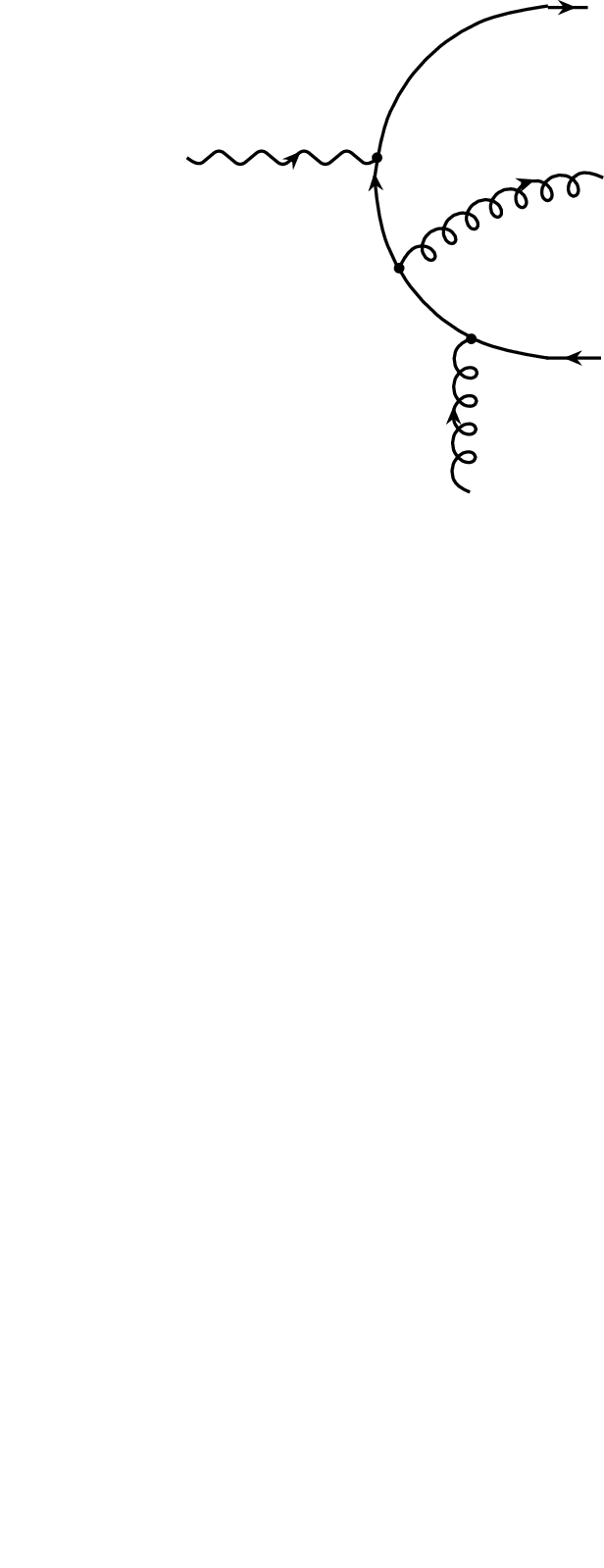}&
\includegraphics[scale=0.35]{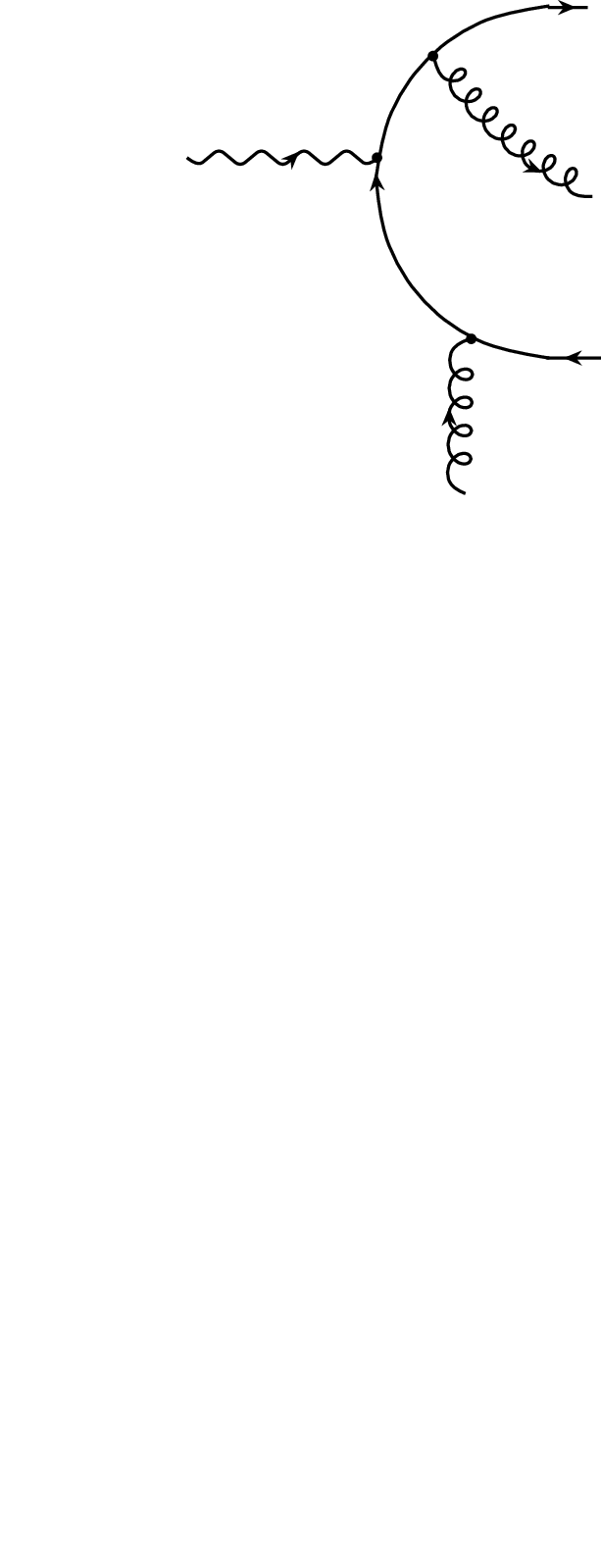}&
\includegraphics[scale=0.35]{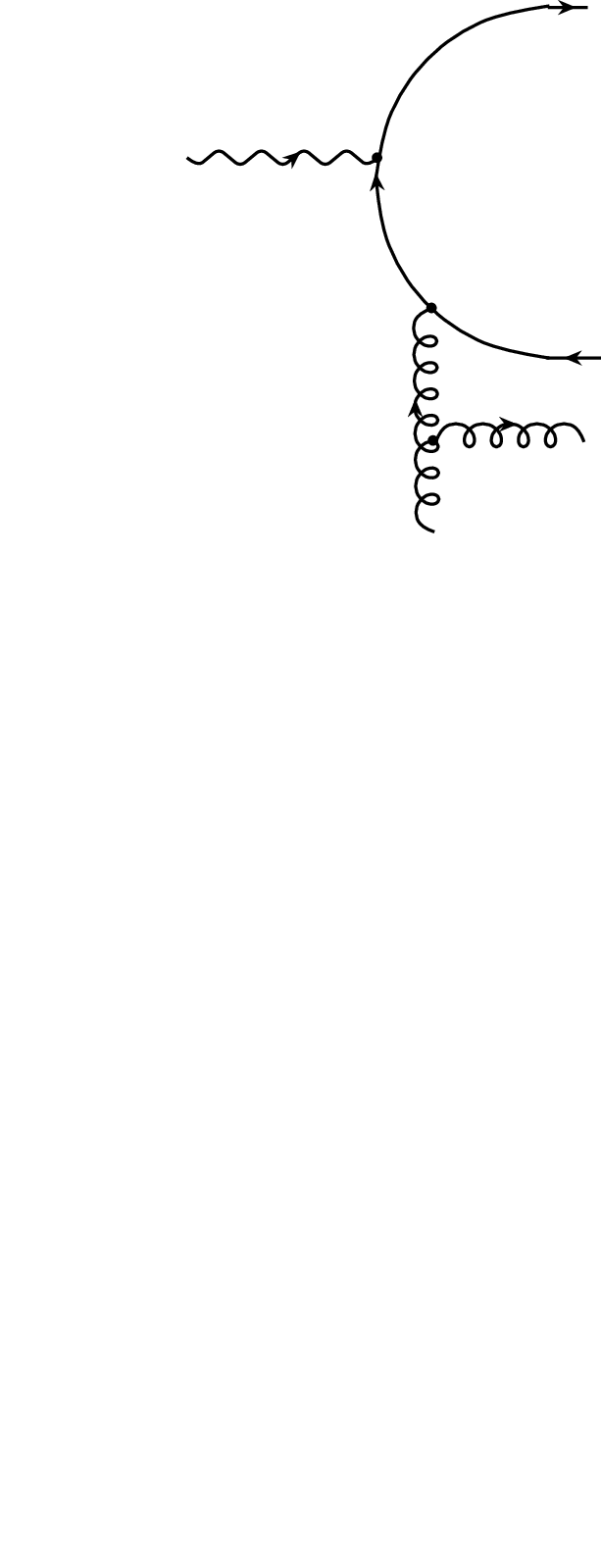}\\ $ L_1$ & $L_2$ & $L_3$ & $L_4$\\[1mm]
\includegraphics[scale=0.35]{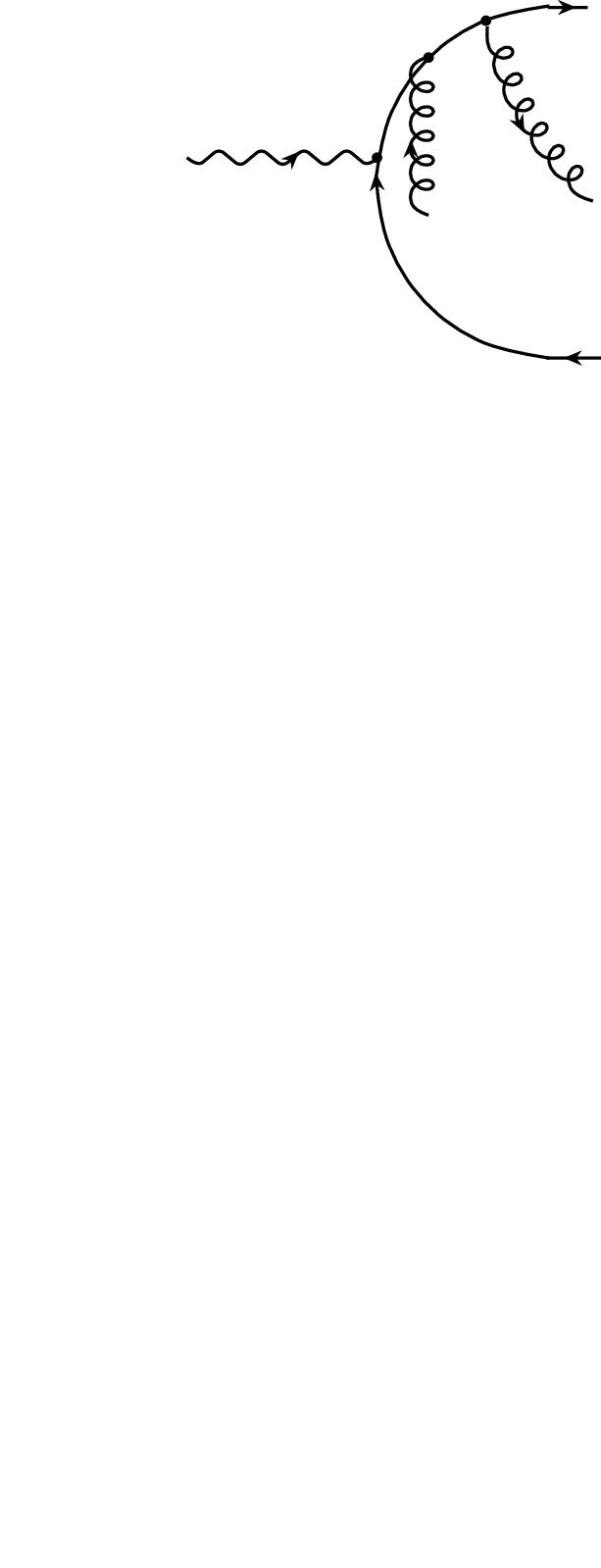}&
\includegraphics[scale=0.35]{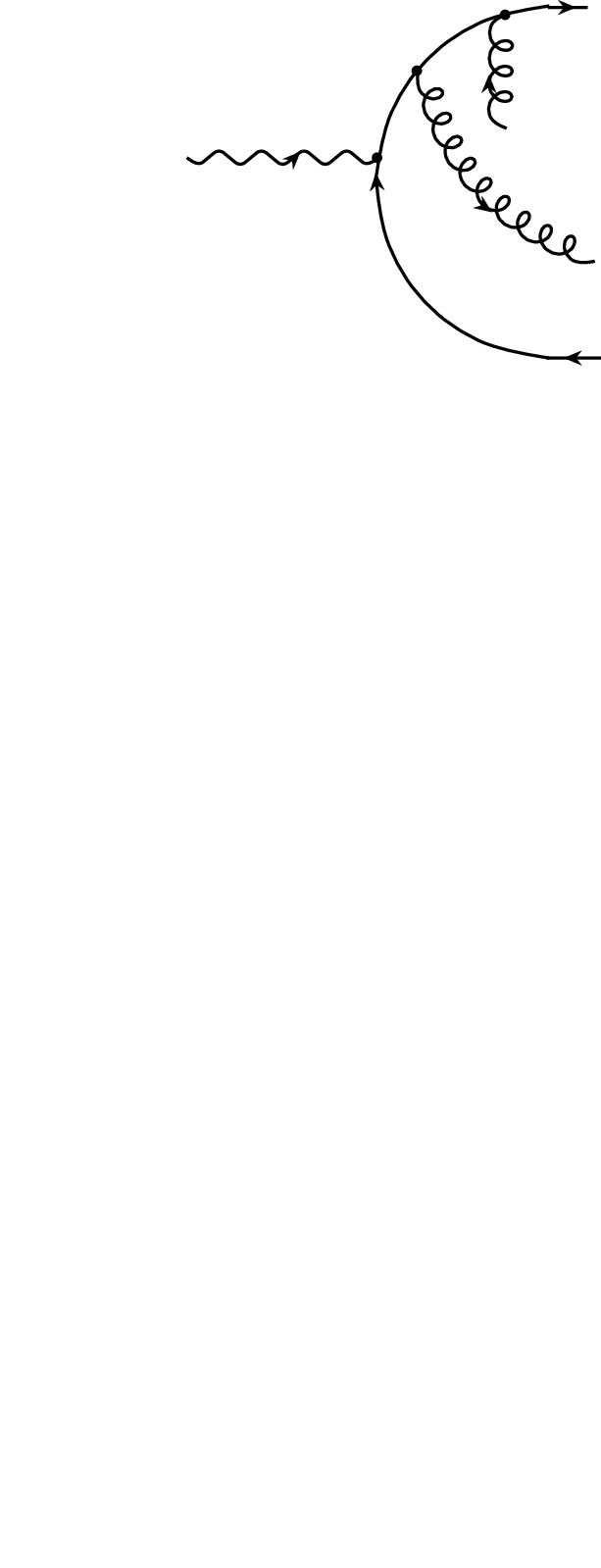}&
\includegraphics[scale=0.35]{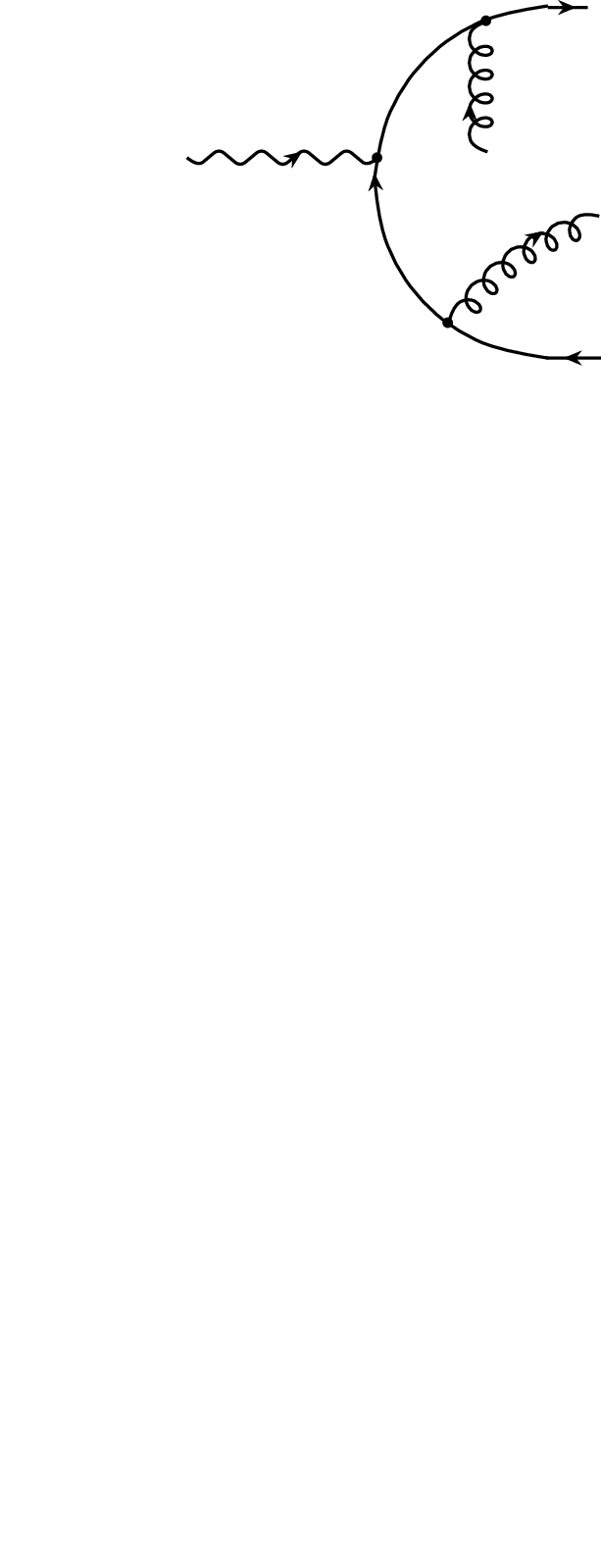}&
\includegraphics[scale=0.35]{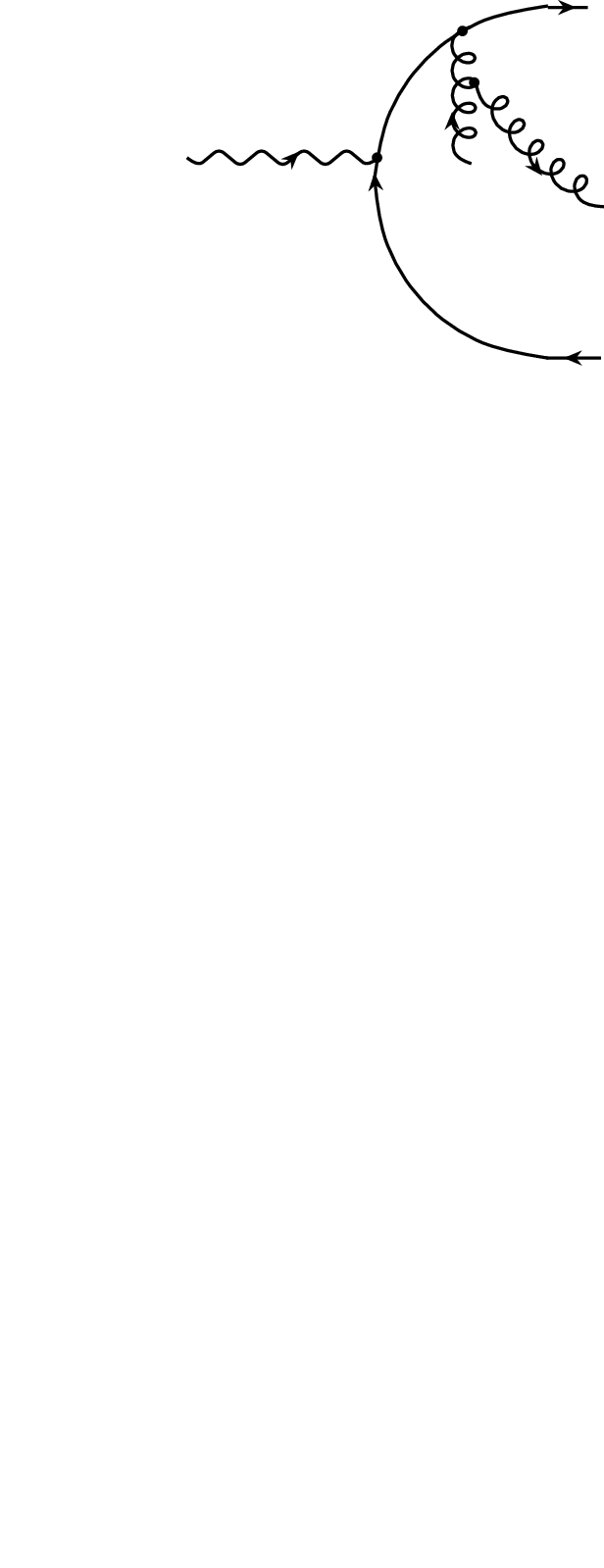}\\ $ L_5$ & $L_6$ & $L_7$ & $L_8$
\end{tabular}\\[1mm]
\begin{tabular}{ccc}
\includegraphics[scale=0.35]{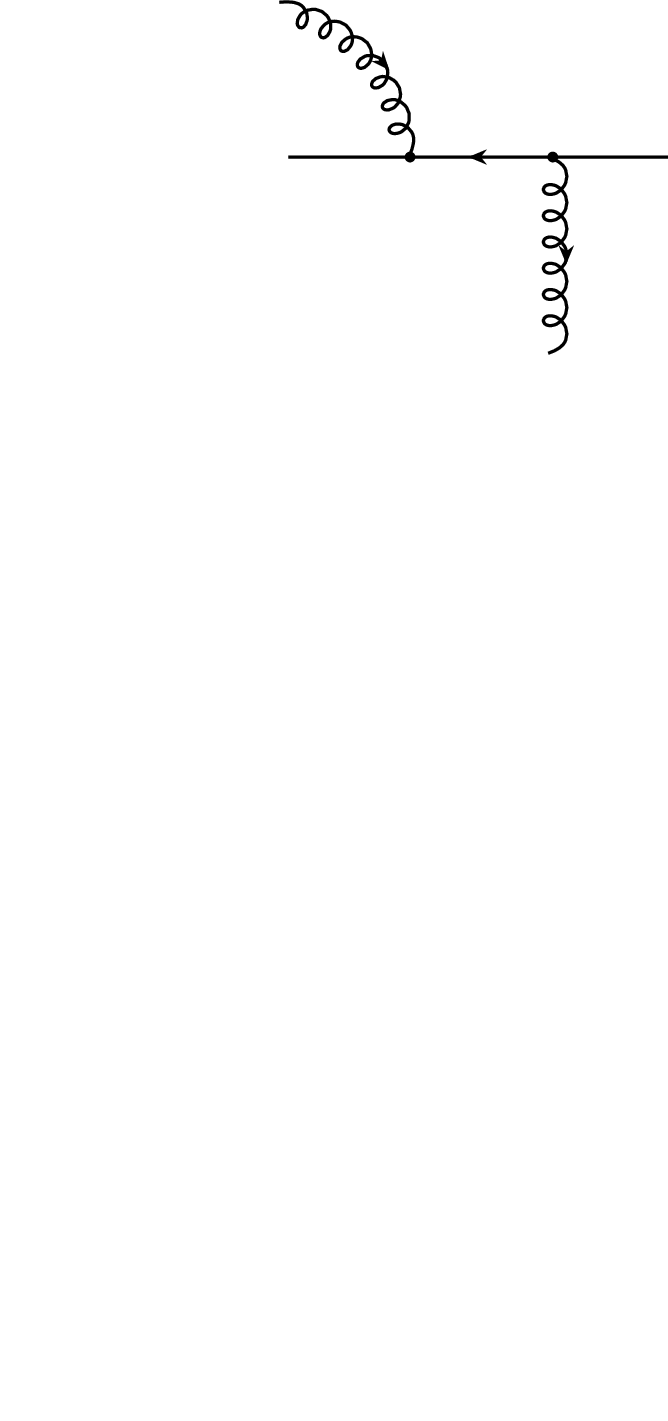}&
\includegraphics[scale=0.35]{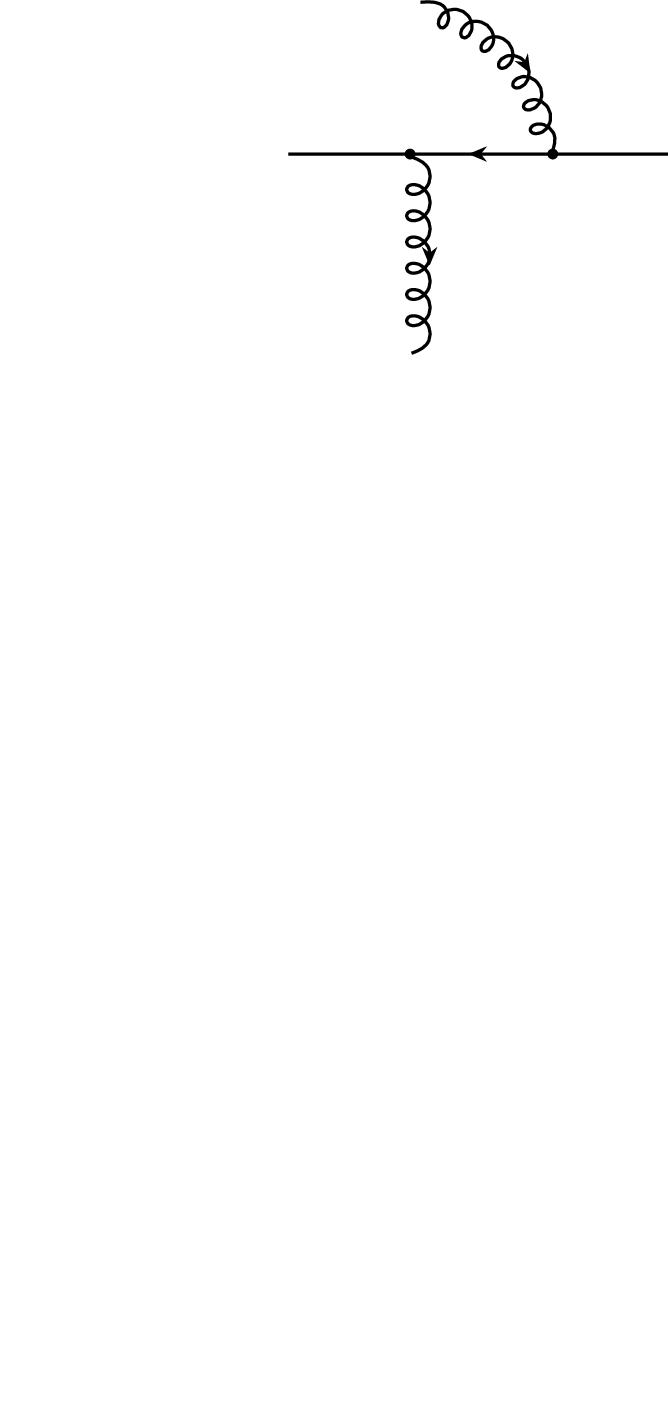}&
\includegraphics[scale=0.35]{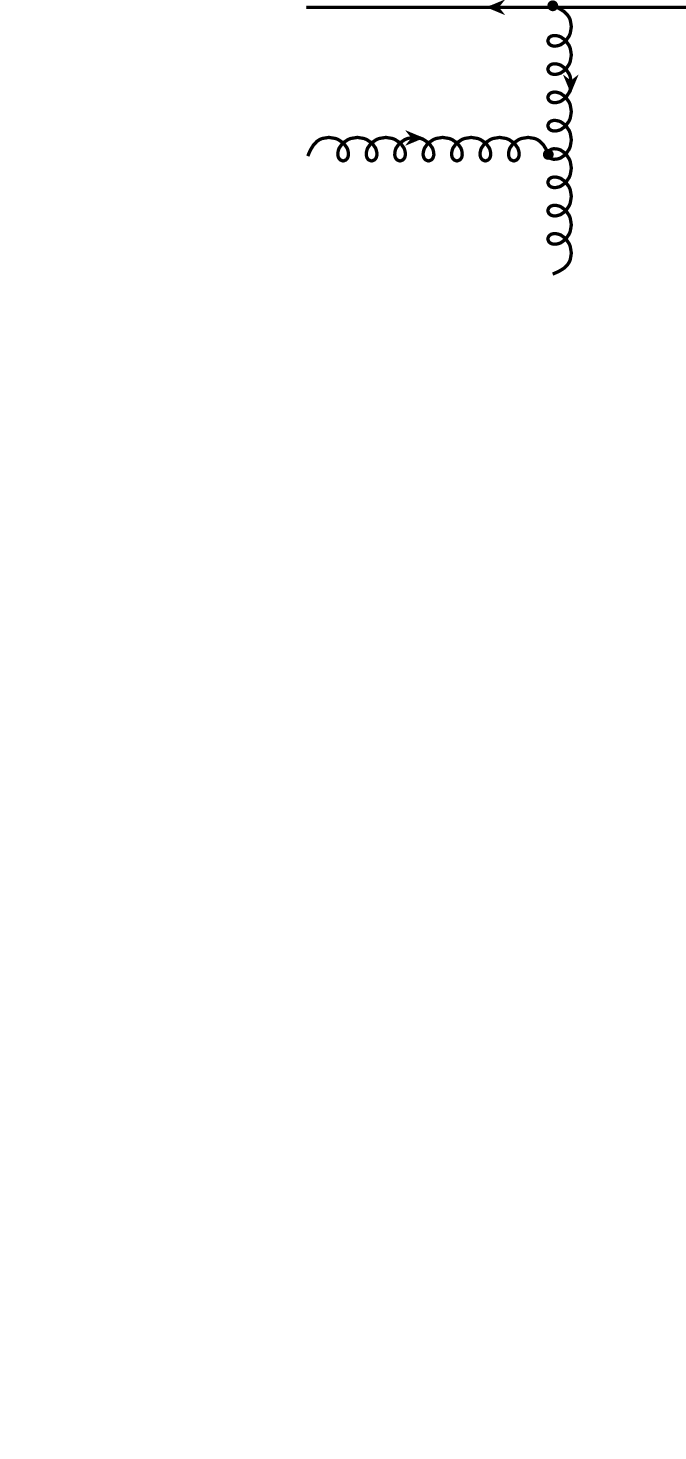}\\ $R_1$ & $R_2$ & $R_3$
\end{tabular}
\end{center}
\caption[]{\small
The left and the right effective vertices for the $\bar Q g$-cut.
 }
\label{fig:9}
\end{figure}

Eqs.~(\ref{IQQ},\ref{IgQq}) 
define the r.h.s. of eq.~(\ref{three1}). To finish our consideration of the
gluon contribution one still needs to
evaluate  ${\cal I}_g(0)$, the one-loop amplitude describing the 
emission of a soft gluon. 

\subsubsection{The emission of a soft gluon} 

The idea of our method is inspired by the famous result of Low 
\cite{Low:1958sn},
known as low energy theorem for radiation of a photon.
The arguments of Low may be used to constrain the amplitude describing the
emission of a soft gluon. In a non-abelian case, 
due to the confinement phenomenon,
the corresponding result has not such a fundamental meaning as in QED.
Nevertheless,
it can be useful, for problems treatable by  
perturbative methods. We will first explain  the essential steps of our
approach for a simple example, namely 
the calculation of the LO gluonic amplitude
(\ref{LOg}).
Then we proceed to the evaluation of ${\cal I}_g(0)$.       

Let us consider the  gluonic process
\begin{equation}
\gamma(q)\,  G(x_1 p)\to V(K) \, G(x_2 p) 
\label{ssss}
\end{equation}
at LO in the limit
when the emitted gluon is soft; $x_2\to 0,\,  x_1\to \zeta$.
With respect to the soft gluon the diagrams in
Fig.~\ref{fig:2}
may be divided into three groups. In diagrams {\it a)} and {\it b)} 
the soft gluon is
radiated from the on-shell quark line, in diagrams {\it c)} and 
{\it d)} it is attached
to the on-shell antiquark line, whereas in diagrams {\it e)} and 
{\it f)} the soft gluon 
is emitted from the virtual antiquark and the virtual 
quark lines, respectively. 
In the first two cases 
the quark propagator attached to the soft gluon vertex is close to the mass
shell. We call them pole contributions, contrary to the third non-pole 
case which describes the emission of the soft gluon from 
the internal part of the process.\footnote{ 
Due to color neutrality of the two gluons in the process 
(\ref{ssss}) the emission of gluon $G(x_2 p)$ from the on-shell 
line of gluon $G(x_1 p)$ is forbidden, thus in our case 
there is no gluon pole contribution.} 
Our idea is 
to calculate the 
amplitude of the process (\ref{ssss}) in the soft gluon limit 
considering the pole contributions only. Below we will show how using gauge
invariance 
the non-pole contributions may be
derived from the pole ones. 

Neglecting the proton mass and $\Delta_\perp$ one has
\begin{equation}
p=(1+\xi)\, W\, n_+ \, , \ \ \ q=\frac{W}{2(1+\xi)}\, n_- \, , \ \ \
K=q+\zeta p \, .
\label{ffff}
\end{equation}
For the photon polarization vector we choose the gauge $(e_\gamma p)=0$, hence
$e_\gamma=e_\gamma^\perp$. Since we are interested in the helicity non-flip 
amplitude the meson polarization vector can also be  chosen transverse,
$e_V=e_V^\perp$.  

For the process (\ref{ssss}) in collinear kinematics it
happens that for the pole contributions a pole 
factor $1/x_2$ coming from the denominator of the
quark propagator is compensated by the factor $x_2$ from the nominator. This
means that contributions of both the pole and the non-pole diagrams 
are regular at $x_2\to 0$ and that both classes of diagrams  
contribute to the amplitude on  equal footing.  
However in order to apply our method we need to have a pole factor in
the pole contributions. For this purpose we change the kinematics of the
process (\ref{ssss}) slightly away from the collinear one 
introducing the small transverse component to 
the momenta of the photon and the soft gluon
\begin{equation}
q\to q^\prime =q+k_\perp \, , \ \ \ x_2 p\to k= x_2 p + k_\perp \, .
\label{jjjj}
\end{equation}  
Note that    
this replacement makes the photon and the soft gluon 
lines slightly virtual, 
$q^{\prime 2} =k^2=k_\perp^2$. But this effect 
is quadratic in $k_\perp$ and, therefore,
it is small and can be safely neglected, as we will always do below. 

The change of the photon momentum 
(\ref{jjjj}) leads to the following replacement in the expression for the
photon polarization vector
\begin{equation}
e_\gamma=e_\gamma^\perp \to e_\gamma^\prime=
e_\gamma^\perp -\frac{(e_\gamma^\perp
k_\perp)}{(p q)}\, p \, , \ \ \
(e_\gamma^\prime q^\prime) =0 \, .  
\label{llll}
\end{equation}
We denote the polarization vectors of the gluons with 
momenta $x_1 p$ and $k$ 
by $e^1_g$ and $e^2_g$,
\begin{equation}
(e^1_g\, p)=0\, , \ \ \ (e^2_g\, k)=0 \, , 
\label{glvect}
\end{equation}
and choose a gauge such that $(e^1_g\, q)=(e^2_g\, q)=0\, $. Thus, the
polarization vector of the first gluon is transverse, $e^1_g=e^{1\perp}_g$,
whereas the polarization vector of the soft gluon contains both a
transverse and a longitudinal component. $e^2_g$ is  transverse
only in the collinear limit:  $e^2_g\to e^{2\perp }_g$ at $k_\perp\to 0\,$.

Let us consider one of the pole diagrams, say, diagram {\it b)}. Its
contribution to the gluonic amplitude reads
\begin{equation} 
A_{b)}=D\, Sp \left[
\not \! e^2_g \frac{\not \! K/2+\not \! k+m}{(K/2+k)^2-m^2}
\not \! e^1_g \frac{\! \not \! K/2+\! \not \! k
-x_1\!\! \not \! p+m}{(K/2+k-x_1 p)^2-m^2}\not \! e^{\, \prime}_\gamma
\not \! e^*_V\, (\! \not \! K+M) 
\right] \, ,
\label{diagramb}
\end{equation}   
here $D$ is some factor which is 
irrelevant for our argumentation. The first propagator on the r.h.s.
of (\ref{diagramb}) is the 
propagator of the quark attached to the
soft gluon vertex. Its denominator, 
$(K/2+k)^2-m^2=(k K)$, vanishes in the soft gluon limit.  
In accordance with the  nominator of this propagator we define two
contributions 
\begin{eqnarray}
&&
A_{b)}=A^{add}_{b)}+A^z_{ b)} \, ;
\ \ \ {\rm where}
\label{twocont} \\
&&
A^{add}_{b)}=D\, Sp \left[
\not \! e^2_g \frac{\not \! K/2+m}{(K/2+k)^2-m^2}
\not \! e^1_g \frac{\! \not \! K/2+\! \not \! k
-x_1\!\! \not \! p+m}{(K/2+k-x_1 p)^2-m^2}\not \! e^{\, \prime}_\gamma
\not \! e^*_V\, (\! \not \! K+M)
\right] \, ,
\nonumber \\
&&
A^z_{b)}=D\, Sp \left[
\not \! e^2_g \frac{\not \! k}{(K/2+k)^2-m^2}
\not \! e^1_g \frac{\! \not \! K/2+\! \not \! k
-x_1\!\! \not \! p+m}{(K/2+k-x_1 p)^2-m^2}\not \! e^{\, \prime}_\gamma
\not \! e^*_V\, (\! \not \! K+M)
\right] \, .
\nonumber
\end{eqnarray}
Commuting in the first term the factors $\not \! e^2_g$ and $(\not \!
K/2+m)$ we obtain 
\begin{equation}
A^{add}_{b)}=D\, \frac{( e^2_g K)}{(k K)}\, Sp \left[
\not \! e^1_g \frac{\! \not \! K/2+\! \not \! k
-x_1\!\! \not \! p+m}{(K/2+k-x_1 p)^2-m^2}\not \! e^{\, \prime}_\gamma
\not \! e^*_V\, (\! \not \! K+M)
\right] \, .
\label{case1}
\end{equation}   
The trace on the r.h.s. of eq.~(\ref{case1}) vanishes linearly in
$k$. Indeed, using the 
properties of the polarization vectors discussed above it is easy to see that 
\begin{equation}
A^{add}_{b)}=-\frac{D}{m} \,
\frac{( e^2_g K)}{(k K)}  \, Sp \left[
\not \! e^1_g  \not \! k
\not \! e^{\, \prime}_\gamma
\not \! e^*_V
\right] +{\cal O}(k) \, .
\label{case11}
\end{equation}     
Similarly, for the second term we obtain
\begin{equation}
A^{z}_{b)}=-\frac{D}{m\, (k K)} \, Sp \left[
\not \! e^2_g\not \! k\not \! e^1_g 
\not \! e^{\, \prime}_\gamma
\not \! e^*_V\! \not \! K
\right] +{\cal O}(k) \, .
\label{case2}
\end{equation}
$A^{add}_{ b)}$ 
vanishes in the collinear kinematics since 
at $k_\perp\to 0$: $k\to x_2 p$, $e^2_g\to e^{2\perp}_g$ and
$(e^2_g K)\to 0\, $. 
However, for $k_\perp\neq 0$ both contributions to $A_{ b)}$ are of the
same order. Note that $A^{z}_{ b)}$ is finite for 
$k_\perp=0$ and $x_2\to 0$.

The consideration of the other pole diagrams follows the same
lines. We calculated, similar to eqs. (\ref{case11},\, \ref{case2}), the
corresponding contributions to $ A^{add}$ 
and $A^{z}$ of each pole diagram.    
The total gluonic amplitude is
\begin{equation}
A\equiv e^{2,\, \mu}_{g}A_{\mu}=e^{2,\, \mu}_{g}
\left[
A^{add}_{\mu}+A^{z}_{\mu}+ A^{n-pole}_{\mu}
\right] \, ,
\label{formgluonic}
\end{equation}  
where the first two terms represent the pole contributions,
and 
the third term stands for the contribution of the non-pole diagrams.
The later can be obtained from $A^{add}_{\mu}$ using gauge
invariance. Due to current conservation we have
\begin{equation}
k^{\mu}A_{\mu}=0 \, .
\label{gaugeinv}
\end{equation} 
Since
\begin{equation}
k^{\mu} A^z_{ \mu}=0 
\label{gaugeinv1}
\end{equation}
by  construction, see eqs.~(\ref{twocont},\, \ref{case2}), we obtain 
\begin{equation}
k^{\mu} A^{add}_{ \mu}=-k^{\mu} A^{n-pole}_{ \mu} \, .
\label{deduce}
\end{equation}
In its turn $A^{add}_{ \mu}$ has the form
\begin{equation}
A^{add}_{ \mu}=\frac{  K_\mu}{(k K)}(k P) \, , \ \ \
P^\mu=\sum_i P_i^\mu \, .
\label{P}
\end{equation}
The vector $P$ receives contributions from the pole diagrams enumerated
by the index $i$. For instance, according to eq.~(\ref{case11}), 
the contribution of diagram {\it b)} is
\begin{equation}
P_{b)}^\mu=-\frac{D}{m}   \, Sp \left[
\not \! e^1_g  \not \! \gamma^\mu
\not \! e^{\, \prime}_\gamma
\not \! e^*_V
\right] \, .
\label{Pb}
\end{equation} 
Similarly, we denote the contributions of separate pole diagrams to
$A^{z}$ as $A^{z}_i$,  
\begin{equation}
A^{z}=\sum_i A^{z}_i \, . 
\label{Azis}
\end{equation}
From eqs.~(\ref{deduce}) and (\ref{P}) we deduce that 
\begin{equation}
A^{n-pole}_{\mu}=-P_\mu \, .
\label{A-nonpole}
\end{equation}  
Thus we have shown how in the soft gluon limit 
the contribution of the non-pole diagrams 
can be derived without explicit calculations. 

Returning to the collinear kinematics, we have 
\begin{equation}
A|_{k_\perp\to 0}=e^{2\, \perp \, ,\, \mu}_{g}
\left[
A^{z}_{\mu}-P_\mu
\right]|_{k_\perp\to 0} \,\, ,
\label{backtocoll}
\end{equation} 
here we used that $A^{add}$ vanishes at $k_\perp\to 0$. Thus, the first 
term in eq.~(\ref{backtocoll}) represents the contribution of the pole diagrams
in the collinear limit whereas the second term, $\sim P_\mu$, restores
the non-pole contribution to the gluonic amplitude in this limit.

Finally, to obtain the gluon hard scattering amplitude ${\cal
A}^{(0)}_{g}(y=0)$ one needs to
perform 
the summation over  $2+2\epsilon$
transverse polarizations of the gluons ($e^{1\, \perp \, ,\,
\mu}_{g\, ,\, \lambda}=
e^{2\, \perp \, ,\, \mu}_{g\, ,\, \lambda}, \ \lambda=1,\dots,2+2\epsilon$)
in the amplitude of the gluonic process $A|_{k_\perp\to 0}$, 
and then take the limit $x_2\to 0$.  

Proceeding separately for each pole diagram 
with these  steps,  including the
summation over the gluon polarizations,  
we find 
\begin{equation}
{\cal A}^{(0)}_{g}(y=0)=\sum_i  
D_i \, , \ \ \
D_i=D^{z}_{ i}
+D^{add}_{i} \, ,
\label{LOgluonA}
\end{equation}
where $D_i$ stands for the contribution to the gluon hard scattering
amplitude
of the individual diagram.
$D^{z}_{ i}$ and 
$D^{add}_{ i}$ corresponds to the contribution of the pole diagram $i$
to $A^z_\mu$ and $P_\mu$ respectively. After a simple calculation we find
\begin{equation}
D^{z}_{ b)}=D^{z}_{ d)}=D^{add}_{ b)}=D^{add}_{ d)}=\frac{\alpha_S
(1+\epsilon)}{4}\, , \ \ \
D^{z}_{ a)}=D^{z}_{ c)}=D^{add}_{ a)}=D^{add}_{ c)}=0 \, .
\label{LOdiagrR} 
\end{equation}
Thus we confirm eq.~(\ref{LOg}). 
Calculating contributions of the non-pole 
diagrams {\it e)} and {\it f)} directly
one can check that the
non-pole contribution is indeed correctly restored by 
$\sum_i D_i^{add}$. 
Although it is very simple this  calculation 
contains all essential points of our method.   

\begin{figure}
\begin{center}
\kar{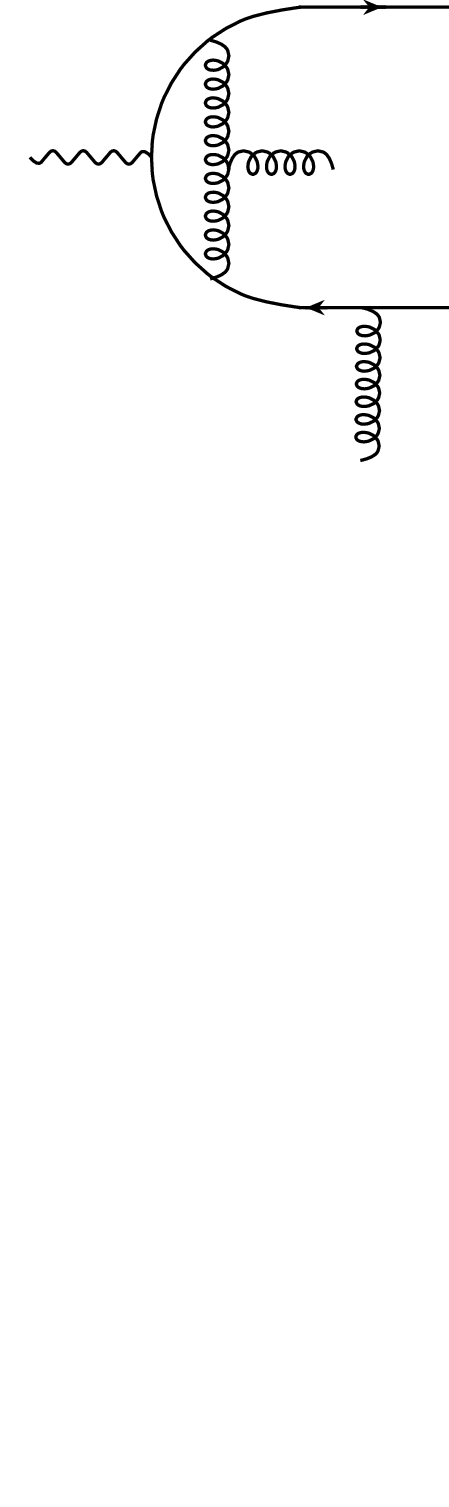}{$D_{1}$}
\kar{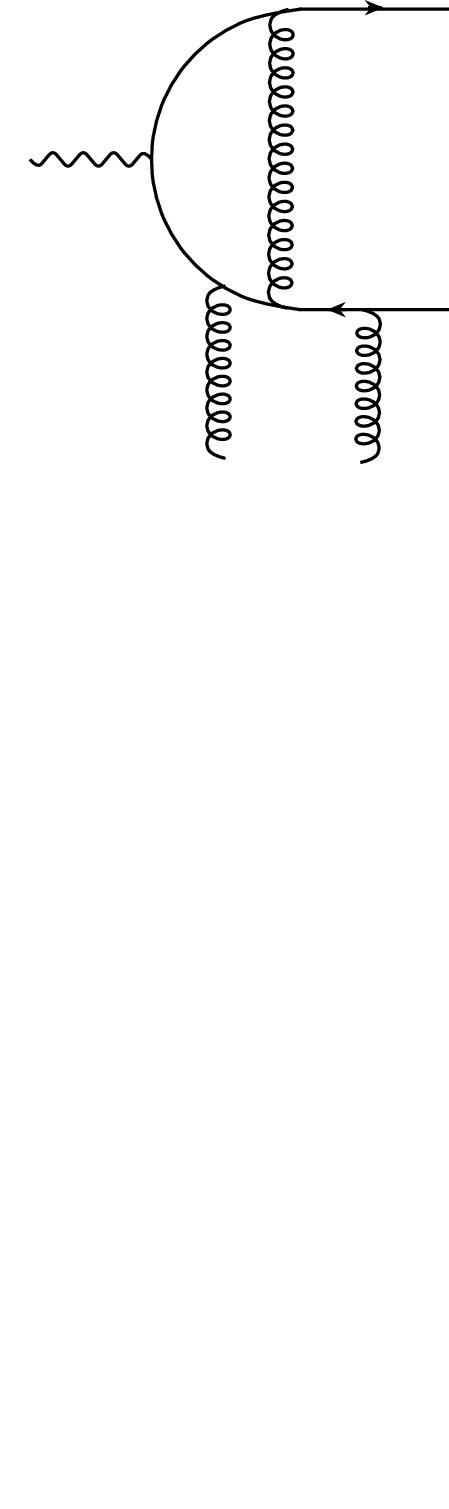}{$D_{2}$}
\kar{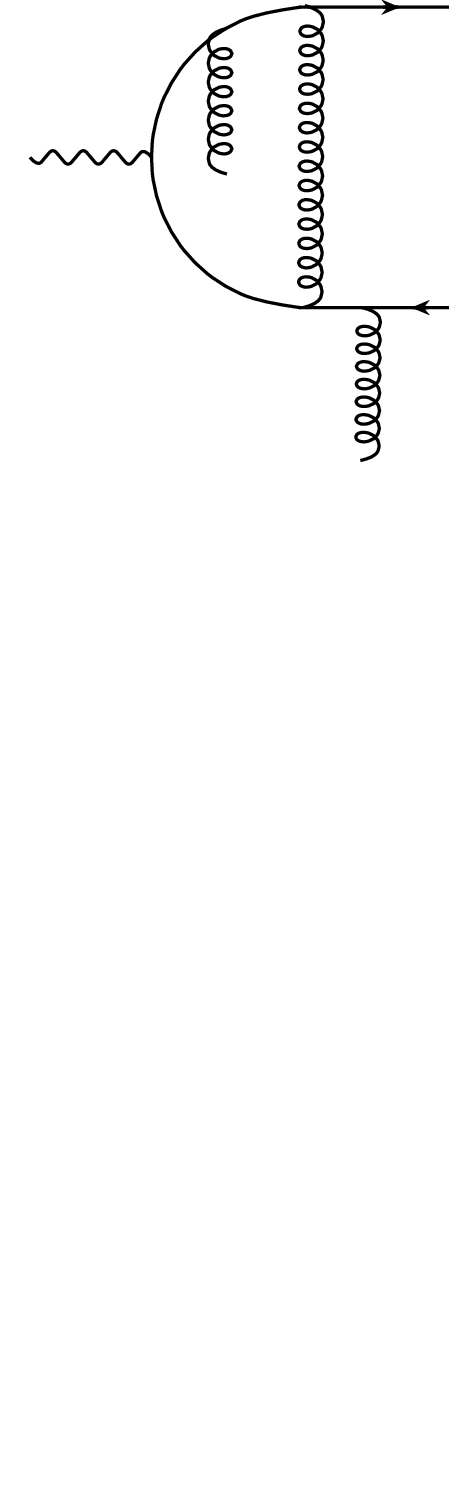}{$D_{3}$}
\kar{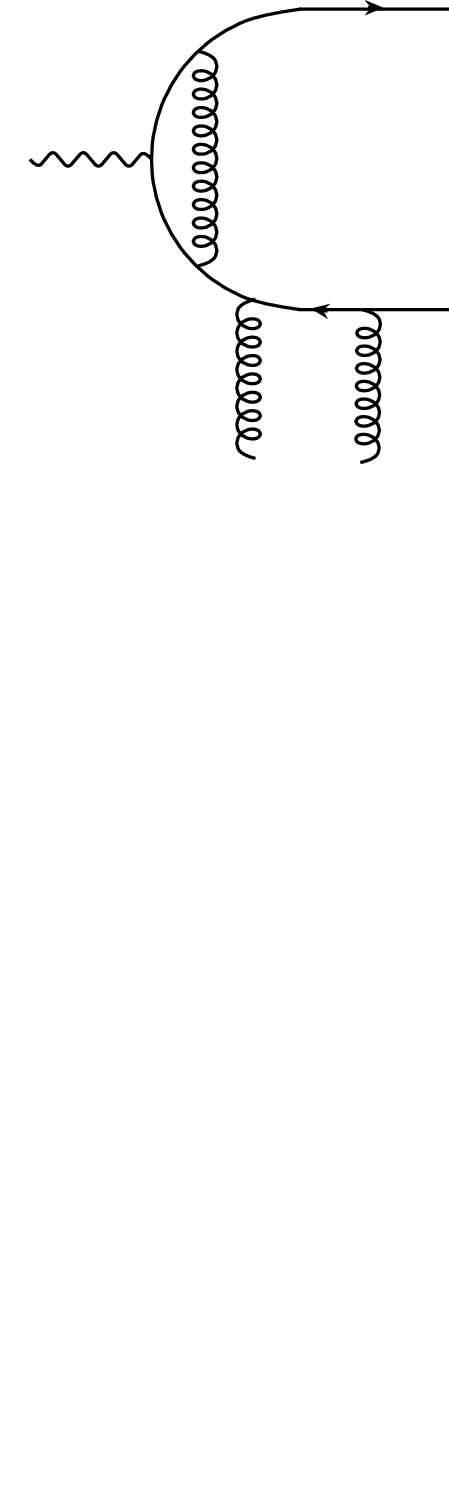}{$D_{4}$}
\kar{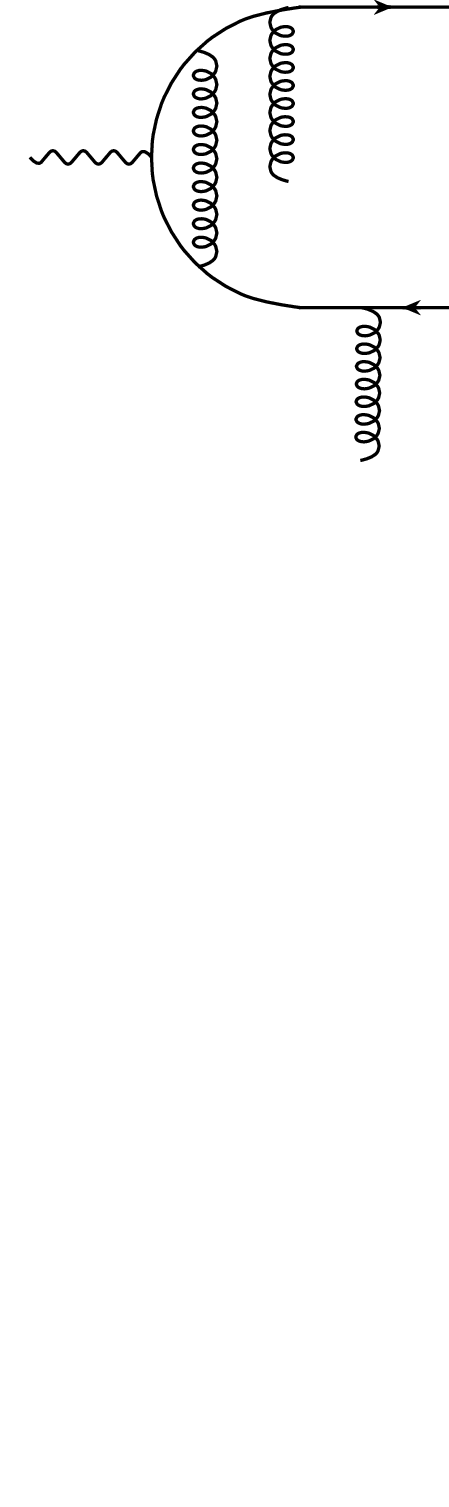}{$D_{5}$}
\kar{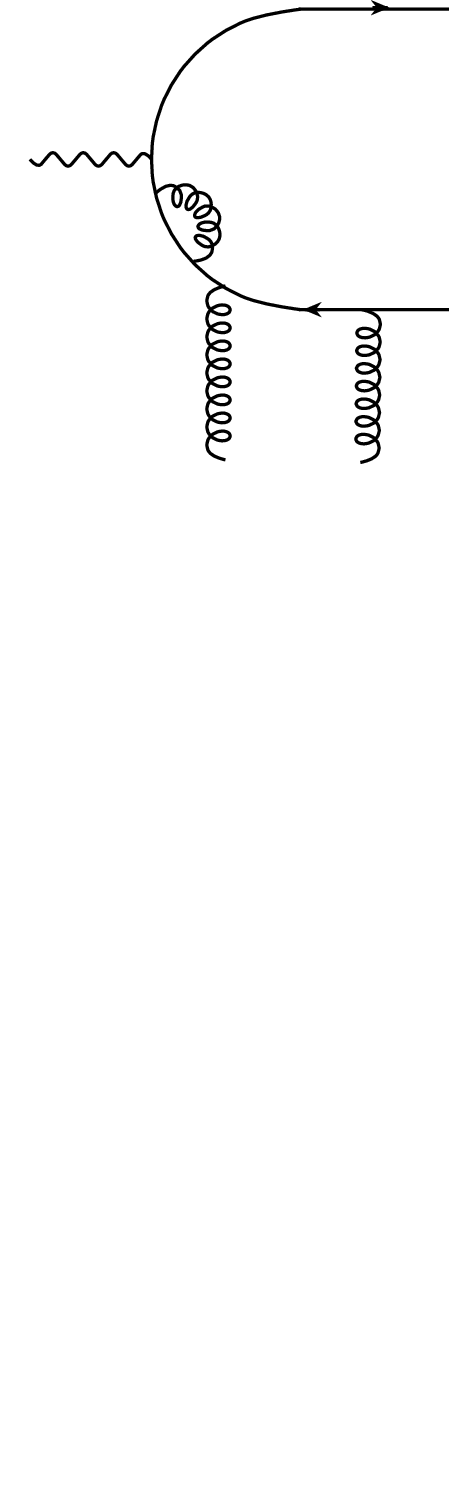}{$D_{6}$}
\kar{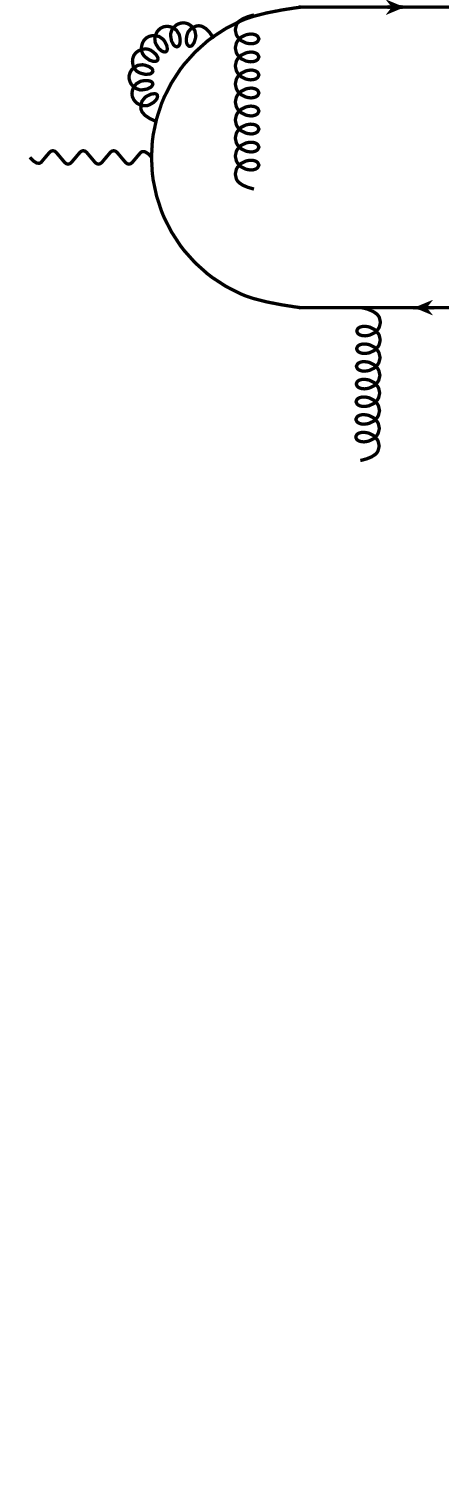}{$D_{7}$}
\kar{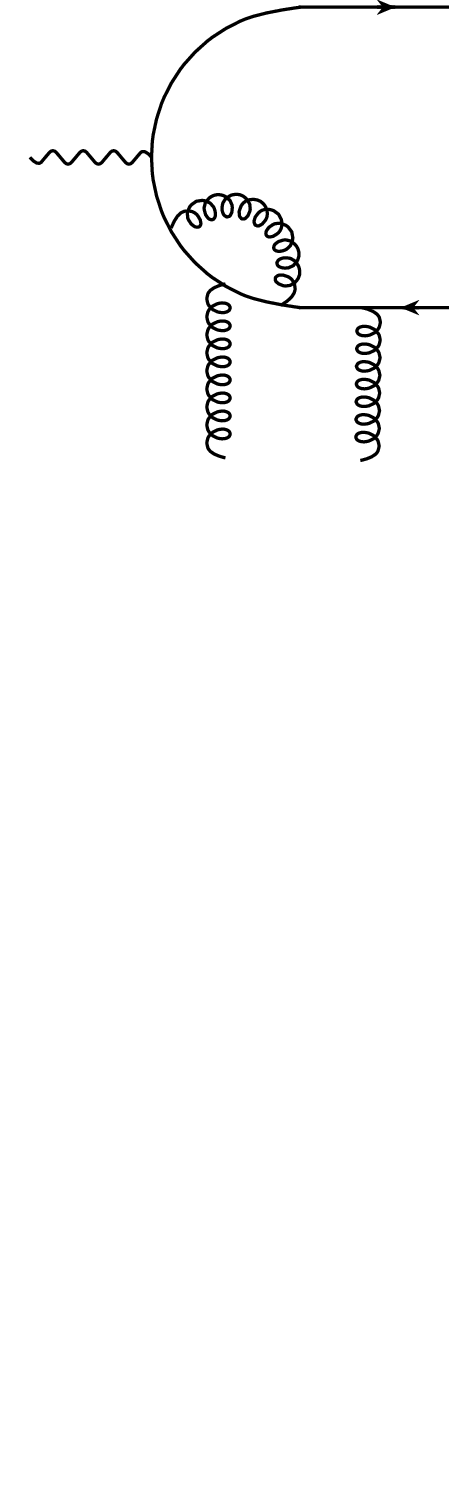}{$D_{8}$}
\kar{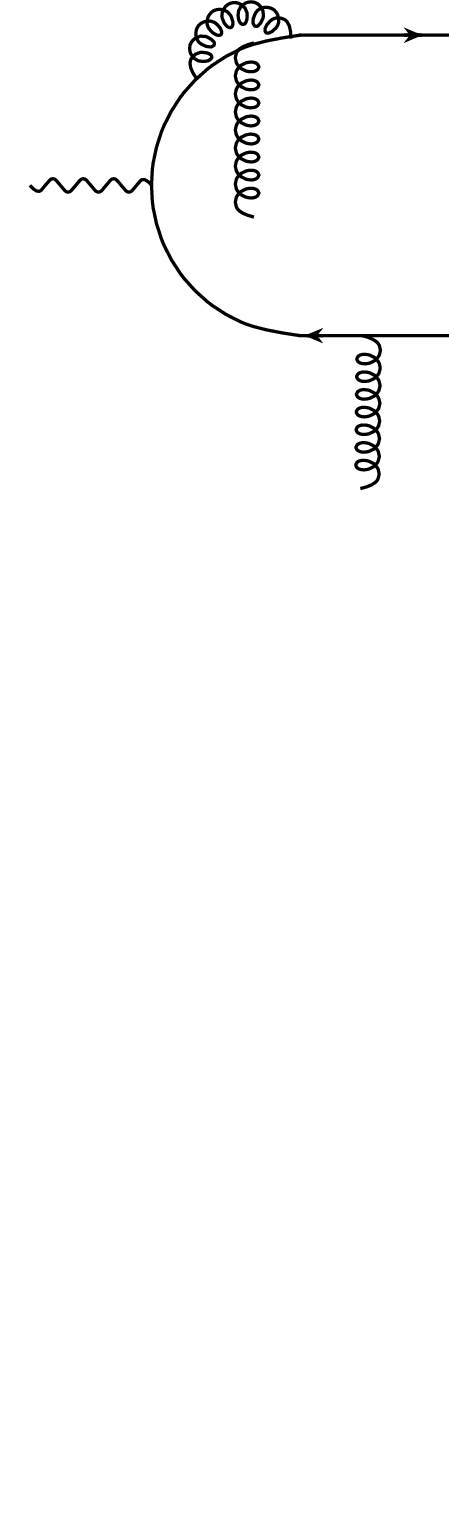}{$D_{9}$}
\kar{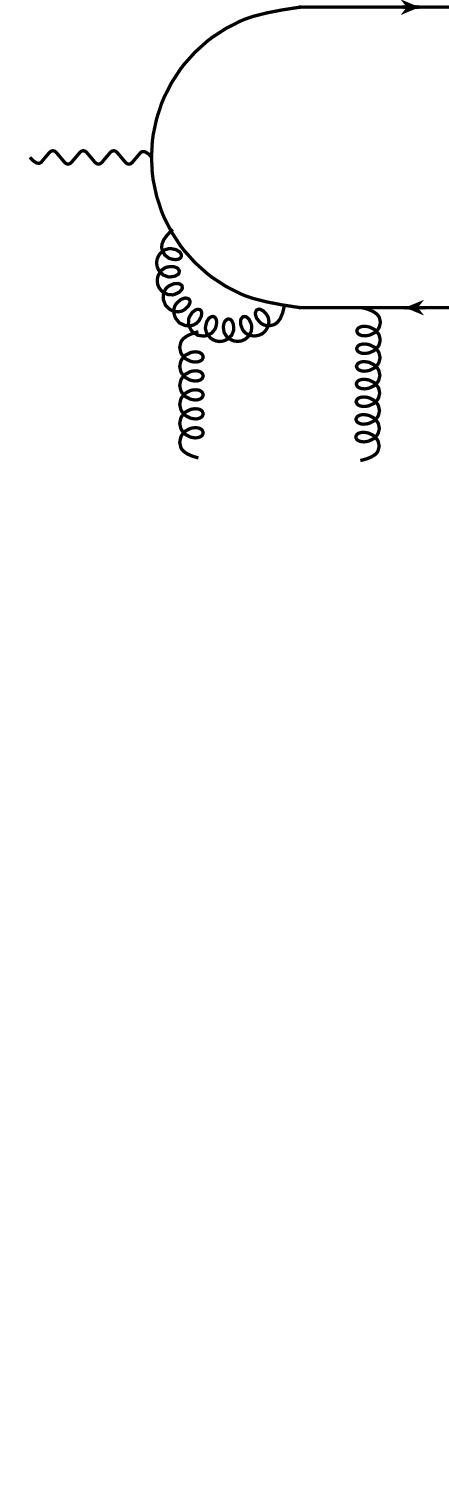}{$D_{10}$}
\kar{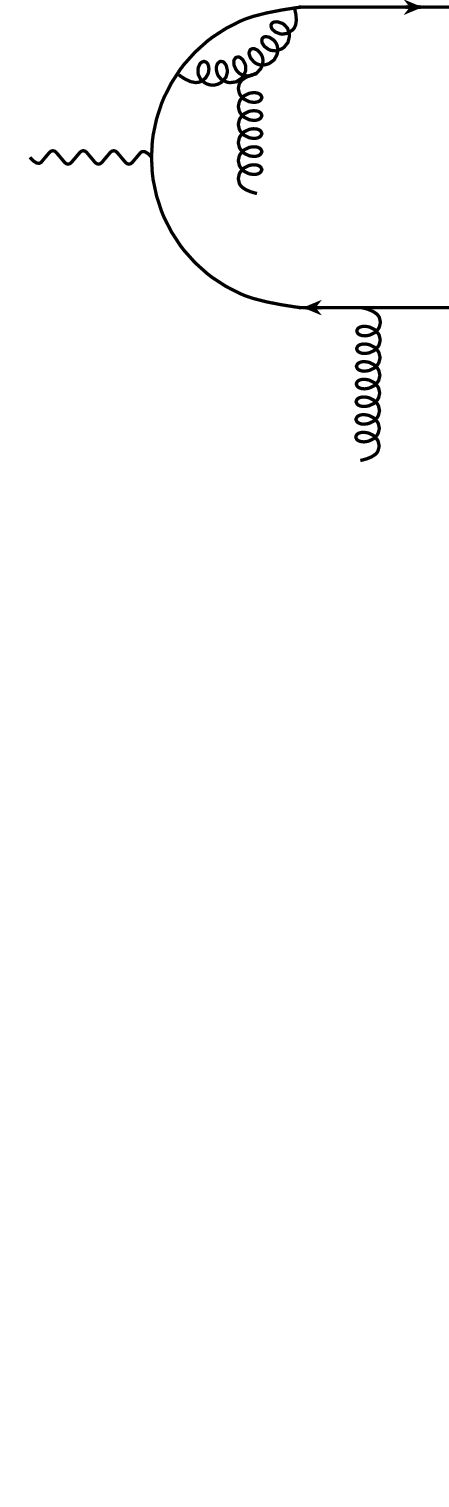}{$D_{11}$}
\kar{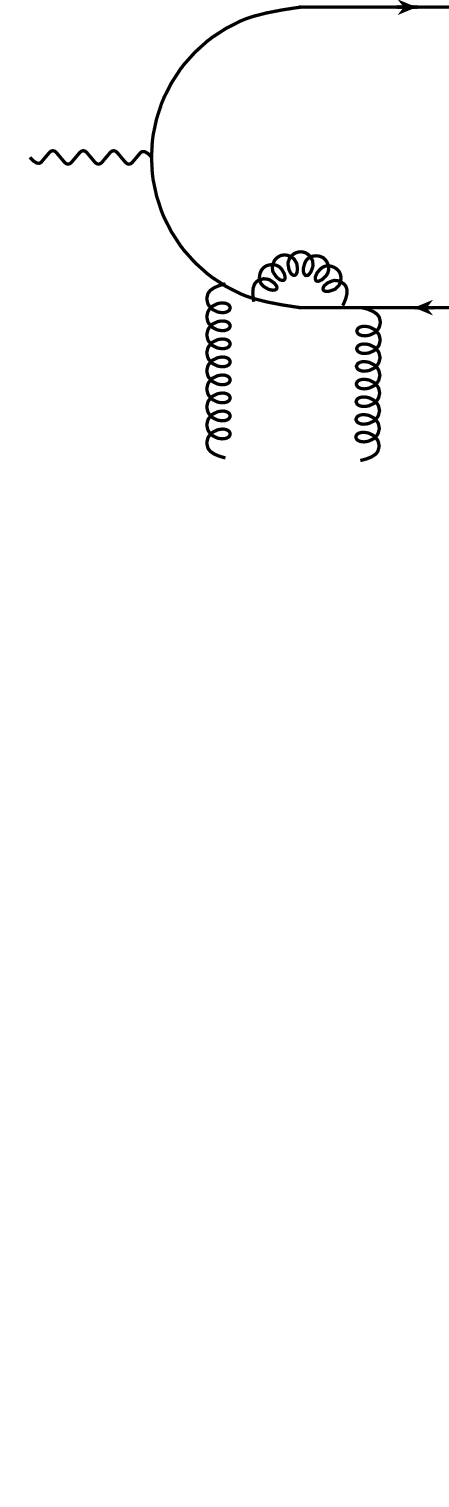}{$D_{12}$}
\kar{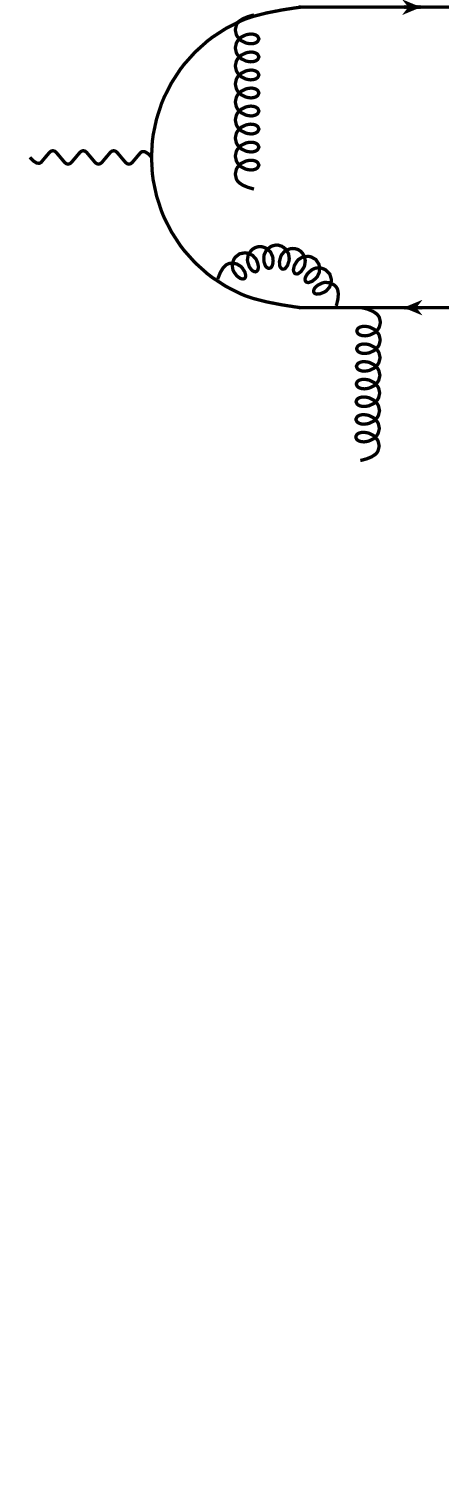}{$D_{13}$}
\kar{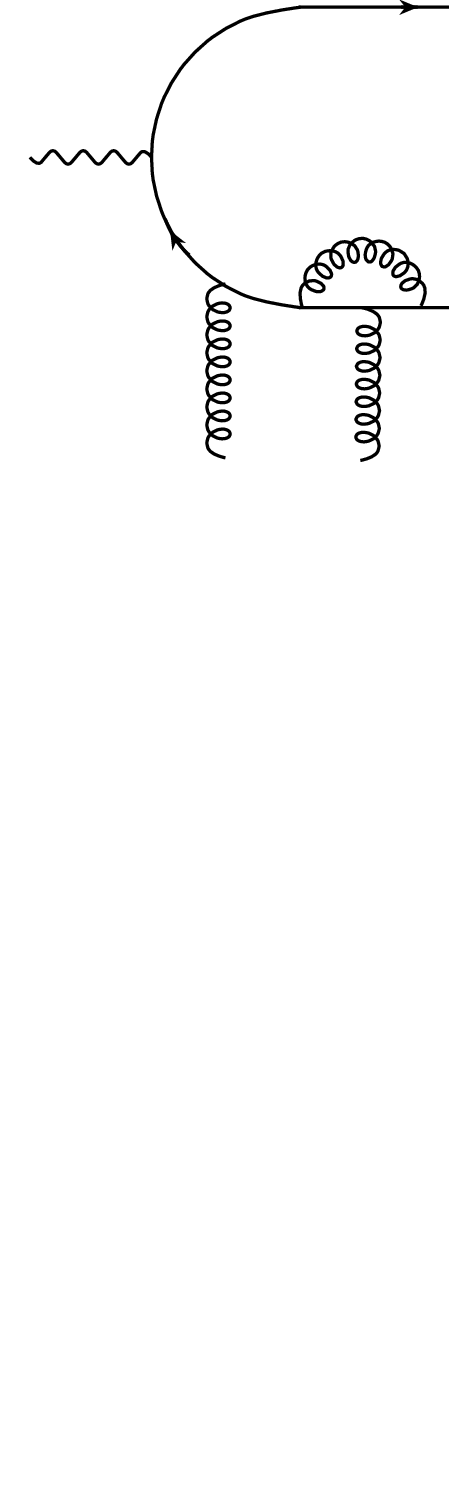}{$D_{14}$}
\kar{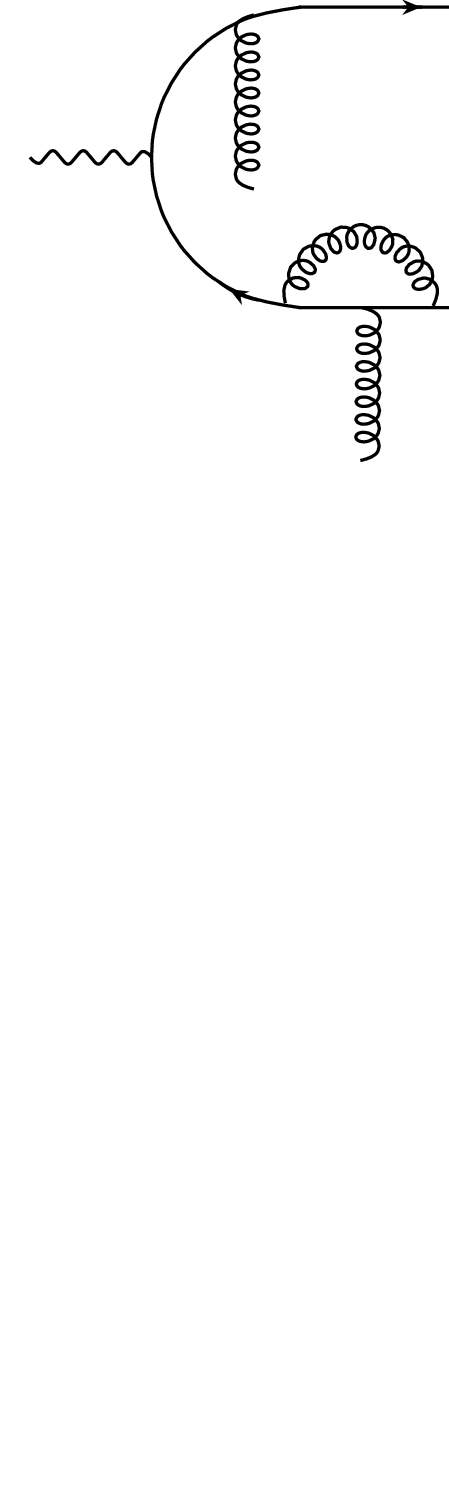}{$D_{15}$}
\kar{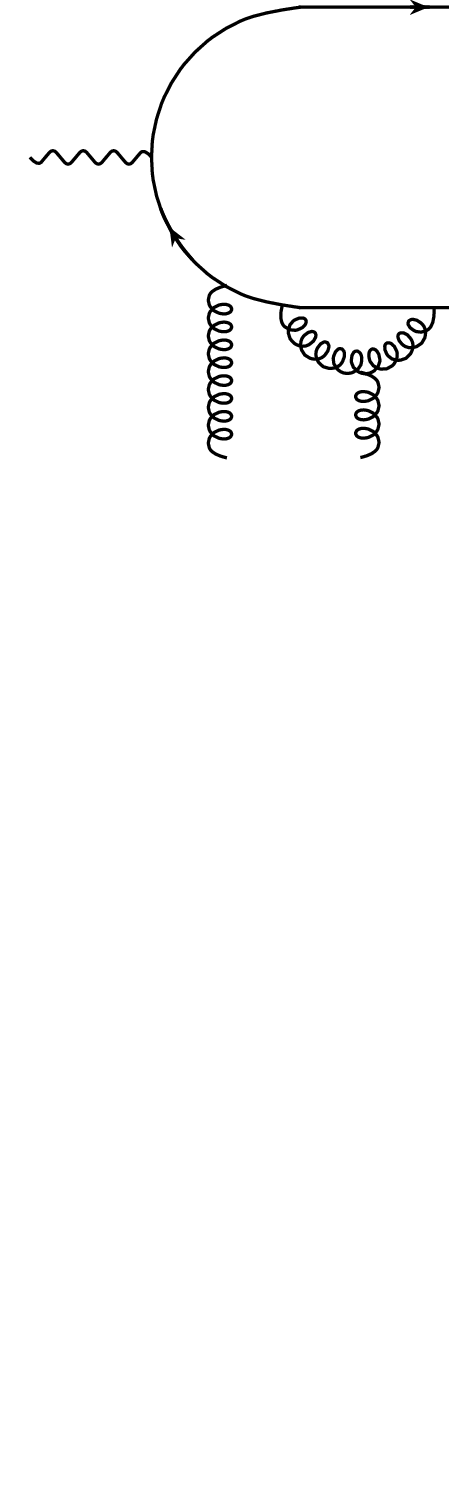}{$D_{16}$}
\kar{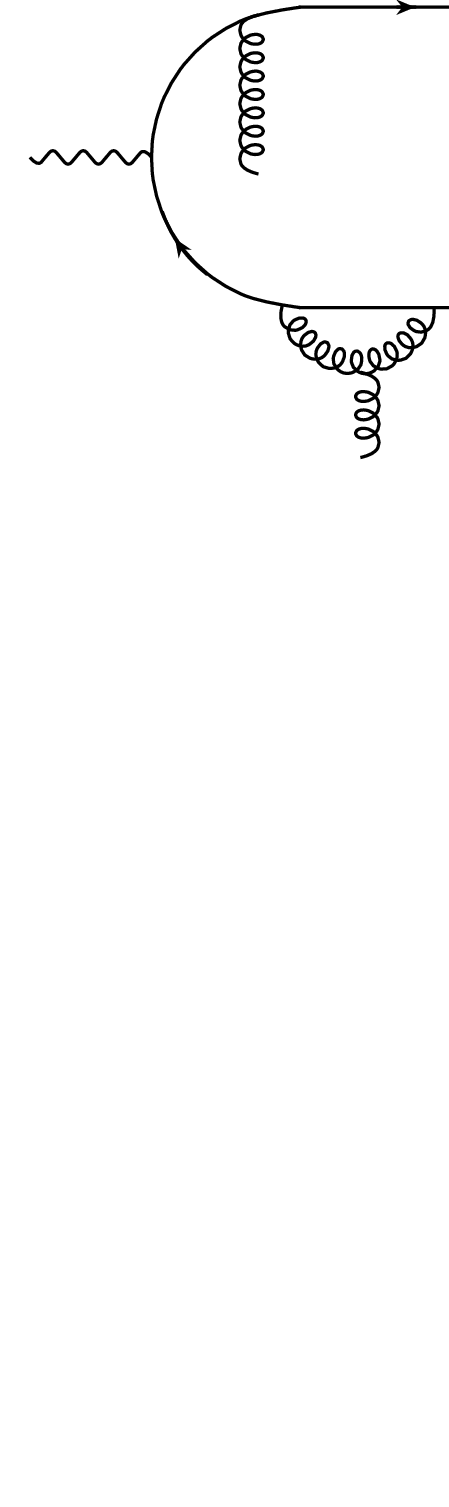}{$D_{17}$}
\end{center}\caption{ Diagrams $D_i$, 
describing the radiation of a soft
gluon from the on-shell antiquark line.}
\label{fig:mmm}
\end{figure}

Now we proceed with this method to the 
evaluation of ${\cal I}_g(0)$. 
The one-loop diagrams describing the radiation of a soft gluon from the
on-shell antiquark line are shown in Fig.~\ref{fig:mmm}. 
A similar set of diagrams can be drawn for the radiation of the soft gluon
from the on-shell quark line. Since these two sets of diagrams transform
into  one
another under the charge parity transformation it is enough to 
calculate one of
them, say, those in Fig.~\ref{fig:mmm} and then to double the result. 

The results of our calculation of the contributions of 
individual antiquark pole diagrams are summarized in Tables 1 and 2. 
Using  our  procedure we obtained for each diagram $D_1,\dots ,D_{11}$ two
quantities $D_i^z$ and $D_i^{add}$.\footnote{  
The diagrams $D_2$, $D_3$ include the instantaneous
Coulomb exchange which we treated in dimensional regularization  
as discussed above.} 

Besides soft and collinear 
singularities the one-loop gluonic amplitude contains also 
ultraviolet poles which have to be subtracted in the on-shell
scheme. 
The full renormalization procedure includes  mass counterterm
diagrams, the renormalization of the heavy quark field and the
renormalization of the strong coupling constant. 
The field and the coupling renormalization  
will be discussed later,
together with the factorization of collinear singularities.

\begin{figure}
\begin{center}
\kar{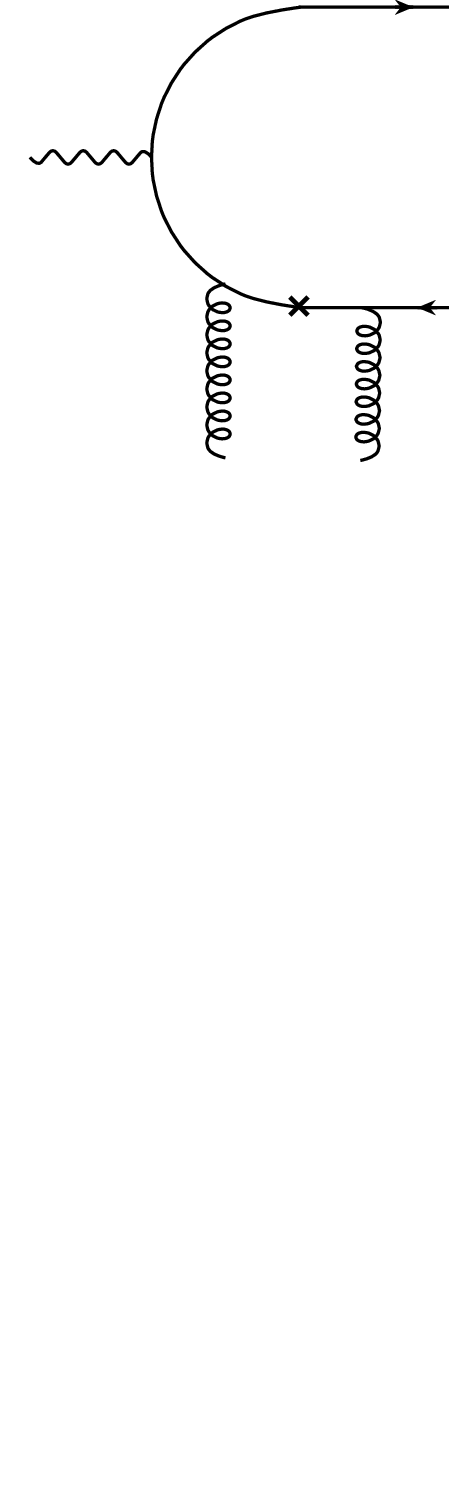}{$C_1$}
\kar{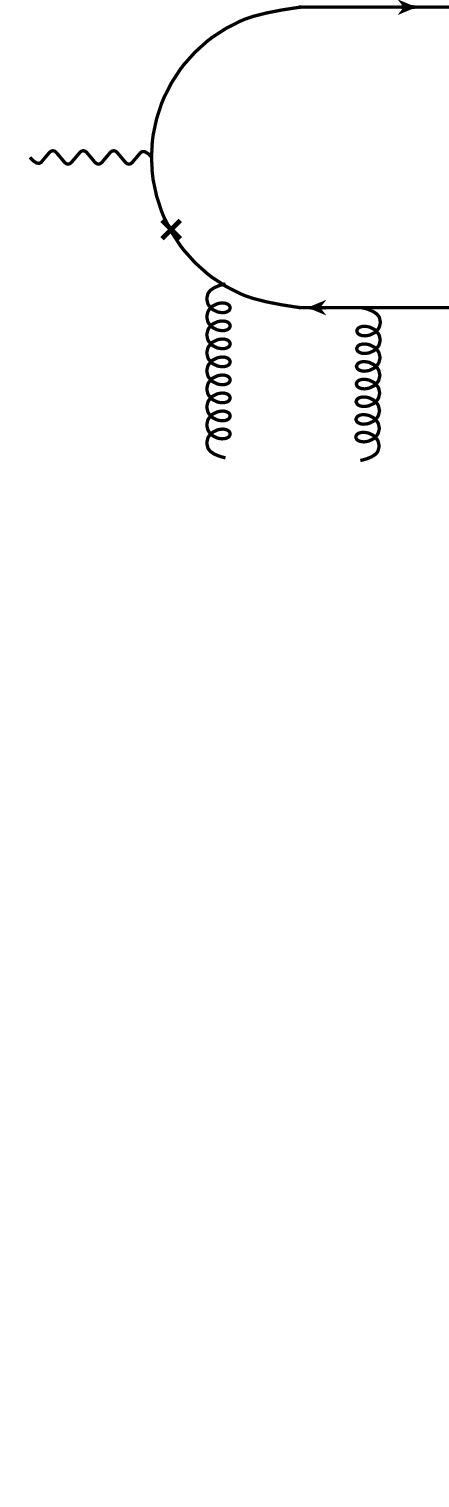}{$C_2$}
\kar{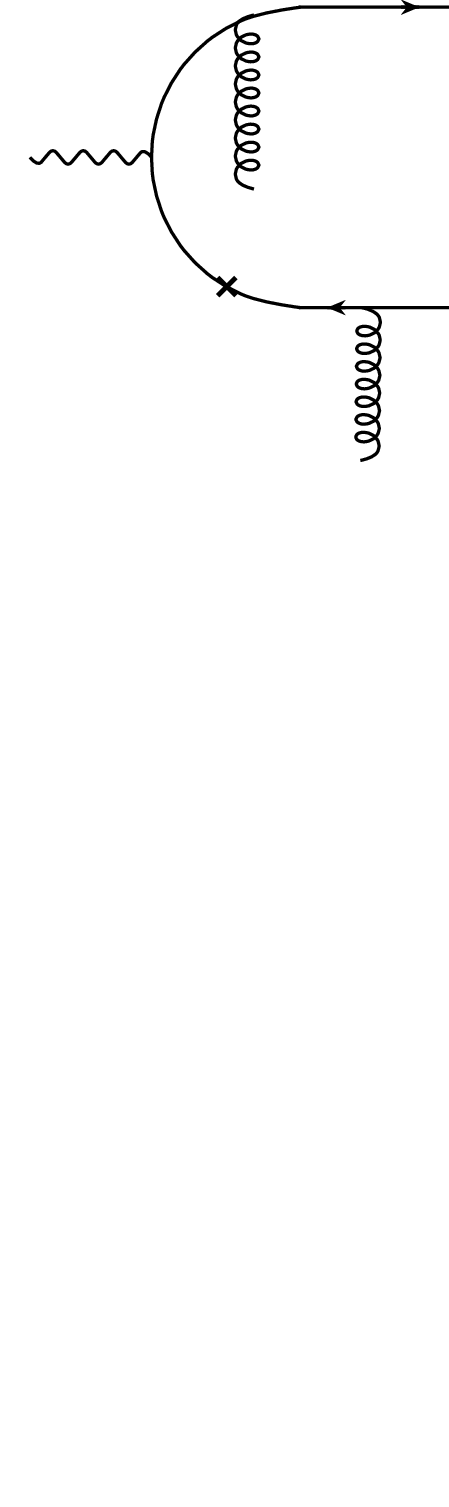}{$C_3$}
\kar{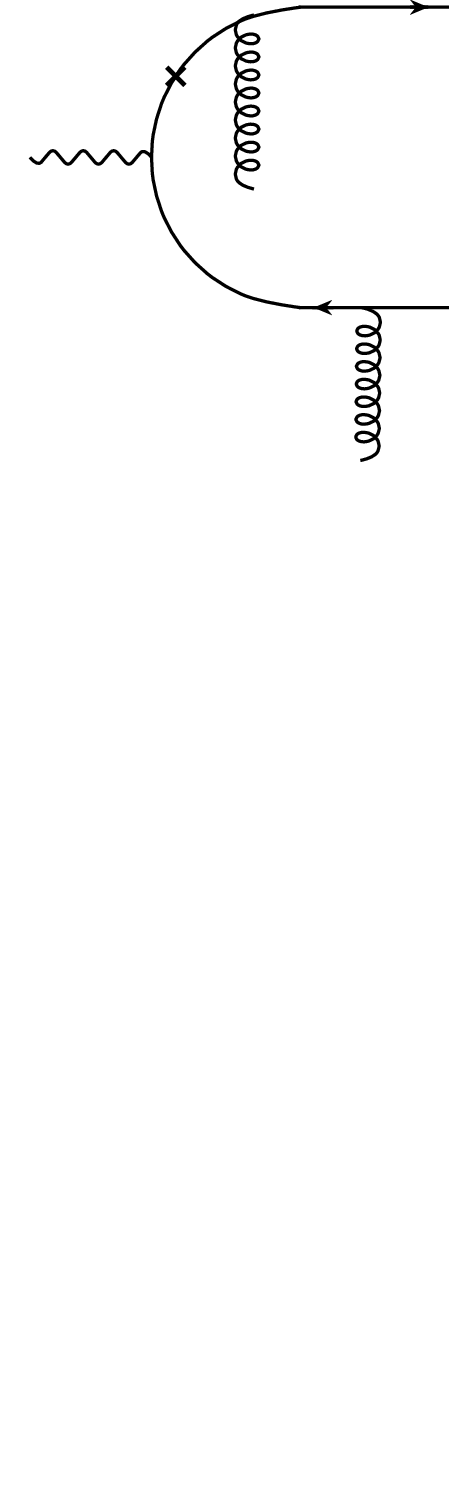}{$C_4$}
\end{center}\caption{Mass counterterm diagrams which have an antiquark pole in
the soft gluon
limit.}
\label{mcounter}
\end{figure}

Here we will consider the mass counterterm diagrams.
This can be done in our method by considering
only mass counterterm diagrams having an antiquark pole in
the soft gluon limit. They are shown in Fig.~\ref{mcounter}.
Thus, similar to $D_i^z$ and $D_i^{add}$, we have in Table 2
two contributions for the diagrams $C_2$ and $C_4$. 
Below we show that the sets of
diagrams $D_{12}, D_{14}, D_{16}$ and $D_{13}, D_{15},
D_{17}$ together with the mass counterterm diagrams $C_1$ and $C_3$ 
add up to two combinations which are gauge invariant.
These sets are the separate gauge invariant 
contributions which can be calculated, similar to $D_i^z$, directly in the 
collinear limit.

\begin{figure}
\begin{center}
\kar{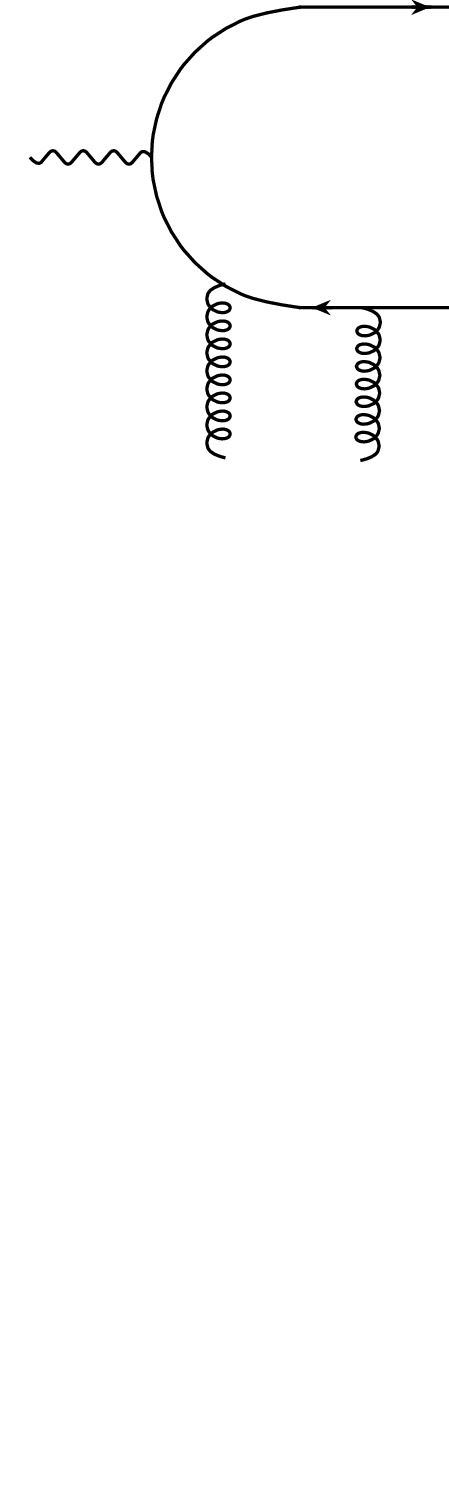}{$B_1$}
\kar{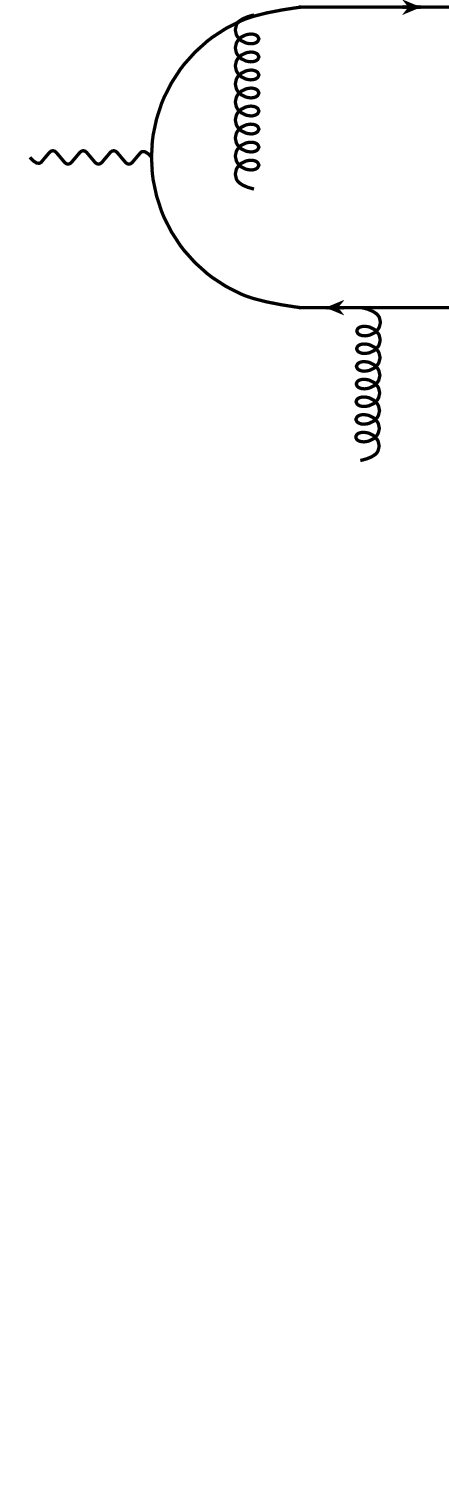}{$B_2$}
\end{center}\caption{LO antiquark pole diagrams.}
\label{LOpole}
\end{figure}

At the one-loop level the mass and  quark field  renormalization 
constants are equal \cite{BGS}
\begin{equation}
 \frac{\delta m}{m} = \delta Z_2 =
-\frac{\alpha_S\,  C_F }{(4\pi)^{1+\epsilon}}
\left(\frac{m^2}{\mu^2}\right)^\epsilon
\left(\frac{3+2\epsilon}{1+2\epsilon}\right)\Gamma[-\epsilon] \ .
\label{Z2}
\end{equation}
Mass counterterm diagrams  are
multiplied by $\delta m /m$.
Let us consider $(D_{12}+C_1 \frac{\delta m}{m}+ D_{14}+D_{16})$ and 
$(D_{13}+C_3 \frac{\delta m}{m} + D_{15}+D_{17})$, which  
represent the one-loop correction to the
soft gluon vertex
\begin{equation}
(i g t^a )\to
(i g t^a )\left(1+\frac{\alpha_S\, 
\Gamma[1-\epsilon]}
{(4\pi)^{1+\epsilon}}\left(\frac{m^2}{\mu^2}\right)^\epsilon
 w\right)\! \not \! e_g^2 \ , \quad \mbox{where} \quad
w=w_1+w_2+w_3\;,\ 
\label{softGvertex}
\end{equation} 
multiplied by the LO  antiquark 
pole diagrams $B_1$ and $B_2$ shown in Fig. \ref{LOpole}. 
After a straight\-for\-ward calculation we obtain\footnote{Note that $w_1$
is finite for $k\to 0$ only if the mass counterterm
diagram is included.}
\begin{eqnarray}
w_1=\ \  & \includegraphics[scale=0.4]{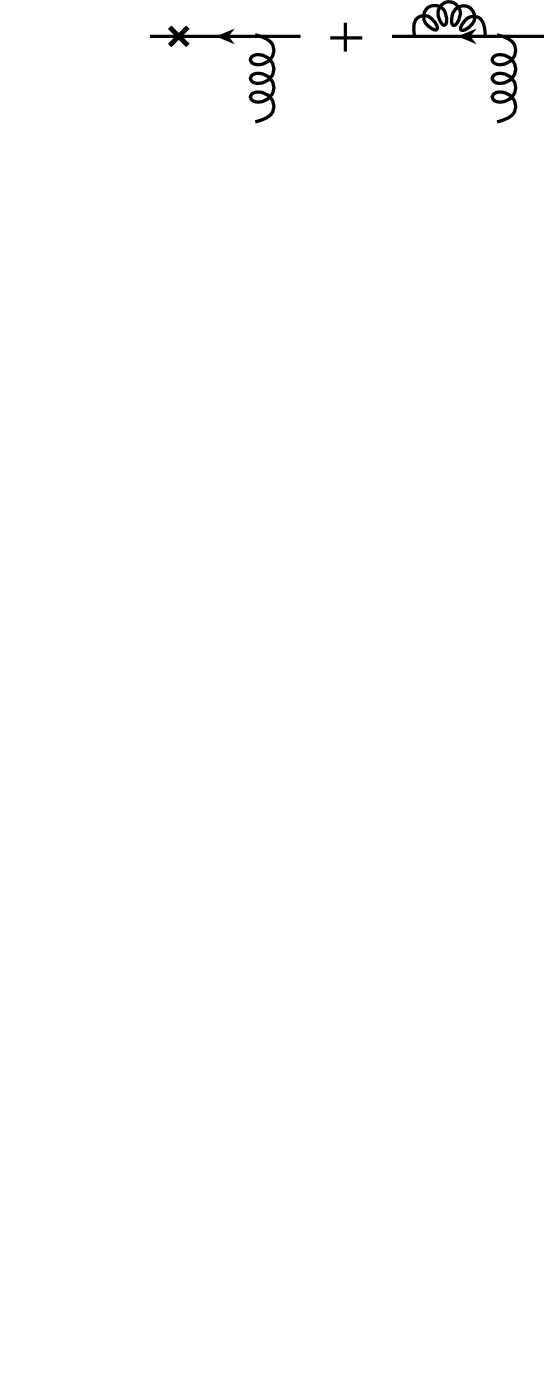} &
\ \ \ \ \ \ = c_1
\left[\frac{3+2\epsilon}{\epsilon (1+2\epsilon)}
\right] \ ,
    \\
w_2=\ \  & \includegraphics[scale=0.4]{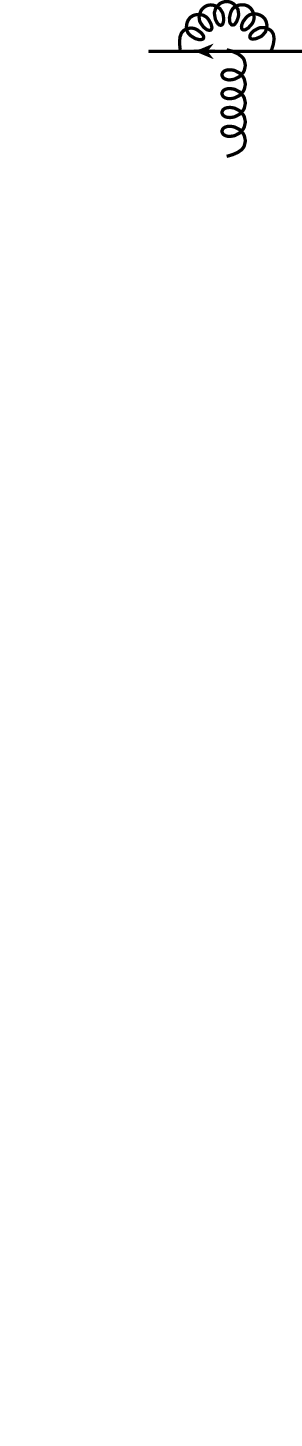} &
\ \ \ \ \ \ = 
c_2\left[-\frac{3+2\epsilon}{\epsilon (1+2\epsilon)}
-\frac{\not \! k}{m}
\left(\frac{1-2\epsilon}{1+2\epsilon}\right)\right] \ ,
\\
w_3=\ \  & \includegraphics[scale=0.4]{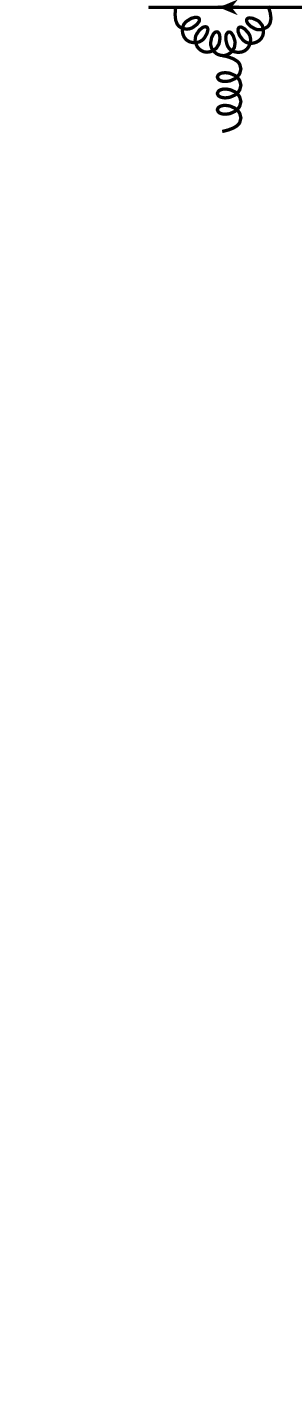} &
\ \ \ \ \ \ = (c_1 - c_2)
\left[-\frac{3+2\epsilon}{\epsilon (1+2\epsilon)}
+\frac{\not \! k}{m}
\left(\frac{1-\epsilon}{\epsilon(1+2\epsilon )}\right)
\right] \ .
\label{vertexterm}
\end{eqnarray}
The sum of these contributions equals 
\begin{equation}
 w = 
 \left(
 \frac{c_1-c_2}{\epsilon} - 3 c_1 +2 c_2+ {\cal O}(\epsilon)
 \right)
 \frac{\not \! k}{m} \ .
\label{w}
\end{equation}
We found that for the one-loop correction to the soft gluon vertex 
the contribution $\sim \not \! e_g^2 $ cancel. What is left  
is $\sim \not \! k \!\! \not \!\! e_g^2 $, which means that 
the part $\sim \alpha_S$
of (\ref{softGvertex}) is gauge invariant. Inserting it into the LO
diagrams $B_1$ and $B_2$ we obtained the results presented in the first two
lines of Table 2.

Note that in the abelian case $c_1,c_2=1$, and, according to (\ref{w}),
the correction to the soft vertex is finite at $\epsilon\to 0$. 
It corresponds to the
contribution of a fermion anomalous magnetic moment, $\alpha/(2\pi)$, 
which is in accordance
with the general statement of Low's theorem in QED. In the 
non-abelian case this contribution has no such clear physical meaning
since it is infrared divergent.

Finally, summing all contributions in Tables 1 and 2 and multiplying the
result by a factor 2 (thus taking into account the quark pole 
diagrams) we arrive at 
the following expression for the  gluonic amplitude  at $x_2=0$
\begin{equation}
 {\cal I}_g(0) =
  -2\frac{c_1 - c_2}{ \epsilon^2}
  -\frac{7c_1 - c_2}{2 \epsilon } +
  c_1 \left( - \frac{3}{2}  + \frac{3 {\pi }^2}{4} + 10 \ln(2) \right)
  - c_2 \left( 5   + \frac{5 {\pi }^2}{8} + 2\ln (2) \right) 
 \, . 
\label{0gluonic}
\end{equation}

\subsection{NLO results}

Let us discuss the structure of singularities of the  parton
amplitudes. 
First of all note that, according to
(\ref{three1}), the double pole terms which are present in 
eqs.~(\ref{0gluonic}) and (\ref{IgQq}) cancel.
Thus, the gluonic amplitude as well as the
quark one contains only single poles in
$\epsilon$.   
This means that soft singularities present in the individual contributions 
cancel in the final expressions for the NLO parton amplitudes.
What is left are the single poles in $\epsilon$ which as we show below
represent the
ultraviolet and collinear singularities. 

To demonstrate the
validity of factorization one needs to check that the ultraviolet poles
are removed by the heavy quark
field and the strong coupling renormalization, and that the collinear poles are
absorbed into the quark and gluon GPDs.  
For this purpose let us recall the structure of the factorization formula
\begin{equation}
{\cal M}\sim \int\limits_{-1}^1 dx 
\left[
\left(
\tilde T_g^{(0)}(x,\xi)+\tilde T_g^{(1)}(x,\xi)
\right)\tilde F^g(x,\xi,t)+
\tilde T_q^{(1)}(x,\xi)\tilde F^{q,S}(x,\xi,t)
\right] \, ,
\label{recall}
\end{equation}
where  the tilde indicates that the 
renormalization and the
separation of the collinear singularities has yet not been performed.  
The bare hard-scattering amplitudes are
\begin{eqnarray}
&&
\tilde T_g^{(0)}(x,\xi)=\frac{ \xi 
}{(x-\xi)(x+\xi)(1+\epsilon)}{\cal A}_g^{(0)}\left(\frac{x-\xi}{2\xi}\right)
=\frac{ \xi
}{(x-\xi)(x+\xi)(1+\epsilon)}\alpha_S(1+\epsilon)\, ,
\nonumber \\
&&
  \tilde T_g^{(1)}(x,\xi)=\frac{ \xi
}{(x-\xi)(x+\xi)(1+\epsilon)}
{\cal A}_g^{(1)}\left(\frac{x-\xi}{2\xi}\right)\, ,
\quad
\tilde T_q^{(1)}(x,\xi)=
{\cal A}_q^{(1)}\left(\frac{x-\xi}{2\xi}\right)\, ,
\label{tildeTs}
\end{eqnarray}
here ${\cal A}_q^{(1)}$ is defined by eqs.~(\ref{conv}), (\ref{three1}),
and the NLO gluonic amplitude ${\cal A}_g^{(1)}$ 
by eqs.~(\ref{conv}), (\ref{three1}), (\ref{IQQ}),
(\ref{IgQq})
and  (\ref{0gluonic}). In (\ref{tildeTs})
and in some equations below we  suppress for shortness 
the $i\varepsilon$ prescriptions, they are easily restored by the
replacement $\xi\to \xi-i\varepsilon$.  

The factorization of the collinear singularities corresponds to
the substitution, in accordance with the definition of
GPDs, of the bare quantities $\tilde F^{q,S}(x,\xi,t)$, 
$\tilde F^g(x,\xi,t)$ by the renormalized ones.
 In the modified minimal-subtraction
$(\overline{\rm{MS}})$ scheme one has at the one-loop level
\begin{eqnarray}
&&
\tilde
F^{q,S}(x,\xi,t)=F^{q,S}(x,\xi,t,\mu_F)-\frac{\alpha_S(\mu_F)}{2\pi}
\left(\frac{1}{\hat \epsilon}+
\ln\left(\frac{\mu_F^2}{\mu^2}\right)\right)\times
\nonumber \\
&&
\times\int\limits^1_{-1}dv \left[V_{qq}(x,v) F^{q,S}(v,\xi,t,\mu_F)+
V_{qg}(x,v) F^{g}(v,\xi,t,\mu_F)\right] \, ,
\label{renqGPD} \\
&&
\tilde
F^{g}(x,\xi,t)=F^{g}(x,\xi,t,\mu_F)-\frac{\alpha_S(\mu_F)}{2\pi}
\left(\frac{1}{\hat \epsilon}+
\ln\left(\frac{\mu_F^2}{\mu^2}\right)\right)\times
\nonumber \\
&&
\times\int\limits^1_{-1}dv \left[V_{gg}(x,v) F^{g}(v,\xi,t,\mu_F)+
V_{gq}(x,v) F^{q,S}(v,\xi,t,\mu_F)\right] \, ,
\label{regGPD} 
\end{eqnarray}   
where $V_{qq}, V_{gg}$, $V_{gq}, V_{qg}$ denote the one-loop evolution
kernels. 
\begin{equation}
\frac{1}{\hat \epsilon}=\frac{1}{\epsilon}+\gamma_E -\ln(4\pi) \, ,
\end{equation}
$\gamma_E$ is Euler's constant. Inserting (\ref{renqGPD}) and (\ref{regGPD}) 
into eq.~(\ref{recall}) and
truncating 
the series at the order $\alpha_S^2$ we found
the following collinear counterterms to the gluon and quark hard-scattering
amplitudes
\begin{eqnarray}
&&
\Delta^{col}_g(x,\xi)=-\frac{\alpha_S}{2\pi}
\left(\frac{1}{\hat \epsilon}+
\ln\left(\frac{\mu_F^2}{\mu^2}\right)\right)\int\limits^1_{-1}dv
\, \tilde T^{(0)}_g(v,\xi)\, V_{gg}(v,x) \, ,
\label{countergl}
\\
&&
\Delta^{col}_q(x,\xi)=-\frac{\alpha_S}{2\pi}
\left(\frac{1}{\hat \epsilon}+
\ln\left(\frac{\mu_F^2}{\mu^2}\right)\right)\int\limits^1_{-1}dv
\, \tilde T^{(0)}_g(v,\xi)\, V_{gq}(v,x) \, .
\label{counterqu}
\end{eqnarray}  
Note that, since $\tilde T^{(0)}_q=0$ for our process, the renormalization
of the quark GPD (\ref{renqGPD}) does not generate contributions ($\sim
V_{qq}, V_{qg}$) to the collinear
counterterms.  
Calculating the 
integrals (\ref{countergl}), (\ref{counterqu}) with these kernels
we obtain
\begin{eqnarray}
&&
\Delta^{col}_g(x,\xi)=-\frac{\alpha_S^2}{2\pi}
\frac{\xi}{(x-\xi)(x+\xi)}
\left(\frac{1}{\hat \epsilon}+
\ln\left(\frac{\mu_F^2}{\mu^2}\right)\right)
\left[N_c\, 
{\cal C}_g\left(\frac{x-\xi}{2\xi}\right)+\frac{\beta_0}{2}\right]\, ,
\label{cougl}
\\
&&
\Delta^{col}_q(x,\xi)=-\frac{\alpha_S^2}{2\pi}
\left(\frac{1}{\hat \epsilon}+
\ln\left(\frac{\mu_F^2}{\mu^2}\right)\right)C_F\, 
{\cal C}_q\left(\frac{x-\xi}{2\xi}\right) \, ,
\label{couqu}
\end{eqnarray}
where 
\begin{equation}
\beta_0=\frac{11 \, N_c}{3}-\frac{2\, n_f}{3} \, ,
\label{betafunc}
\end{equation}
$n_f$ is an effective number of light quark flavors,
\begin{eqnarray}
&&
{\cal
C}_g(y)=(1+2y(y+1))\left(\frac{\ln(-y)}{1+y}-\frac{\ln(1+y)}{y}\right)\, ,
\nonumber \\
&&
{\cal C}_q(y)=(1+2y)\left(\frac{\ln(-y)}{1+y}-\frac{\ln(1+y)}{y}\right)\, .
\label{calCs}
\end{eqnarray}

\begin{figure}[]
\begin{center}
\scalebox{0.8}{
\input{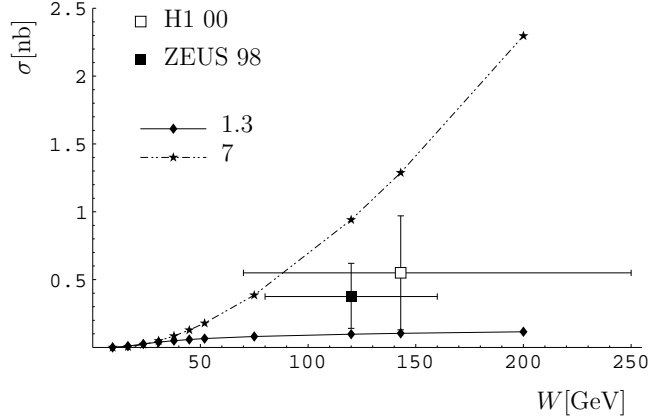}
}
\end{center}
\caption[]{\small  
The cross section for $\Upsilon$ photoproduction;
the theoretical predictions at LO 
for the scales 
$\mu_F=\mu_R=[1.3, 7]\,$GeV 
(ranging from bottom to top), and the data are from ZEUS \cite{Breitweg:1998ki} 
and H1 \cite{Adloff:2000vm}.}  
\label{LOUpsilon}
\end{figure}

For the renormalization of the strong coupling one has to 
substitute the bare coupling constant $\alpha_S$ by the running coupling 
$\alpha_S(\mu_R)$ in the 
$\overline{\rm{MS}}$ scheme,
\begin{equation}
\alpha_S=\alpha_S(\mu_R)\left[
1+\frac{\alpha_S(\mu_R)}
{4\pi}\beta_0\left(\frac{1}{\hat \epsilon}+
\ln\left(\frac{\mu_R^2}{\mu^2}\right)\right)
\right]\, .
\label{alphaSS}
\end{equation}
This substitution generates the following counterterm to the gluon
hard-scattering amplitude
\begin{equation}
\Delta_g^{\alpha_S}(x,\xi)=\frac{\alpha_S^2}{4\pi}\frac{\xi}{(x-\xi)(x+\xi)}
\left(\frac{1}{\hat \epsilon}+
\ln\left(\frac{\mu_R^2}{\mu^2}\right)\right)\beta_0 \, .
\label{couAl}
\end{equation}
To account for the heavy quark field renormalization effect one has to add the
counterterm
\begin{equation}
\Delta_g^{Z_2}(x,\xi)=\delta Z_2\, \tilde T^{(0)}_g(x,\xi)  \, ,
\label{couZ2}
\end{equation} 
with $\delta Z_2$ given in eq.~(\ref{Z2}).

In the sum of the bare hard-scattering amplitudes and the 
counterterms described above 
all poles in $\epsilon$ cancel. Thus, we can  now take the limit
$\epsilon\to 0$
\begin{eqnarray}
&&
T_g(x,\xi)=\left[\tilde
T_g(x,\xi)+\Delta_g^{col}(x,\xi)+\Delta_g^{\alpha_S}(x,\xi)
+\Delta_g^{Z_2}(x,\xi)\right]_{\epsilon\to 0}
\, ,\nonumber \\
&&
T_q(x,\xi)=\left[\tilde
T_q(x,\xi)+\Delta_q^{col}(x,\xi)\right]_{\epsilon\to 0} \, ,
\label{sumcount}
\end{eqnarray}
and arrive at  finite results for the 
hard-scattering amplitudes: 
\begin{equation}
T_q(x,\xi)=\frac{\alpha_S^2(\mu_R) C_F}{2\pi}  
f_q\left(\frac{x-\xi+i\varepsilon}
{2\xi}\right) \, ,
\label{TTq}
\end{equation}
\begin{multline}
 f_q(y) =
 \ln\big(\frac{4m^2}{\mu_F^2}\big)
  \,
  (1+2y)\left(
   \frac{\ln(-y)}{1+y}-\frac{\ln(1+y)}{y}
  \right)-
  \pi^2\frac{13(1+2y)}{48y(1+y)}+\frac{2\ln2}{1+2y}\\
 + \frac{\ln(-y)+\ln(1+y)}{1+2y}+
  (1+2y)\left(
   \frac{\ln^2(-y)}{1+y}-\frac{\ln^2(1+y)}{y}
  \right)\\
+  \frac{3-4y+16y(1+y)}{4y(1+y)}Li_2(1+2y)-
  \frac{7+4y+16y(1+y)}{4y(1+y)}Li_2(-1-2y) \, ,
\label{fqq}
\end{multline}
for the quark, and
\begin{equation}
T_g(x,\xi) =
 \frac{\xi}{(x-\xi+i\varepsilon)(x+\xi-i\varepsilon)}
 \left[
  \alpha_S(\mu_R) + \frac{\alpha_S^2(\mu_R)}{4\pi}
f_g\left(\frac{x-\xi+i\varepsilon}
{2\xi}\right)
 \right] \, ,
\label{TTg}
\end{equation}
\begin{multline}
 f_g(y)=
 4(c_1-c_2)\big(1 + 2y(1 + y)\big)
 \big(
    \frac{\ln(-y)}{1 + y} - \frac{\ln(1 + y)}{y}
 \big)
 \big(
 \ln\frac{4m^2}{\mu_F^2}-1
 \big)+\beta_0 \ln\frac{\mu_R^2}{\mu_F^2}
 \\
+4(c_1-c_2)\big(1+2y(1+y)\big)
    \left(
     \frac{\ln^2(-y)}{1+y}-\frac{\ln^2(1+y)}{y}
    \right)-8 c_1\\
- \pi^2\left(
 \frac{2+y(1+y)(25+88y(1+y))}{48y^2(1+y)^2}c_1+
\frac{10+y(1+y)(7-52y(1+y))}{24y^2(1+y)^2}c_2
 \right)\\ 
-\left[
  c_1 \frac{1 + 6 y (1 + y) (1 + 2 y (1 + y))}{y (1 + y)(1 + 2 y)^2}+
  c_2\frac{(1 + 2 y)^2}{y (1 + y)}
  \right]\ln(2)\\
+\pi\frac{\sqrt{-y(1+y)}}{y (1 + y)}\left(
   \frac{7}{2} c_1-3 c_2
 \right)\\
+ 2 c_2 \frac{\sqrt{-y(1+y)}}{y (1 + y)}\left(
  \frac{1+4y}{1+y}\arctan \sqrt{\frac{-y}{1+y}}+
  \frac{3+4y}{y}\arctan \sqrt{\frac{1+y}{-y}}
  \right)\\
 -\frac{\arctan^2 \sqrt{\frac{-y}{1+y}}}{2y(1+y)} \left(
 (7+4y)c_1 - 2\frac{1+2y-2y^2}{1+y}c_2\right)\\
-
 \frac{\arctan^2 \sqrt{\frac{1+y}{-y}}}{2y(1+y)} \left(
 (3-4y)c_1 - 2\frac{3+6y+2y^2}{y}c_2\right) \\
+2\, a_1(y)\ln(-y) + 2\, a_1(-1-y)\ln(1+y) \\
+2\, a_2(y)Li_2(1+2y) + 2\, a_2(-1-y)Li_2(-1-2y) \, ,
\label{fgl}
\end{multline}
for the gluon. $a_1(y), a_2(y)$ are defined in 
eqs.~(\ref{a1y}), (\ref{a2y}). 
The expressions in (\ref{TTq})-(\ref{fgl})
represent the main result of this paper.

At high energies, $W^2\gg M^2$, the imaginary part of the amplitude
dominates. The leading contribution to the NLO correction comes from the
integration region $\xi\ll |x|\ll 1$. Simplifying the gluon (\ref{fgl})
and the quark (\ref{fqq}) hard-scattering
amplitudes in this limit we obtain the estimate
\begin{eqnarray}
&&
 {\cal M}
\approx \frac{-4\, i\, \pi^2 \sqrt{4\pi\alpha}  \, e_q (e^*_V e_\gamma
)}
{N_c \, \xi}\left(\frac{\langle O_1 \rangle_V}{m^3}\right)^{1/2}
\times
\nonumber \\
&&
\times
\left[\alpha_S(\mu_R)  F^g(\xi,\xi,t)  +\frac{\alpha_S^2(\mu_R)
N_c}{\pi}\ln\left(\frac{m^2}{\mu_F^2}\right)
\int\limits^1_{\xi} \frac{dx}{x} F^g(x,\xi,t)
\right.
\nonumber \\
&&
\left.
+\frac{\alpha_S^2(\mu_R)
C_F}{\pi}\ln\left(\frac{m^2}{\mu_F^2}\right)
\int\limits^1_{\xi} dx \left( F^{q,S}(x,\xi,t)- F^{q,S}(-x,\xi,t) \right)
\right] \, .
\label{appro}
\end{eqnarray}
Given the behavior of the gluon and the quark GPDs at small $x$,
 $F^g(x,\xi,t)\sim
const$ and $F^{q,S}(x,\xi,t)\sim
1/x$, we see from (\ref{appro}) that the relative value of 
the NLO correction
is parametrically large at small $\xi$,
\begin{equation}
\sim \frac{\alpha_S(\mu_R)
N_c}{\pi}\ln\left(\frac{1}{\xi}\right)\left[
\ln\left(\frac{m^2}{\mu_F^2}\right)
+ \frac{
C_F}{N_c}
\ln\left(\frac{m^2}{\mu_F^2}\right)
\frac{F^{q,S}(\xi,\xi,t)-F^{q,S}(-\xi,\xi,t)}
{F^g(\xi,\xi,t)}
\right] \, .
\label{ux}
\end{equation}
The gluon correction in (\ref{ux}) is negative unless one chooses a 
value of the factorization scale
$\mu_F<m$, which is substantially smaller than the kinematic scale $M=2m$.  
The quark correction is also parametrically large at high energies. It is
expected to be 
sizable since it collects the contributions of all the light quarks and
antiquarks. 
These qualitative observations are supported by the numerical analysis.

\section{Numerical analysis}

\begin{figure}[]
\begin{center}
\scalebox{0.8}{
\input{nlo.pstex_t}
}\end{center}\caption[]{\small  
The cross section of the $\Upsilon$ photoproduction;
theoretical predictions at NLO
for the scales
$\mu_F=\mu_R=[1.3,7]\mbox{ GeV}$ (ranging from top to bottom), 
and the data from ZEUS \cite{Breitweg:1998ki} and H1 \cite{Adloff:2000vm}.
}
\label{NLOUpsilon}
\end{figure}
 
We assume as values of the quark pole masses: 
$m_c=1.5 \mbox{ GeV},\, m_b=4.9\mbox{ GeV}$. $\langle O_1 \rangle_V$ was
evaluated using eq.~(\ref{decay}) with $\alpha_S=\alpha_S(\mu_R)$. 
For the generalized parton distributions we adopt the  parametrizations,
evolved both in LO and NLO, of  
\cite{Freund:2002qf} 
that are based on the CTEQ6 set of forward
distributions \cite{Pumplin:2002vw}. 
We neglect the contributions proportional to 
 ${\cal E}^q(x,\xi,t)$
and ${\cal E}^g(x,\xi,t)$.
In the numerical calculations we 
use  LO strong running coupling and  LO GPDs and  NLO coupling 
and  NLO GPDs for  LO and  NLO observables correspondingly.

\begin{figure}[]
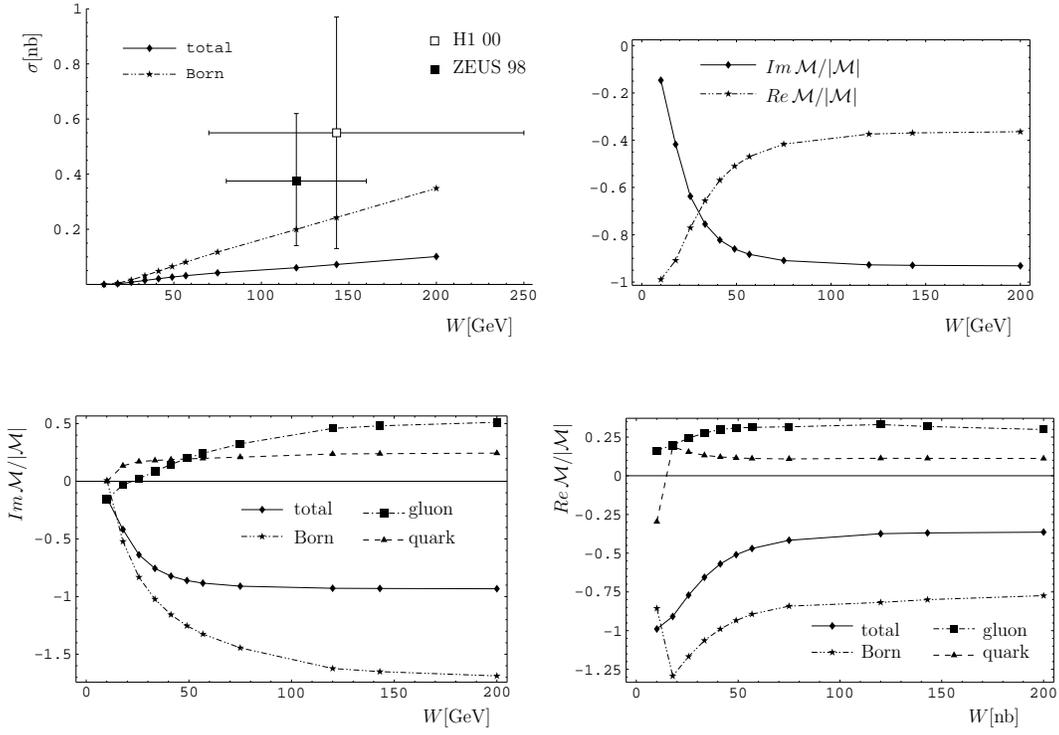

\begin{center}
\scalebox{0.6}{
\input{set1.pstex_t}
}\,\,\,\,\,\,
\scalebox{0.6}{
\input{set2.pstex_t}
}\\[1cm]
\hspace*{1mm}\scalebox{0.6}{
\input{set3.pstex_t}
}\,\,\,\,
\scalebox{0.6}{
\input{set4.pstex_t}
}
\end{center}
\caption[]{\small 
$\Upsilon$ photoproduction, NLO prediction for $\mu_F=\mu_R =4.9\mbox{
GeV}$ and its decomposition into different contributions, see text.}
\label{UpsilonM}
\end{figure}

Let us start with $\Upsilon$ photoproduction. We calculate with our formulas
the forward amplitude and the forward differential
cross section,  $d\sigma/d\Delta_\perp^2$ at $\Delta_\perp=0$. For the
$\Delta_\perp$ dependence we assume, in accordance with the measurements at 
HERA, 
the simple exponential
\begin{equation} 
\frac{d\sigma}{d \Delta_\perp^2}=\left(
\frac{d\sigma}{d \Delta_\perp^2}|_{\Delta_\perp=0}\right)
e^{-b\vec \Delta_\perp^2}\, , \quad \sigma=\frac{1}{b}\left(
\frac{d\sigma}{d \Delta_\perp^2}|_{\Delta_\perp=0}\right) \, . 
\label{dsdt}
\end{equation} 
For the slope parameter we use $b=4.4\mbox{ GeV}^{-2}$. 

In Fig.~\ref{LOUpsilon}  
the LO predictions for the total cross section of 
$\Upsilon$ photoproduction
are shown as a function of energy, the data points 
are from ZEUS \cite{Breitweg:1998ki}
and H1 \cite{Adloff:2000vm}. The curves correspond to different
values of the factorization scale $\mu_F$ which is chosen equal to $\mu_R$.
The experimental uncertainties are large. We find that for the broad
interval of scales, $\mu_F=\mu_R=1.3\div 7\mbox{ GeV}$,
our LO predictions lie 
within the experimental error bars. 
 The strong dependence of the predictions on the
factorization scale is related to the well known fact that
scaling violation is large for
small $x$. At small $x$ the gluon density increases rapidly with growing
$\mu_F$ which leads to an increase of the LO predictions with $\mu_F$. 
In Fig.~\ref{NLOUpsilon} we present the results of the NLO calculations 
for the same set of scales.
For meson production 
in NLO this effect is partially compensated, as it should
be, due to the 
dependence of the gluon and the quark hard-scattering NLO amplitudes on
$\mu_F$, see eqs.~(\ref{TTq})-(\ref{fgl}).
As a result we observe a substantial reduction of the scale ambiguity of the 
theoretical predictions  in  NLO in comparison with  LO. 

The NLO predictions are generally smaller than  the LO ones. The reason
is twofold. First, according to the parametrizations we use, in this
kinematic region  
the gluon GPD in NLO is about a factor of two smaller than the gluon GPD in
the LO. This is  another manifestation of the large scaling violation
effects at small $x$. 
Second, we find, in accordance with the estimate (\ref{appro}),  that
 the part  of the gluon NLO hard-scattering 
amplitude (\ref{TTg}) $\sim \alpha_S^2$
leads to a contribution which is large and has at $\mu_F\gtrsim m$ the  
opposite sign  
 as the  contribution  $\sim
\alpha_S$ induced by the Born 
term of (\ref{TTg}).
The last statement is illustrated in Fig.~\ref{UpsilonM} where  the
different contributions to the NLO result for 
$\mu_F=\mu_R=4.9\mbox{ GeV}$ are shown.  In the left upper
panel we present 
the cross section. Here the curve labeled Born represents the results 
 calculated using only the 
Born term of (\ref{TTg}),  the other
curve, labeled total, is the cross section calculated with the complete
result for the 
NLO hard-scattering amplitudes, including  the 
quark  contribution. To avoid misunderstanding, in both calculations 
the NLO GPDs were used.  
We see that  the 
parts of the 
NLO hard-scattering amplitudes
$\sim \alpha_S^2$ 
make the cross section significantly smaller.       
On the bottom left and the bottom right panels of
Fig.~\ref{UpsilonM} we present the
decomposition into different contributions
of the imaginary and the real parts of the NLO amplitude
divided by its absolute value, $Im\, {\cal M}/|{\cal M}|$ and $Re\,  {\cal
M}/|{\cal M}|$. The curves labeled total represent the results calculated with 
the complete NLO hard-scattering amplitudes (\ref{TTg}) and (\ref{TTq}). 
In these figures
the quark contribution and
the decomposition of the 
gluon contribution into the Born  
and the part induced by the term$\sim \alpha_S^2$
of (\ref{TTg}) are shown separately. The corresponding curves are labeled
 as 'quark',  'Born'
and 'gluon'. In the right upper panel we show  $Im\, {\cal
M}/|{\cal M}|$ and $Re\,  {\cal M}/|{\cal M}|$
together. We see that, despite the fact that
the value of a strong coupling constant is small at this scale,
$\alpha_S(\mu_R)/(2\pi)\sim 0.033$ at $\mu_R=4.9\mbox{ GeV}$, the gluon correction
constitutes $\sim 30\%$ of the Born contribution with the opposite sign. 
The quark contribution is about $\sim 15\%$, with the opposite sign with
respect to the Born contribution. Note also that at high energies the imaginary
part of the amplitude is about twice the real part.      

\begin{figure}[]
\begin{center}
\scalebox{0.8}{
\input{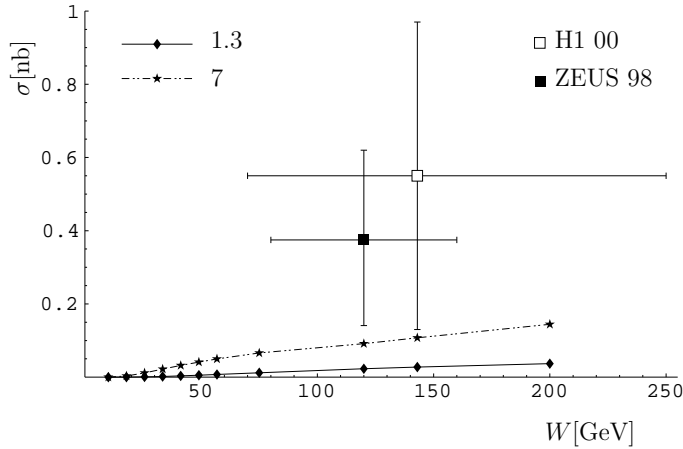}
}
\end{center}
\caption[]{\small  
The cross section for $\Upsilon$ photoproduction, and
NLO predictions  
for the scales $\mu_F = [1.3,7]\mbox{ GeV}$, $\mu_R = 5.9\mbox{ GeV}$.
}
\label{blm}
\end{figure}

The reduction of the ambiguity for the theoretical predictions 
due to a variation of $\mu_F$
is even more pronounced if one chooses
a fixed value of renormalization scale, see Fig.~\ref{blm}. 
In this case
the  value of the cross section is predicted to be smaller than for  
equal scales, $\mu_R=\mu_F$, compare
Figs.~\ref{blm} and \ref{NLOUpsilon}.               

To summarize our results for  $\Upsilon$ photoproduction we conclude that 
 the NLO corrections stabilize the theoretical 
predictions with respect to variation of the factorization 
scale. The NLO corrections 
are numerically important. They make the NLO cross sections smaller than  
the LO ones, and for the GPD-model used our results seem to lie somewhat 
below the data.      

\begin{figure}[]
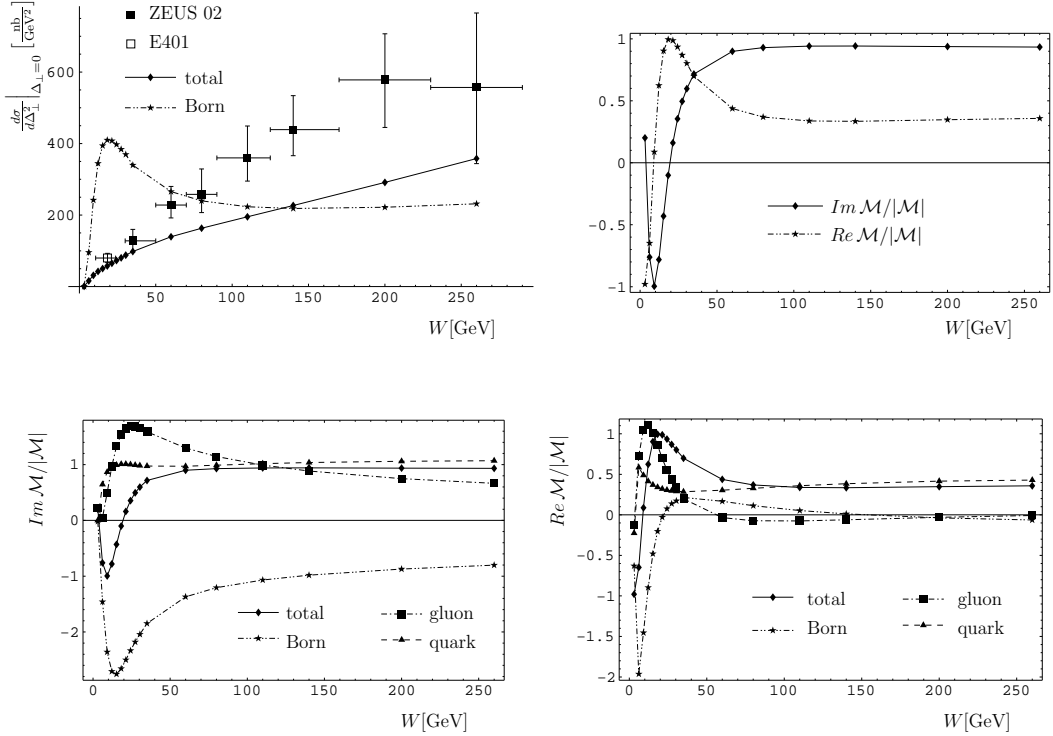

\begin{center}
\scalebox{0.6}{
\input{sj1.pstex_t}
}\,\,\,\,\,\,
\scalebox{0.6}{
\input{sj2.pstex_t}
}\\[1cm]
\hspace*{1mm}\scalebox{0.6}{
\input{sj3.pstex_t}
}\,\,\,\,
\scalebox{0.6}{
\input{sj4.pstex_t}
}
\end{center}
\caption[]{\small  
The differential  cross section for $J/\psi$ photoproduction,
NLO predictions for $\mu_F=\mu_R =1.52\mbox{ GeV}$,  and the data from 
E401
\cite{Binkley:1981kv} and 
ZEUS \cite{Chekanov:2002xi}. 
The labeling of the curves is the same as in Fig.~\ref{UpsilonM}.}
\label{figPSI}
\end{figure}

For the photoproduction of $J/\Psi$ the situation is different than
for
$\Upsilon$ production since in this case  a value of the hard scale, the
quark mass, is smaller. The NLO corrections are much larger than in the case
of $\Upsilon$ production for the following two reasons.  
First, the value of the QCD running coupling is larger
at smaller scales. Second, the value of $\xi$ and, consequently, 
the effective values of $x$ in the factorization formula is about two orders 
of magnitude smaller than for  $\Upsilon$ production.
Therefore the effect of the enhancement of the NLO correction at 
small $x$, see
eqs.~(\ref{appro}), (\ref{ux}), is much larger for $J/\Psi$. This is 
illustrated
in Fig.~\ref{figPSI} where the labeling of the curves is the same as in
Fig.~\ref{UpsilonM}. The data are from E401
\cite{Binkley:1981kv}
and ZEUS \cite{Chekanov:2002xi}. Note that contrary to  Fig.~\ref{UpsilonM},
in the left upper panel of Fig.~\ref{figPSI} the results for the forward
differential cross section are shown. We see that although 
the predictions for the cross section are in  reasonable 
agreement with the data, the absolute value of the NLO correction 
is very large.
Note also 
that in this case the quark GPD makes a significant contribution.
The sum of the gluon and the quark NLO corrections is as much as 
twice of the Born contribution and of opposite sign. Therefore in 
NLO the total amplitude has the opposite sign 
as in LO.
Note also that the imaginary part of the NLO amplitude
goes through zero  at $W\sim 25\mbox{ GeV}$, which is unnatural.
Thus, we conclude that for the $J/\Psi$
photoproduction the higher order corrections are  
not under control.  

\section{Summary}

We have shown by an explicit calculation of the partonic one-loop
amplitudes that in the heavy quark limit 
the  collinear factorization and the nonrelativistic QCD approach
applied to quarkonium photoproduction are compatible and lead to the 
unambiguous predictions (\ref{TTq})-(\ref{fgl}) for the hard-scattering 
amplitudes in the NLO. 
Presumably such a factorization scheme can be generalized to all
orders of the strong coupling expansion. The study of this issue, although   
very interesting, goes beyond the scope of the present paper.  

The numerical analysis for the $\Upsilon$
photoproduction shows that in comparison to LO
the NLO corrections to the hard-scattering amplitudes 
reduce significantly the ambiguity
of the predictions related to the choice of factorization scale. 
The NLO corrections are large, at HERA energies they constitute about $\sim
40\%$ of the Born contribution at the amplitude level and are of the opposite
sign compared to the Born contribution.     

Contrary to that, we find that for the photoproduction
of $J/\psi$ in HERA kinematics
the magnitude of the NLO correction is about two times larger than
the Born contribution. Also we observe a very strong dependence of the
theoretical predictions on $\mu_F$. 
That forces us to conclude that at high energies for 
$J/\psi$ photoproduction these corrections are not under theoretical control 
if one works in  NLO, i.e. 
in the collinear factorization scheme truncated at the second order 
of the strong coupling expansion. 

Note that all steps of the dispersion method developed in this paper can be 
applied directly to NLO    
electroproduction, the process of a heavy vector meson 
being produced by a
virtual photon. 
In this case the calculations may be much more involved 
due to the presence of an additional parameter, 
namely $m/Q$. However, for
electroproduction 
the photon virtuality 
shifts the hard scale  to the higher values in
comparison to photoproduction. This gives  hope that the 
factorization approach may be reliable  for  electroproduction
of $J/\Psi$ starting from some, not too high, values of $Q$.

We show, see eqs.~(\ref{appro}), (\ref{ux}), that convolution of the NLO
hard-scattering amplitudes with GPDs produces contributions which are 
parametrically enhanced at high energies, $\sim \alpha_S^2\ln(1/\xi)$.
These contributions originate from the diagrams of partonic subprocess
with gluon 
exchange in the $t$-channel and are related to the
$s-$channel radiation of an intermediate parton in the wide interval of rapidity, away
from the photon fragmentation region. In higher orders such radiation 
generates contributions $\sim \alpha_S(\alpha_S\ln(1/\xi))^n$.
The $k_\perp$- factorization approach allows  to sum this class of
logarithmic corrections to all orders in $\alpha_S$, 
consistently with the fixed-order       
factorization of the collinear singularities \cite{Catani:1994sq}. 
It would be very interesting to perform such studies for  heavy 
vector meson production. We believe that a
resummation of these contributions will lead to  
much more stable theoretical predictions at high energies.

\vspace*{1.2cm}

\noindent
{\bf Acknowledgments}

\vspace*{0.5cm}

\noindent
We are grateful to A. Freund for providing with the code for the generalized
parton distributions.
Work of D.I. is supported in part by Alexander von Humboldt Foundation, by BMBF
and by DFG 436 grant and by RFBR 02-02-17884. 
L.~Sz. is partially supported by the French-Polish scientific
agreement Polonium.

\begin{table}[h]
\caption{Contributions to ${\cal I}_g(0)$ of 
diagrams $D_1$, $\dots$, $D_{11}$.}
\begin{center}
\begin{tabular}{lrrr}
\hline \hline
Diagram & $\epsilon^{-2}$ & $\epsilon^{-1}$ & $\epsilon^0$ \\
\hline
$D_1^{z}$ &
 $
 -\frac{3}{8}(c_1-c_2)$&$
 -\frac{3}{8}(c_1-c_2) $&$
 (c_1-c_2) \left( \frac{{\pi }^2}{16} + \frac{5 \ln (2)}{4} \right)
 $\\
$D_1^{add}$ &
 $
 -\frac{3}{8}(c_1-c_2) $&$
 -\frac{3}{8}(c_1-c_2)$&$
 (c_1-c_2)\left( -\frac{1}{2} + \frac{{\pi }^2}{4} + \frac{7\,\ln (2)}{2} 
\right)
 $\\
$D_2^{z}$ &
 $ 0 $&$
 -\frac{1}{4} c_2 $&$
 c_2\,\left( - \frac{3}{4} + \frac{{\pi }^2}{32} + \frac{\ln (2)}{2} \right)
 $\\
$D_2^{add}$ &
 $ 0 $&$
 -\frac{3}{4} c_2 $&$
 c_2 \left( -\frac{9}{4} - \frac{3 {\pi }^2}{32} + \frac{13 \ln (2)}{4} \right)
 $\\
$D_3^{z}$ &
 $ 0 $&$ 0 $&$
 c_2\left(-\frac{1}{4}- \frac{{\pi }^2}{32}+\ln(2)\right)
 $\\
$D_3^{add}$ &
 $ 0 $&$ 0 $&$
 c_2\left(\frac{1}{4}+ \frac{3{\pi }^2}{32}-\frac{\ln (2)}{4}\right)
 $\\
$D_4^{z}$ &
 $ 0 $&$
 -\frac{1}{4} c_1 $&$
 c_1 \left( - \frac{1}{4}   + \frac{{\pi }^2}{16}\right)
 $\\
$D_4^{add}$ &
 $ 0 $&$
 -\frac{1}{4} c_1 $&$
 c_1 \left( - \frac{1}{8} + \frac{{\pi }^2}{32} - \frac{\ln(2)}{2} \right)
 $\\
$D_5^{z}$ &
 $ 0 $&$ 0 $&$
 0
 $\\
$D_5^{add}$ &
 $ 0 $&$ 0 $&$
 c_1 \left( \frac{1}{8} - \frac{{\pi }^2}{32} - \frac{\ln (2)}{2} \right)
 $\\
$D_6^{z}$ &
 $ 0 $&$
 \frac{5}{8} c_1$&$
 c_1 \left( - \frac{1}{8} - \frac{\ln (2)}{4} \right)
 $\\
$D_6^{add}$ &
 $ 0 $&$
 \frac{13}{16} c_1 $&$
 c_1 \left( - \frac{1}{8} - \frac{\ln (2)}{8} \right)
 $\\
$D_7^{z}$ &
 $ 0 $&$ 0 $&$
 c_1  \frac{3}{8}
 $\\
$D_7^{add}$ &
 $ 0 $&$
 \frac{3}{16}  c_1$&$
 c_1 \left( - \frac{3}{8} + \frac{\ln (2)}{8} \right)
 $\\
$D_8^{z}$ &
 $ 0 $&$
 -\frac{1}{4} c_2  $&$
 c_2 \left( - \frac{1}{4}  + \frac{{\pi }^2}{16} - \frac{\ln (2)}{2} \right)
 $\\
$D_8^{add}$ &
 $ 0 $&$
 -\frac{1}{4} c_2  $&$
 c_2\,\left( \frac{1}{4} -\frac{{\pi }^2}{16} - \frac{3\,\ln (2)}{4} \right)
 $\\
$D_9^{z}$ &
 $ 0 $&$ 0 $&$
 0
 $\\
$D_9^{add}$ &
 $ 0 $&$ 0 $&$
 - c_2 \frac{\ln (2)}{4}
 $\\
$D_{10}^{z}$ &
 $
 \frac{1}{8}(c_1-c_2)  $&$
 -\frac{3}{8}(c_1-c_2)  $&$
 (c_1-c_2)\left( \frac{1}{2} - \frac{ \ln (2)}{2} \right)
 $\\
$D_{10}^{add}$ &
 $
 -\frac{3}{8}(c_1-c_2)   $&$
 -\frac{3}{8}(c_1-c_2)  $&$
 -(c_1-c_2)\left( \frac{1}{2} +  \frac{3 \ln (2)}{8} \right)
 $\\
$D_{11}^{z}$ &
 $ 0 $&$ 0 $&$
 0
 $\\
$D_{11}^{add}$ &
 $ 0 $&$ 0 $&$
 -(c_1-c_2)\frac{3 \ln (2)}{8}
 $\\
\hline\hline
\end{tabular}
\end{center}
\end{table}
%
\begin{table}[h]
\caption{Contributions to ${\cal I}_g(0)$ of
 $D_{12}$, $\dots$, $D_{17}$ and the mass counterterm
diagrams. 
}
\label{tab:Di}
\begin{center}
\begin{tabular}{lrr}
\hline \hline
Diagram &  $\epsilon^{-1}$ & $\epsilon^0$ \\
\hline
$
 \left( D_{12}+C_1 \frac{\delta m}{m}+ D_{14}+D_{16} \right)
 $ &
 $
 -\frac{1}{4}(c_1-c_2) $&$
  c_1 \left( \frac{1}{2} + \frac{\ln (2)}{2}\right)
 -c_2 \left( \frac{1}{4} + \frac{\ln (2)}{2}\right)
 $\\
$
 \left(D_{13}+
 C_3 \frac{\delta m}{m} + D_{15}+D_{17} \right)
 $ &
 $ 0 $&$
 -(c_1-c_2)\frac{1}{4}
 $\\
$C_2^{z} \frac{\delta m}{m} $ &
 $
 -\frac{3}{8} c_1 $&$
 c_1 \left( \frac{1}{8} + \frac{3 \ln (2)}{4} \right)
 $\\
$C_2^{add} \frac{\delta m}{m}$ &
 $
 -\frac{9}{16} c_1 $&$
 c_1  \frac{9 \ln (2)}{8}
 $\\
$C_4^{z} \frac{\delta m}{m} $ &
 $ 0 $&$
 -c_1  \frac{3}{8}
 $\\
$C_4^{add} \frac{\delta m}{m} $ &
 $
 -\frac{3}{16} c_1 $&$
 c_1 \left( \frac{1}{4} + \frac{3 \ln (2)}{8} \right)
 $\\
\hline \hline
\end{tabular}
\end{center}
\end{table}
%


\end{document}